\documentclass[11pt,a4paper]{article}
\pdfoutput=1 %% activate this line when submit %%

\usepackage{jheppub} % for details on the use of the package, please % see the JHEP-author-manual

\usepackage{youngtab}
\usepackage{hyperref}
\usepackage{fancybox}
\usepackage{float}
\usepackage{amsfonts}
\usepackage{amsmath}
\usepackage{amsbsy}
\usepackage{graphicx}
\usepackage{amssymb}
\usepackage{amsthm}
\usepackage{bm}
\usepackage{comment}
\usepackage{epsfig}
\usepackage{latexsym}
\usepackage{pdflscape}

\usepackage{color}
\makeatletter
\def\@fpheader{\relax}
\makeatother

\DeclareSymbolFont{AMSa}{U}{msa}{m}{n}
\DeclareSymbolFont{AMSb}{U}{msb}{m}{n}
\DeclareMathSymbol{\fieldR}{\mathalpha}{AMSb}{"52}

\newcommand{\beq}{\begin{eqnarray}}
\newcommand{\eeq}{\end{eqnarray}}
\newcommand{\bea}{\begin{eqnarray}}
\newcommand{\eea}{\end{eqnarray}}
\newcommand{\be}{\begin{equation}}
\newcommand{\ee}{\end{equation}}
\newcommand{\bq}{\begin{equation}}
\newcommand{\eq}{\end{equation}}

\def\enh{e_{\hbox{\tiny NH}}}
\def\egam{e_\gamma}
\def\ec{e_c}
\def\es{e_{\hbox{\tiny S}}}
\def\cQ{\mathcal{Q}}
\def\cM{\mathcal{M}}
\def\eg{{\it e.g. }}
\def\ie{{\it i.e. }}

\def\d{\delta}
\def\6{\partial}

\def\6{\partial}

\title {Phase diagram of the charged black hole bomb system}

\author[a]{Alex Davey,}
\affiliation[a]{STAG research centre and Mathematical Sciences, University of Southampton, UK}

\author[a]{Oscar J. C. Dias,}
\author[a]{and Paul Rodgers}
\emailAdd{amd1g13@soton.ac.uk}
\emailAdd{ojcd1r13@soton.ac.uk}
\emailAdd{pwr1u17@soton.ac.uk}

\begin{document} 
\begin{abstract}
{We find the phase diagram of solutions of the charged black hole bomb system. In particular, we find the static hairy black holes of Einstein-Maxwell-Scalar theory confined in a Minkowski box. We impose boundary conditions such that the scalar field vanishes at and outside a cavity of constant radius. These hairy black holes are asymptotically flat with a scalar condensate floating above the horizon. We identify four critical scalar charges which mark significant changes in the qualitative features of the phase diagram. When they coexist, hairy black holes always have higher entropy than the Reissner-Nordström black hole with the same quasilocal mass and charge.  So hairy black holes are natural candidates for the endpoint of the superradiant/near-horizon instabilities of the black hole bomb system. We also relate hairy black holes to the boson stars of the theory. When it has a zero horizon radius limit, the hairy black hole family terminates on the boson star family. Finally, we find the Israel surface tensor of the box required to confine the scalar condensate and that it can obey suitable energy conditions.}
\end{abstract}

\maketitle
\flushbottom

%%%%%%%%%%%%%%%%%%%%%%%%%%%%%%%%%%%%%%%%%%%%%%%%%%
%%%%%%%%%%%%%%%%%%%%%%%%%%%%%%%%%%%%%%%%%%%%%%%%%%
\section{Introduction}\label{sec:intro}
%%%%%%%%%%%%%%%%%%%%%%%%%%%%%%%%%%%%%%%%%%%%%%%%%%
%%%%%%%%%%%%%%%%%%%%%%%%%%%%%%%%%%%%%%%%%%%%%%%%%%

The black hole bomb setup was designed by Zel'dovich \cite{Zeldovich:1971} and Press and Teukolsky  \cite{Press:1972zz} (see also \cite{Cardoso:2004nk}) very much in the aftermath of having understood the mathematical theory of black hole perturbations around a Kerr black hole. It emerged naturally from the fact that the wave analogue of the Penrose-Christodoulou process \cite{Penrose:1969pc,PhysRevLett.25.1596} $-$ superradiant scattering $-$ unavoidably occurs in rotating black holes with angular velocity $\Omega$. If a scalar wave with  frequency $\omega$ and azimuthal number $m$ satisfying  $\omega<m\Omega$ is trapped near the horizon by the potential of a box (for example), the multiple superradiant amplifications and reflections at the cavity lead to an instability. The wave keeps extracting energy and angular momentum from the black hole interior and these accumulate between the horizon and the cavity. Press and Teukolsky assumed that this build up of radiation pressure would raise to levels that could no longer be supported by the box and the latter would eventually break apart.
But the black hole bomb system does not necessarily need to have such a dreadful end.
Actually, more often than not, a black hole instability is a pathway to find new solutions that are stable to the original instability, have more entropy (for given energy and angular momentum) and are thus natural candidates for the endpoint or metastable states of the instability time evolution. This is certainly the case for superradiant fields trapped by the anti-de Sitter (AdS) gravitational potential \cite{Dias:2011at,Dias:2015rxy,Choptuik:2017cyd,Ishii:2018oms,Ishii:2020muv,Ishii:2021xmn} or massive fields in asymptotically  flat black holes  \cite{Herdeiro:2014goa}. So we can expect the same in the original black hole bomb system. 
%We just need to certify that the stress tensor of the cavity can handle the interior radiation pressure while satisfying the required physical energy conditions \cite{Wald:106274}.

Motivated by these considerations, we would like to find the full phase diagram of solutions that can exist in the original black hole (BH) bomb system. By this we mean  to find all possible stationary solutions of the theory with boundary conditions that confine the scalar field inside the box. These would be the non-linear version of the floating solutions in equilibrium that are described in \cite{Press:1972zz}. This certainly requires solving PDEs. 
Therefore, in this paper, we start by considering a simpler system that still has a superradiant scalar field trapped inside a box but those properties can be found solving simply ODEs. This is possible if we first place a Reissner-Nordstr\"om black hole (RN BH) with chemical potential $\mu$ inside a box and then perturb it with a scalar field with charge $q$ and frequency $\omega$. As long as $\omega<q \mu$, a superradiant instability will also develop leading to the charged version of the black hole bomb system  \cite{Denardo:1973pyo}. We thus want to find the phase diagram of static solutions of this system, including those with a scalar condensate floating above the horizon. The latter hairy solutions might have higher entropy than the original RN BH for a given energy and charge where  they coexist. If so they would be a natural candidate for the endpoint of the charged black hole bomb instability, as long as we check that we can build boxes $-$ with an Israel stress tensor  \cite{Israel:1966rt,Israel404,Kuchar:1968,Barrabes:1991ng}  that satisfies the relevant energy conditions \cite{Wald:106274} $-$ that holds the internal radiation pressure without breaking apart. This will further guarantee that we can insert this boxed system in an exterior Reissner-Nordstr\"om background, as required by Birkhoff's theorem  \cite{WILTSHIRE198636,inverno:1992}.

Looking into the details of this programme we immediately find new physics. Indeed, a linear perturbation analysis of the Klein-Gordon equation in an RN BH finds that the system is not only unstable to superradiance but also to the near-horizon scalar condensation instability \cite{Dias:2018zjg}. These two instabilities are typically entangled for generic RN BHs but there are two corners of the phase space where they disentangle and reveal their origin. Indeed, extremal RN BHs with arbitrarily small horizon radius only have the superradiant instability since the near-horizon instability is suppressed as inverse powers of the horizon radius. On the opposite corner, RN BHs with a horizon radius close to the box radius only have the near horizon instability. Essentially, this instability is triggered by scalar fields that violate the near horizon $AdS_{2}$ Breitenl\"ohner-Freedman (BF) bound \cite{Breitenlohner:1982jf} of the extremal RN BH. It was originally  found by Gubser \cite{Gubser:2008px} in planar AdS backgrounds (in a study that initiated the superconductor holographic programme) but it exists in other BH backgrounds (independently of the cosmological constant sign) with an extremal (zero temperature) configuration (see \eg \cite{Dias:2010ma,Dias:2011tj}).

Analysing the setup of the black hole bomb system  leads to the observation that the theory also has horizonless solutions if we remove the RN BH but leave the scalar field inside a box with  a Maxwell field. Indeed, we can certainly perturb a Klein-Gordon field in a cavity and  the frequencies that can fit inside it will be naturally quantized and real. This suggests that, within perturbation theory, we can then back-react this linear solution to higher orders where it will source non-trivial gravitoelectric fields that are regular inside (and outside) the box  \cite{Dias:2018yey}. These are the boson stars of the theory, also known  as solitons (depending on the chosen $U(1)$ gauge; see \eg \cite{Liebling:2012fv} for a review on boson stars). This perturbative analysis is bound to capture only small mass/charge boson stars. But a full numerical nonlinear analysis can identify the whole phase space of boson stars  \cite{Dias:2021acy}. This analysis further reveals that the phase diagram of boson stars is quite elaborate with distinct boson star families. In  particular, it finds that the phase diagram of solitons depends non-trivially on a total of four critical scalar field charges. Two of them can be anticipated using simple heuristic arguments on the aforementioned superradiant and near-horizon instabilities, but the two others only emerge after solving the non-linear equations of motion.

Coming back to our main subject of study, an RN BH placed inside a box is also the starting point to discuss and find the hairy BHs  of the theory.  The latter have a scalar condensate floating above the horizon that is balanced against gravitational collapse by electric repulsion. A box with appropriate Israel junction conditions and stress tensor \cite{Israel:1966rt,Israel404,Kuchar:1968,Barrabes:1991ng} should be able to confine the scalar condensate in its interior, and it should then  be possible to place the the whole boxed system  in a background whose exterior solution is the RN solution. In the present paper, we confirm that this is indeed the case and we find the full phase diagram of static solutions of the charged black hole bomb system. It turns out that the aforementioned four critical scalar electric charges play a relevant role also in the phase diagram of hairy BHs. Indeed, this  diagram is qualitatively distinct depending on which one of the four available windows of critical charges the scalar charge $q$ falls into.
Ultimately, the reason for this dependence follows from the fact $-$ that we will establish$-$ that all hairy BHs that have a zero horizon radius limit choose to terminate on the boson star of the theory (which is fully specified once $q$ is given), in the sense that the zero entropy hairy BHs have the same (Brown-York  \cite{Brown:1992br} quasilocal) mass and charge as the boson star.  In our system, this materializes the idea that, often, small hairy BHs can be thought of as a small BH (RN or Kerr BH) placed on top of a boson star, as long as they have the same thermodynamic potential (chemical potential or angular velocity) to have the two constituents in thermodynamic equilibrium.

One of the four hairy BH families that we  find in this paper was already identified in the perturbative analysis of \cite{Dias:2018yey}. This is the only family of hairy BHs that extends to arbitrarily small mass and charge, thus making it prone to be captured by the perturbative analysis about an empty box with an electric field. But the other three families and their intricate properties cannot  be captured by such a theory because they are not perturbatively connected to the zero mass/charge solution.

Perhaps the most important property of the hairy BHs of the charged black hole bomb is that, when both coexist, they {\it always} have higher entropy than the RN BH that has the {\it same} mass and charge. Therefore, we will conclude that hairy BHs are always the preferred thermodynamic phase of the theory in the microcanonical ensemble.  

Very much like black holes confined in a box can be the a starting point to discuss certain aspects of black hole thermodynamics  \cite{Hawking:1976de,Gibbons:1976pt,Hawking:1979ig,Page:1981,Hawking:1982dh,Braden:1990hw,Andrade:2015gja}  they should also be useful to understand generic superradiant systems where distinct (including perhaps some astrophysical) potential barriers confine fields \cite{Press:1972zz}.  These two are related since the hairy solutions describe non-linear systems where the central solution is in thermodynamic equilibrium with the floating scalar radiation.
In particular, we can expect that  hairy solutions of the charged black hole bomb provide a  toy model with {\it some} universal features for the phase diagram of other confined unstable  systems. Actually, we find that the present phase diagram shares many common features with the phase diagram of superradiant hairy blacks holes in global anti-de Sitter  \cite{Basu:2010uz,Bhattacharyya:2010yg,Gentle:2011kv,Dias:2011tj,Arias:2016aig,Markeviciute:2016ivy,Markeviciute:2018cqs,Dias:2016pma}.

The plan of our manuscript is as follows. In section~\ref{sec:summary} we summarize in two figures the main properties of the phase diagram of hairy black holes and boson stars.
In section~\ref{sec:theory} we formulate the exact setup of our system. The discussion only includes aspects that guarantee that our exposition is self-contained and more details can be found in \cite{Dias:2018yey}. In section~\ref{sec:phasediag} we explicitly construct the hairy black hole solutions in the four relevant windows of scalar charge that, together with the boson star study of  \cite{Dias:2021acy}, allow us to arrive to the conclusions summarized in section~\ref{sec:summary}. Finally, in section~\ref{sec:Boxstructure} we explain how data of the hairy solution inside the box can be used to find the Israel stress tensor of the cavity surface layer and be matched with the exterior Reissner-Nordst\"om  solution. 

%%%%%%%%%%%%%%%%%%%%%%%%%%%%%%%%%%%%%
\section{Summary of phase diagram of boson stars and black holes in a cavity}\label{sec:summary}
%%%%%%%%%%%%%%%%%%%%%%%%%%%%%%%%%%%%%

The Einstein$-$Maxwell$-$Klein-Gordon theory, whereby the scalar field is confined inside a box of radius $L$ in an asymptotically flat background, is fully specified once we fix the mass and charge $q$ of the scalar field. We consider massless scalar fields with dimensionless electric charge $e=q L$ (the system has a scaling symmetry that allows us to measure all physical, \ie dimensionless, quantities in units of $L$). By Birkhoff's theorem  \cite{WILTSHIRE198636,inverno:1992} \footnote{Birkhoff's theorem for Einstein-Maxwell theory states that the unique spherically symmetric solution of the Einstein-Maxwell equations with non-constant area radius function $r$ (in the gauge \eqref{fieldansatz}) is the Reissner-Nordstr\"om solution. If $r$ is constant then the theorem does not apply since one has  the Bertotti-Robinson ($AdS_2\times S^2$) solution.}, outside the cavity the hairy solutions we search for are necessarily described by the RN solution. Thus, we just need to find the hairy solutions inside the box and then confirm that the Israel junctions conditions required to confine the scalar condensate inside the cavity, while having an exterior RN solution, correspond to an Israel energy-momentum stress tensor (proportional to the extrinsic curvature jump across the box layer \cite{Israel:1966rt,Israel404,Kuchar:1968,Barrabes:1991ng}) that is physical, \ie that satisfies  relevant energy conditions  \cite{Wald:106274}. 

Since the solution outside the box is described by the RN solution, we cannot use the Arnowitt-Deser-Misner (ADM) mass $M$ and charge $Q$ \cite{Arnowitt:1962hi} to differentiate the several solutions of the theory. However, we can use the Brown-York quasilocal mass $\mathcal{M}$ and charge $\mathcal{Q}$ \cite{Brown:1992br}, computed at the box location and normalized in units of $L$, and associated phase diagram $\mathcal{Q}$-$\mathcal{M}$ to display and distinguish the solutions of the theory. These quasilocal quantities obey their own first law of thermodynamics that is used to (further) check the results. In the quasilocal phase diagram, the extremal RN 1-parameter family of BHs (with horizon inside the box) provides a natural reference to frame our discussions. 
In particular, because distinct solutions often pile-up in certain regions of the phase diagram, for clarity we will find it useful to plot  $\Delta\mathcal{M}/L$ {\it vs} $\mathcal{Q}/L$ where $\Delta\mathcal{M}=\mathcal{M}-\mathcal{M}\big|_{\rm ext\, RN}$ is the mass difference between the hairy solution and the extremal RN that has the {\it same} $\mathcal{Q}/L$.
Therefore, in this phase diagram $\mathcal{Q}$-$\Delta\mathcal{M}$, the horizontal line with $\Delta\mathcal{M}=0$ represents the extremal RN BH solution. Its horizon at $R_+$  fits inside the box of radius $L$ if $R_+\leq 1$ (which corresponds to $\mathcal{Q}/L\leq 2^{-1/2}$) and non-extremal RN BHs exist above this line.
 However,  horizons of non-extremal RN BHs fit inside the box ($R_+\leq 1$)  if and only if  their quasilocal charges are to the left of the red dashed line that we will  display in our $\mathcal{Q}$-$\Delta\mathcal{M}$ diagrams. Actually, it turns out that this line also represents the maximal quasilocal charge that hairy solutions enclosed in the box can have.

We  find that the spectrum of hairy black holes and boson stars of the theory is qualitatively distinct depending on whether $e$ is smaller or bigger than four pivotal critical scalar field charges ---  $e_{\hbox{\tiny NH}}$, $e_\gamma$, $ \ec$ and $e_{\hbox{\tiny S}}$ ---  which obey the relations $0< e_{\hbox{\tiny NH}} < e_\gamma<  \ec < e_{\hbox{\tiny S}}$.  

\begin{figure}[th]
\centerline{
\includegraphics[width=.505\textwidth]{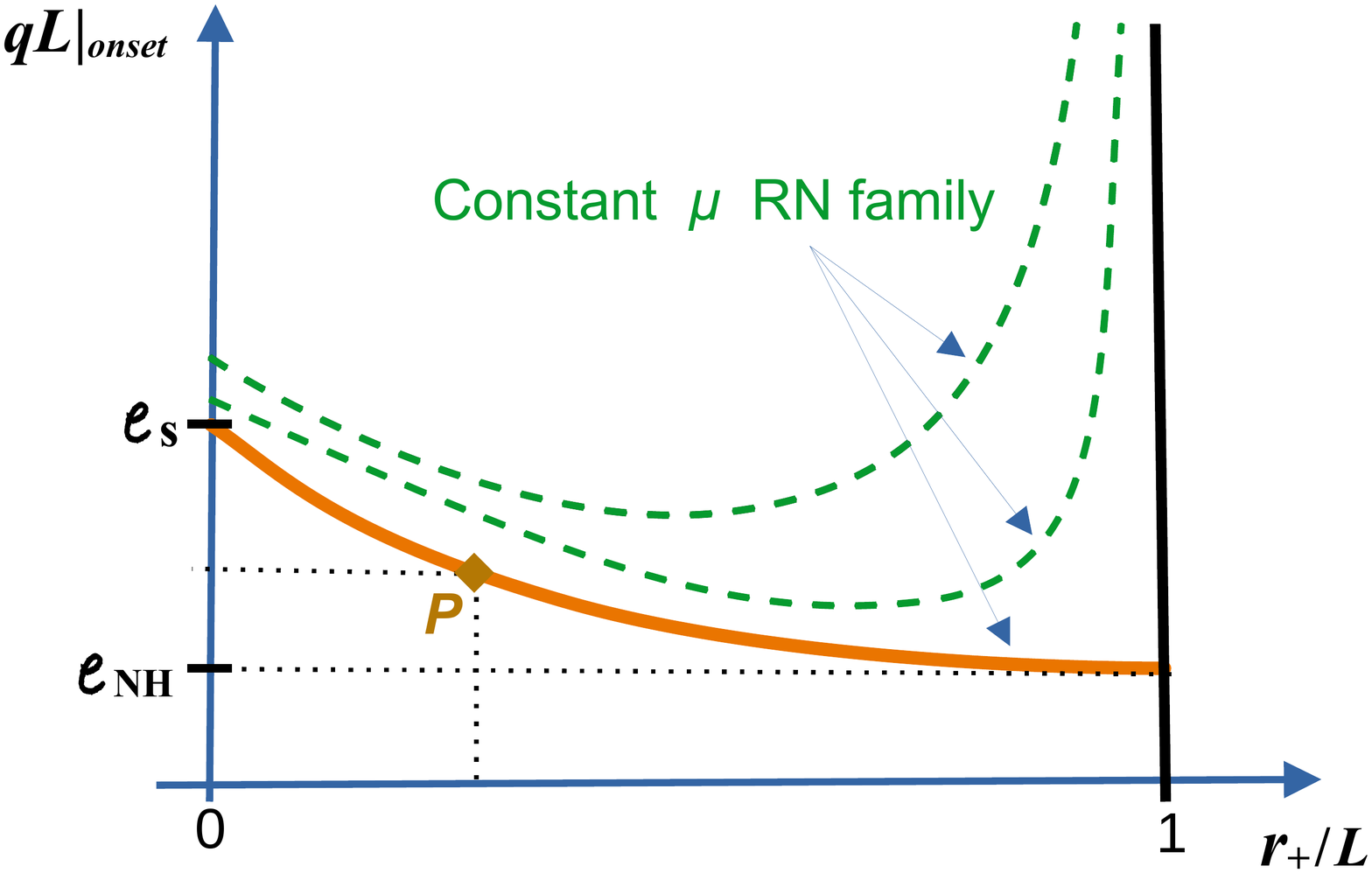}
\hspace{0.1cm}
\includegraphics[width=.5\textwidth]{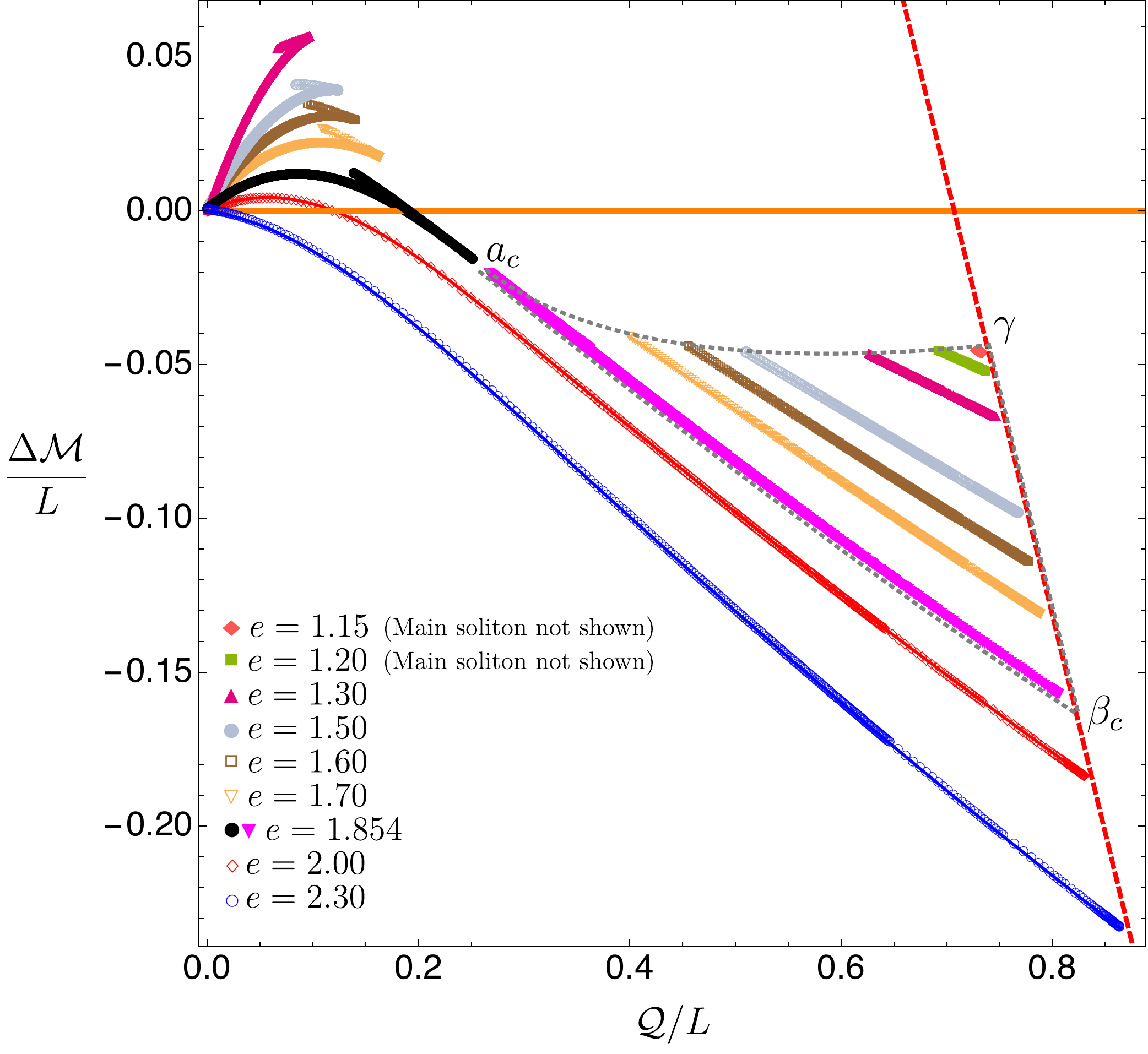}
}
\caption{{\bf Left panel:} Sketch of the scalar field electric charge $e_{\hbox{\tiny onset}} = q L |_{\hbox{\tiny onset}}$ as a function of the horizon radius $R_+ = r_+/L$ of RN BHs in a box (sketched from Fig. 2 of \cite{Dias:2018zjg}). The orange curve in the bottom $-$ that starts at $(R_+,e)=(0,e_{\hbox{\tiny S}})$ and terminates at $(R_+,e)=(1,e_{\hbox{\tiny NH}})$  $-$ describes the {\it minimal} onset charge (which occurs at extremality, $\mu=\mu_{\rm ext}$). That is, (non-)extremal RN BHs can be unstable if an only if $e$ is higher than the one identified by this orange onset curve. On the other hand, if we pick an RN family with constant $\mu<\mu_{\rm ext}$, for instability, we need $e$ to be higher than the associated green dashed line $e(R_+)|_{\hbox{\tiny const $\mu$}}$ also shown. In particular, we see that if we chose a charge in the range $e_{\hbox{\tiny NH}} \leq e \leq e_{\hbox{\tiny S}}$, RN BHs are unstable if and only if they are between the orange minimal onset curve and the horizontal line to the right of point $P$ (gold diamond). {\bf Right panel:} a survey of  boson stars for different values of $e$, for $e\geq e_\gamma$ \cite{Dias:2021acy}.  For a given $e \in [e_{\gamma},e_c[$  we have the main (perturbatively connected to $(0,0)$) and secondary (non-pertubative) solitons (which only exist in the region bounded by the auxiliary grey dashed closed curve $a_c \beta_c \gamma$). The secondary soliton curve with $e$ just above $e_\gamma$ is close to the point $\gamma$, while the soliton with $e=1.854$, just below $e_c$, is the magenta curve (very close to $a_c\beta_c$). Note that the gap in $\mathcal{Q}/L$ between the main soliton and the secondary one starts very large at $e=e_\gamma$ but then decreases and goes to zero precisely at $e=e_c$.
}
\label{FIG:Summary_onset}
\end{figure}

Two of these critical charges, $e_{\hbox{\tiny NH}}$ and $e_{\hbox{\tiny S}}$, can be identified simply studying linear scalar field perturbations about an RN BH in a box. Such RN BHs can be parametrized by the chemical potential $\mu$ and dimensionless horizon radius $R_+=r_+/L$. These parameters are constrained to the intervals $0 \leq \mu\leq \mu_{\rm ext}$ (with the upper bound being the extremal configuration) and $0<R_+\leq 1$. Boxed RN BHs become unstable $-$ the black hole bomb system $-$ if $e$ is above the instability onset charge $e_{\hbox{\tiny onset}}(\mu,R_+)$. Instead of displaying $e_{\hbox{\tiny onset}}(\mu,R_+)$, it proves to be more clear  to display the 2-dimensional plot  $e_{\hbox{\tiny onset}}(R_+)$ for fixed values of $\mu$. A sketch of this plot is given in the left panel of Fig.~\ref{FIG:Summary_onset} (which reproduces the exact results in Fig. 2 of \cite{Dias:2018zjg}). The minimal onset charge is attained for extremal RN black holes ($\mu=\mu_{\rm ext}$): this is the orange curve that connects points $(0,e_{\hbox{\tiny S}})$ and $(1,e_{\hbox{\tiny NH}})$. For completeness, in the left panel of Fig.~\ref{FIG:Summary_onset} we also  sketch the onset charge curves (green dashed)  for two non-extremal RN BHs at fixed $\mu< \mu_{\rm ext}$. Naturally, this onset charge increases as we move away from extremality. Moreover, we see that the (extremal) minimal onset curve terminates at two critical charges that, actually, can be computed analytically:
\begin{description}

\item $\bullet$ $e=e_{\hbox{\tiny NH}}=\frac{1}{2 \sqrt{2}}\sim 0.354$. This is the charge above which scalar fields can trigger a violation of the near horizon $AdS_{2}$ Breitenl\"ohner-Freedman (BF) bound \cite{Breitenlohner:1982jf,Gubser:2008px,Dias:2010ma,Dias:2011tj} of the extremal RN black hole whose horizon radius approaches, from below, the box radius. For details on how to derive this critical charge please refer to Section III.B of \cite{Dias:2018zjg}.

\item   $\bullet$ $e=e_{\hbox{\tiny S}}=\frac{\pi }{\sqrt{2}}\sim 2.221$. This is the critical charge above which scalar fields can drive arbitrarily small RN BHs unstable via superradiance. For a detailed analysis that leads to this critical charge, please see Section III.A of \cite{Dias:2018yey}. 
\end{description} 

The system has two other critical charges, $e_\gamma$ and $ \ec$, that are uncovered when we do a detailed scan of the boson stars (a.k.a. solitons) of the theory. This task was completed in detail in the companion paper \cite{Dias:2021acy}. A phase diagram that summarizes the relevant properties for the present study is displayed in the right panel of Fig.~\ref{FIG:Summary_onset} \cite{Dias:2021acy}. Note that it stores in a single plot the solitons for {\it different theories}, \ie for several distinct values of $e$.
Two families of ground state boson stars (\ie with smallest energy for a given charge) where found in \cite{Dias:2021acy}. One is the {\it main or perturbative boson star family} which can be found within perturbation theory if we back-react a normal mode of a Minkowski cavity to higher orders. In the right panel of Fig.~\ref{FIG:Summary_onset}, these are the solitons that are continuously connected to $(\mathcal{Q},\Delta\mathcal{M})=(0,0)$. The other one is the {\it secondary or non-perturbative soliton family}. In Fig.~\ref{FIG:Summary_onset}, these are the solitons that exist only above a critical $\mathcal{Q}$ and terminate at the red dashed line.
The main/perturbative soliton family exists for any value of $e>0$ (the system has a symmetry that allows us to consider only $e>0$). However, the secondary/non-perturbative soliton only exists for scalar field charges that are in the window  $e_\gamma \leq e < \ec$. 
These are the solitons enclosed in the region $ a_c\beta_c \gamma$ (\ie inside the auxiliary grey dashed closed line with these vertices). They only exist  above $e_\gamma\sim 1.13$  (see point $\gamma$) and below $ \ec\simeq 1.854\pm 0.0005$ (see line $ a_c\beta_c$ just below the magenta line). Below $e_\gamma$ the non-perturbative solitons do not exist because they no longer fit inside the box. Above $ \ec$, the gap between the two soliton families ceases to exist, \ie the non-perturbative soliton merges with the perturbative soliton, and the ground  state boson  stars of the theory extend  from the origin all the away to the red dashed line (see \eg the red diamond curve with $e=2$ or the blue circle curve with $e=2.3$ in Fig.~\ref{FIG:Summary_onset}). 
Summarizing, the main/perturbative soliton has  a Chandrasekhar limit for $0<e< \ec$ but extends from the origin all the way to the red dashed line for $e \geq \ec$. On the other hand, the ground state secondary/non-perturbative soliton only exists in the window  $e_\gamma \leq e < \ec$.

In the following sections we will find the hairy black holes of the Einstein$-$Maxwell$-$Klein-Gordon theory. We will conclude that, whenever the hairy black holes have a zero horizon radius limit, they terminate on a soliton. Accordingly, the phase diagram of solutions depends on the above four critical scalar field charges. Our main findings are summarized in the phase diagram sketches of Fig.~\ref{FIG:Summary_sketch} and the properties of these phase diagrams depend on the following 5 windows of scalar charge $e$:  

\begin{figure}[th]
\centerline{
\includegraphics[width=.505\textwidth]{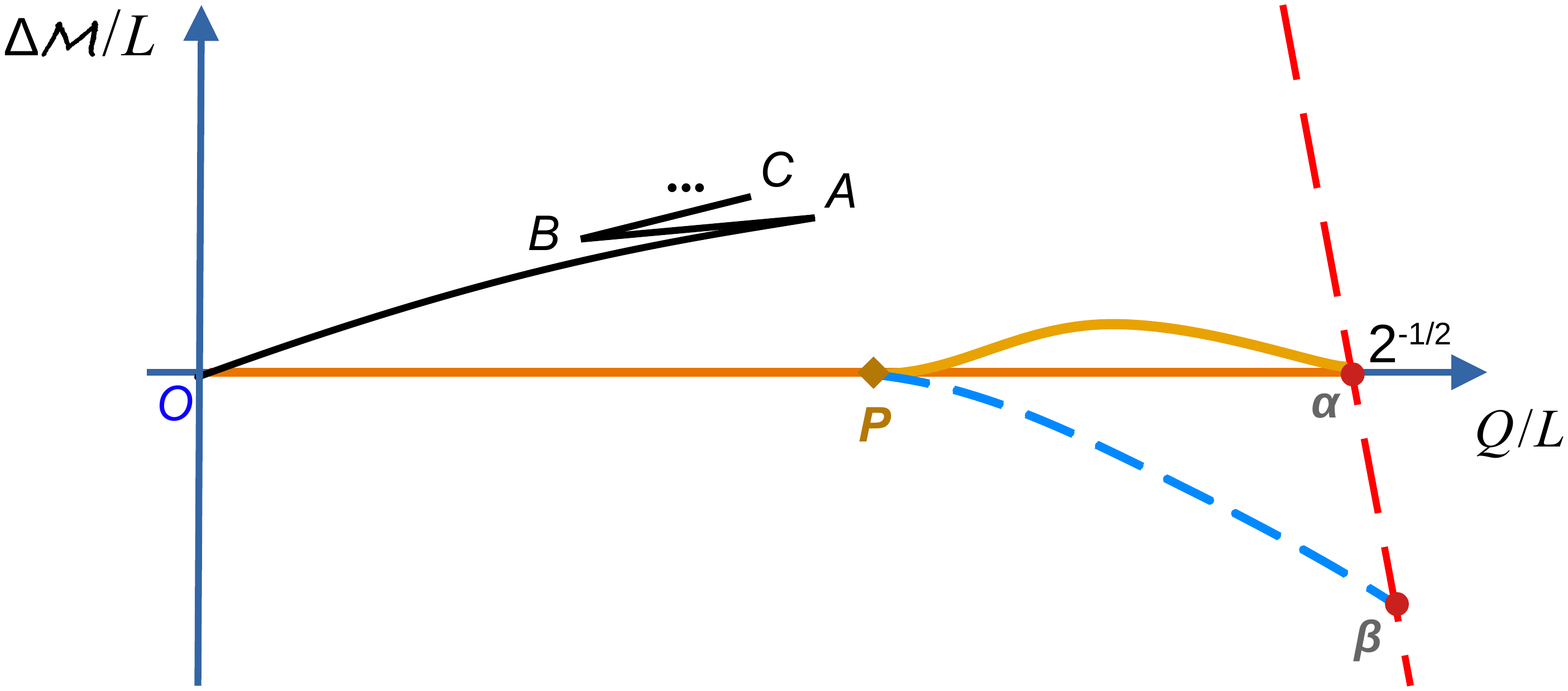}
\hspace{0.3cm}
\includegraphics[width=.50\textwidth]{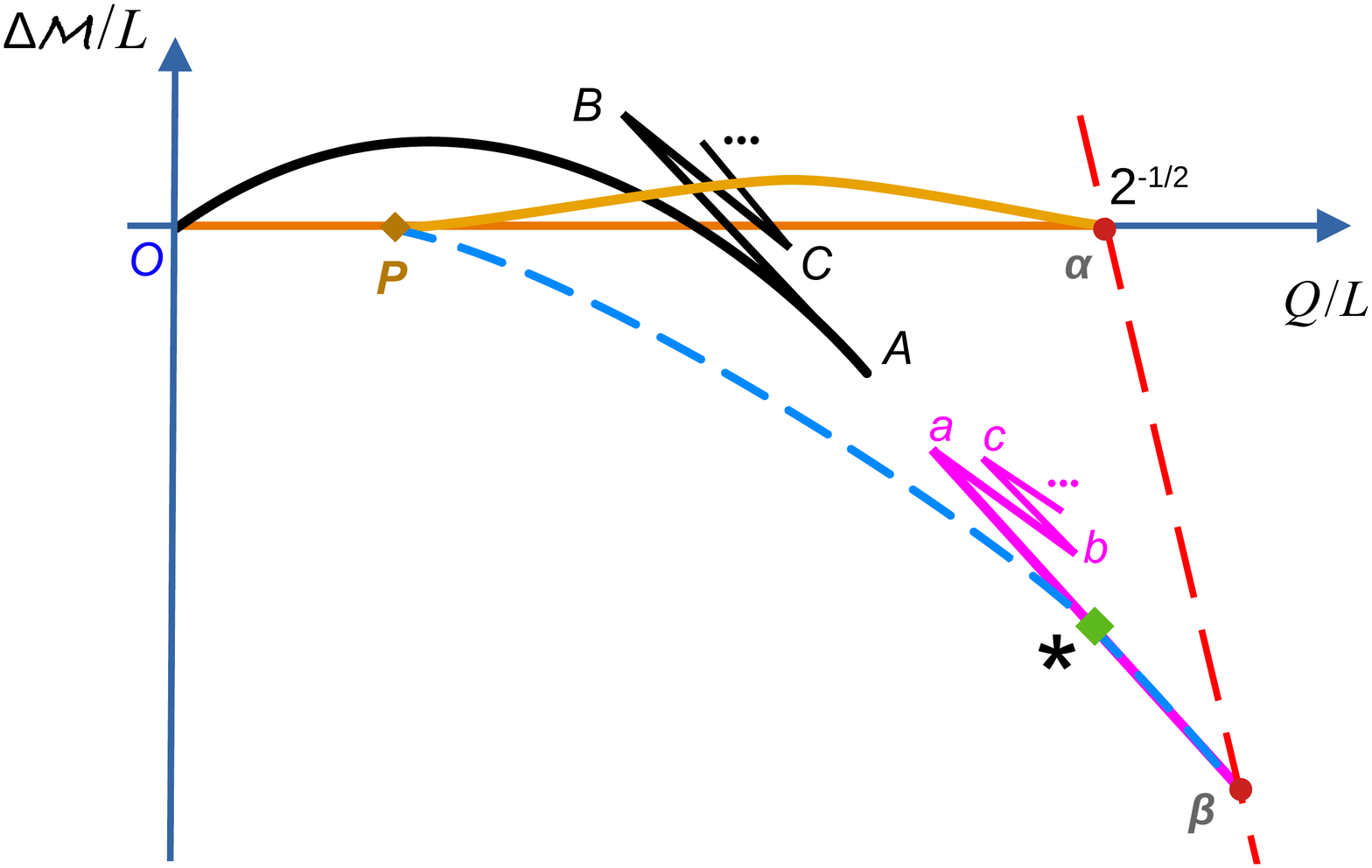}
}
\centerline{
\includegraphics[width=.505\textwidth]{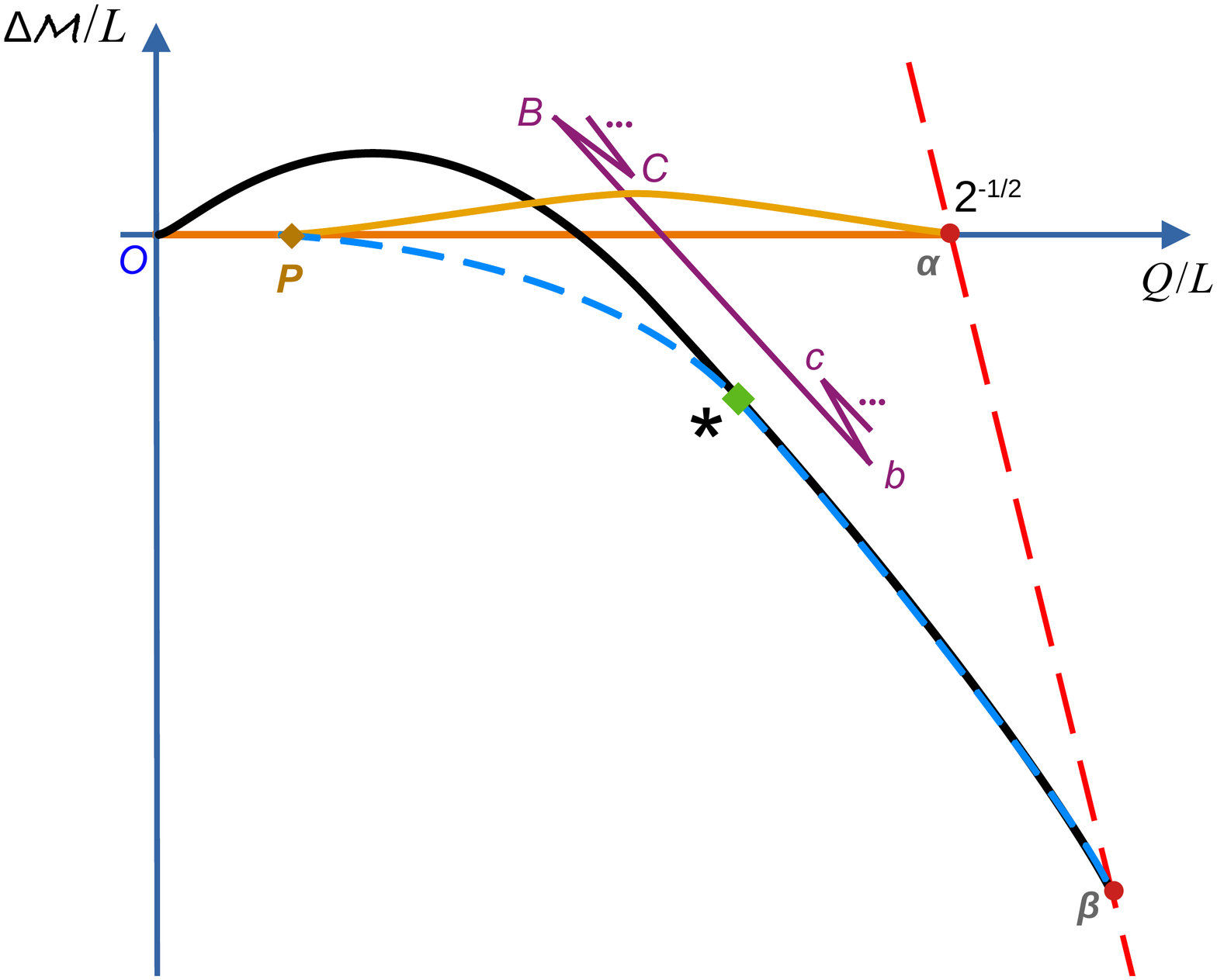}
\hspace{0.3cm}
\includegraphics[width=.505\textwidth]{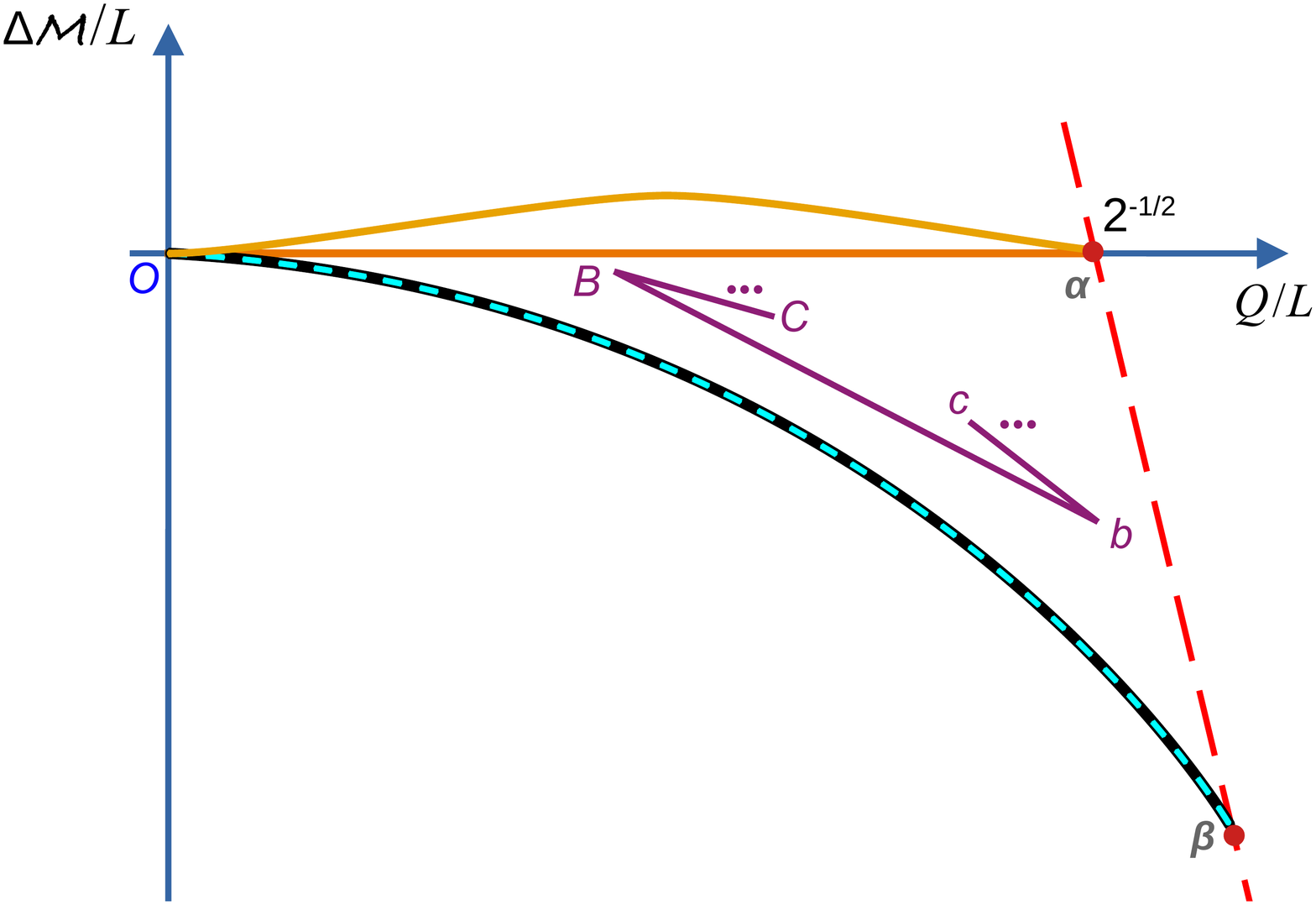} }
\caption{Sketch of the quasilocal phase diagram for black holes and solitons and  as we span relevant windows of scalar field charge $e$.  The critical charges are such that $0< e_{\hbox{\tiny NH}} < e_{\gamma}<  \ec < e_{\hbox{\tiny S}}$.  The quantity $\Delta{\cal M}$ is the quasilocal mass difference between a given solution and an extremal RN BH that has the $\textit{same}$ quasilocal charge $\mathcal{Q}/L$. Hence the orange line at $\Delta{\cal M} = 0$ describes the extremal RN solution that must have $\mathcal{Q}/L\leq 2^{-1/2}$ to fit inside the box.
The red dashed line represents the maximal quasilocal charge of solutions that can fit inside the box. It intersects the extremal RN line at  $\mathcal{Q}/L=2^{-1/2}$. Non-extremal RN BHs confined in the box have $\Delta\mathcal{M}>0$ and fill the  triangular region bounded by $\mathcal{Q}=0$ and by the orange and red dashed lines (not shown completely). The main soliton family is always given by black curves that start at $O$. The secondary soliton family is given either by magenta or purple curves. Hairy black holes exist in the region $P\alpha \beta$ enclosed by the yellow merger line $P\alpha$ (between hairy and RN BHs), the blue curve $P\beta$ where the curvature grows large and the red line $\alpha\beta$. {\bf Top-left panel:} case $e_{\hbox{\tiny NH}}<e<e_{\gamma}$. {\bf Top-right panel:} $e_{\gamma}\leq e<  \ec$. {\bf Bottom-left panel:} case $ \ec\leq e<e_{\hbox{\tiny S}}.$ {\bf Bottom-right panel:} case $e \geq e_{\hbox{\tiny S}}.$}
\label{FIG:Summary_sketch}
\end{figure} 

\begin{enumerate}

\item  $e<e_{\hbox{\tiny NH}}=\frac{1}{2 \sqrt{2}}\sim 0.354$. From  the left panel of Fig.~\ref{FIG:Summary_onset}, one concludes that RN is stable for $e<e_{\hbox{\tiny NH}}$ and thus no hairy BHs exist.  The only non-trivial solutions of the theory are the RN BH and the main/perturbative boson star which has a Chandrasekhar limit (see details in  \cite{Dias:2021acy}).

\item  $e_{\hbox{\tiny NH}}\leq e<e_{\gamma}\sim 1.13$. The phase diagram $\mathcal{Q}$-$\Delta\mathcal{M}$ of solutions for this window is sketched in the top-left panel of Fig.~\ref{FIG:Summary_sketch}. The only boson star of the theory is the main/perturbative family $OABC\cdots$ (already present for $e<e_{\hbox{\tiny NH}}$) with its Chandrasekhar limit $A$ and a series of cusps $A,B,C,\cdots$ and associated zig-zagged branches whose properties were studied in detail in \cite{Dias:2021acy}. As the left panel of Fig.~\ref{FIG:Summary_onset} indicates, RN BHs are now unstable for sufficiently large $R_+$ (\ie $\mathcal{Q}$) if sufficiently close to extremality (see diamond point $P$ and region bounded by the horizontal line to the right of $P$ and the minimal onset curve). The onset of the instability  translates into the yellow curve $P\alpha$ in the top-left panel of Fig.~\ref{FIG:Summary_sketch} and RN BHs below this onset curve $P\alpha$ (and above the extremal horizontal straight line $OP\alpha$) are unstable. Hairy BHs exist inside the region bounded by the closed curve $P\alpha\beta$. They merge with RN BHs at the onset $P\alpha$ of the instability and they extend for lower masses all the way down to the blue dashed line $P\beta$ where they terminate at {\it finite} entropy and {\it zero} temperature because the Kretschmann curvature at the horizon blows up. They are also constrained to be to the left of the red dashed line $\alpha\beta$ because the  horizon  of the hairy BH must fit inside the cavity with unit normalized radius. Note that for this window of $e$ there is no dialogue between the hairy boson star and the hairy BH families.

A representative example of a black hole bomb system with a charge $e=1$ in this window  $e_{\hbox{\tiny NH}}\leq e<e_{\gamma}$ will be discussed in detail in section~\ref{sec:phasediag1} and Fig.~\ref{FIGe1:MassCharge}.

\item $e_{\gamma}\leq e<  \ec\simeq 1.854\pm 0.0005$. The phase diagram $\mathcal{Q}$-$\Delta\mathcal{M}$ of solutions for this window is sketched in the top-right panel of Fig.~\ref{FIG:Summary_sketch}. Besides the main/perturbative family $OABC\cdots$  of boson stars (black line), the system now has the secondary/non-perturbative family $\beta abc\cdots$ of boson stars (magenta line) and there is a gap $Aa$ between these two families. Precisely at $e_\gamma$, this gap $Aa$ is the largest and the non-perturbative boson star family reduces to a single point $\beta$ on top of the red-dashed line (\ie it coincides with point $\gamma$ in right panel of Fig.~\ref{FIG:Summary_onset}). As $e$ grows beyond $e_\gamma$, the gap $Aa$ decreases and it vanishes precisely at $e= \ec$ where the two ground state soliton families merge into a single one (for details of this merger please see \cite{Dias:2021acy}).  

As before, hairy BHs exist in the region enclosed by the closed line $P\alpha\beta$ with the yellow curve $P\alpha$ being again the instability onset curve where the scalar condensate vanishes and hairy BHs merge with the RN BH family. As before, the hairy BHs also extend  for smaller masses all the way down to the singular blue dashed line $P\beta$ where the  Kretschmann curvature at the horizon diverges. But this time, this singular curve $P\beta\equiv P\star\beta$ splits into two segments. Hairy BHs terminating in the trench $P\star$ do so at {\it finite entropy} and {\it zero temperature}, as all the hairy BHs with $e< \ec$. However, hairy BHs terminating at the trench $\star\beta$ do so at {\it zero entropy} and {\it infinite temperature}. In such a way that in the $\mathcal{Q}$-$\Delta\mathcal{M}$ phase diagram, this hairy BH trench $\star \beta$ {\it coincides} with the  secondary/non-perturbative soliton (magenta line between $\star$ and $\beta$). In this sense, we can say that hairy BHs with large charge ($\mathcal{Q}\geq \mathcal{Q}_\star$) terminate on the non-perturbative soliton. This point $\star$ coincides with $\beta$ in the limit $e\to e_\gamma$ and it diverges away from $\beta$ as $e$ moves away from $e_{\gamma}$ towards $ \ec$.

A representative example of a black hole bomb system with a charge $e=1.85$ in this window   $e_{\gamma}\leq e<  \ec$ will be discussed in detail in section~\ref{sec:phasediag2} and Figs.~\ref{FIGe1.85:MassCharge}$-$\ref{FIGe1.85:entropy}.

\item  $ \ec\leq e<e_{\hbox{\tiny S}}=\frac{\pi }{\sqrt{2}}\sim 2.221$.
The phase diagram $\mathcal{Q}$-$\Delta\mathcal{M}$ of solutions for this window is sketched in the bottom-left panel of Fig.~\ref{FIG:Summary_sketch}. At $e= \ec$ the perturbative and non-perturbative boson star families merge and for $e\geq  \ec$ the boson star ground state is always the  perturbative family  $O\beta$ (black curve) that extends from the origin to the red dashed line. There is also a secondary family of boson stars $\cdots CBbc\cdots$ (purple curve) but it is not the ground state family and it plays no role in the description of hairy BHs. Therefore we do not discuss it further (see \cite{Dias:2021acy} for details).

Hairy BHs exist inside the closed line $P\alpha\beta$. Again, the yellow curve $P\alpha$ is the instability onset curve where hairy BHs merge with the RN BH family. The hairy BHs extend  for smaller masses all the way down to the singular blue dashed line $P\star\beta$ where the  Kretschmann curvature at the horizon diverges. Hairy BHs terminating in the trench $P\star$ do so at {\it finite entropy} and {\it zero temperature}, while hairy BHs terminating at the trench $\star\beta$ do so at {\it zero entropy} and {\it infinite temperature}. In the $\mathcal{Q}$-$\Delta\mathcal{M}$ phase diagram, this hairy BH trench $\star \beta$ {\it coincides} with the perturbative soliton (black line between $\star$ and $\beta$). In this sense, hairy BHs with large charge ($\mathcal{Q}\geq \mathcal{Q}_\star$) terminate on the perturbative soliton. As $e$ increases  above $ \ec$, points $P$ and $\star$  move to the left of the phase diagram, \ie to lower values of $\mathcal{Q}$ and they approach the origin $O$ as $e\to e_{\hbox{\tiny S}}$.

A representative example of a black hole bomb system with a charge $e=2$ in this window   $ \ec\leq e< e_{\hbox{\tiny S}}$ will be discussed in detail in section~\ref{sec:phasediag3} and Figs.~\ref{FIGe2:MassCharge}$-$\ref{FIGe2:entropy}.

\item  $e \geq e_{\hbox{\tiny S}}$. The phase diagram $\mathcal{Q}$-$\Delta\mathcal{M}$ of solutions for this window is sketched in the bottom-right panel of Fig.~\ref{FIG:Summary_sketch}. 
Precisely at $e_{\hbox{\tiny S}}$, the  slope $\mathrm{d}\Delta\mathcal{M}/\mathrm{d}\mathcal{Q}$ of the main/ perturbative boson  star family  (black curve $O\beta$) vanishes at the origin and for $e\geq e_{\hbox{\tiny S}}$, perturbative boson stars always have $\Delta\mathcal{M}<0$. Not less importantly,  at and above $e_{\hbox{\tiny S}}$, all extremal RN BHs are unstable, \ie point $P$ seen in the plots for $e<e_{\hbox{\tiny S}}$ hits the origin $O$. Consequently, hairy BHs now exist for all values of $\mathcal{Q}$ (that can fit inside the cavity), \ie in the 2-dimensional region with boundary  $O\alpha\beta$. And, for any  $\mathcal{Q}$, hairy BHs bifurcate from RN at the instability onset $O\alpha$ (yellow curve) and extend for smaller masses till they terminate $-$ with zero entropy, and divergent temperature and Kretschmann curvature $-$ along a curve (blue dashed line) that {\it coincides} with the boson star curve $O\beta$ (black curve). 

A representative example of a black hole bomb system with a charge $e=2.3$ in this window   $e\geq e_{\hbox{\tiny S}}$ will be discussed in detail in section~\ref{sec:phasediag4} and Figs.~\ref{FIGe2.3:MassCharge}$-$\ref{FIGe2.3:entropy}.

\end{enumerate}

Independently of $e$, a universal property of hairy BHs is that, when they coexist with boxed RN BHs, they always have higher entropy than the boxed RN BH with the same quasilocal mass  and charge. That is to say, in the phase space region where they exist, hairy BHs are always the dominant phase in microcanonical ensemble.  Moreover, hairy BHs are stable  to the superradiant mode that drives the boxed RN BH unstable. It follows from these observations and the second law of thermodynamics that the endpoint of the superradiant/near horizon instability of the boxed RN  BH, when we do a time evolution at constant mass and charge, should be a hairy BH. It would be interesting to confirm this doing time evolutions along the lines of those done in \cite{Sanchis-Gual:2015lje,Sanchis-Gual:2016tcm,Sanchis-Gual:2016ros}.    

The present work can be seen as the final study of a sequel of works on the charged black hole bomb system. Ref. \cite{Dias:2018zjg} started by studying the linear superradiant and near-horizon instabilities of the boxed RN BH. This identified the zero-mode and growth rates  of these instabilities. The hairy boson stars and hairy BHs were found within perturbation theory in  \cite{Dias:2018yey}. As expected, this perturbative analysis is valid only for small condensate amplitudes and small horizon radius and thus it is able to capture only small energy/charge hairy solutions. Therefore, for the solitons, the perturbative analysis can capture the main or perturbative boson star family at small mass/charge. But it misses: 1) the existence of the Chandrasekhar limit and cusps of this family, 2) the existence of the secondary or non-perturbative boson star family, and 3) it misses the existence of two important critical charges $e_\gamma$ and $ \ec$ where the non-perturbative soliton starts existing and merges with the perturbative family. These properties were only identified once the Einstein$-$Maxwell$-$Klein-Gordon equation was solved  fully non-linearly in  \cite{Dias:2021acy}. On the other hand, the perturbative analysis of  \cite{Dias:2018yey} also finds the hairy BHs that, for $e\geq e_{\hbox{\tiny S}}$, are perturbatively connected to a Minkowski spacetime with a  cavity. By construction, these perturbative hairy BHs reduce, in the zero horizon radius limit,  to the boson star of the theory. However, the perturbative analysis of  \cite{Dias:2018yey} says nothing about the hairy BHs of the theory when $e< e_{\hbox{\tiny S}}$. In the present manuscript, we fill this gap.

In the introduction we already mentioned that the  potential barrier that confines the scalar condensate in our boxed or black hole bomb system might be a good toy model for other systems with potential barriers that provide confinement. In particular, we find that the phase diagram  of hairy boson stars and  BHs in the black hole bomb system is qualitatively similar to the one found for asymptotically anti-de Sitter solitons \cite{Basu:2010uz,Bhattacharyya:2010yg,Gentle:2011kv,Dias:2011tj,Arias:2016aig,Markeviciute:2016ivy,Markeviciute:2018cqs,Dias:2016pma}. In this latter case, the AdS boundary conditions act as a natural gravitational box with radius inversely proportional to the cosmological length that provides bound states. In this sense, our work also complements and completes previous AdS studies since the existence range of the secondary/non-perturbative boson star family, its merger with the main/perturbative soliton at $e= \ec$, and the fact that hairy BHs can also terminate on this soliton family for $e_\gamma \leq e< \ec$ was not established in detail in  \cite{Basu:2010uz,Bhattacharyya:2010yg,Gentle:2011kv,Dias:2011tj,Arias:2016aig}. 

%%%%%%%%%%%%%%%%%%%%%%%%%%%%%%%%%%%%%%%%%%%%%%%%%%
%%%%%%%%%%%%%%%%%%%%%%%%%%%%%%%%%%%%%%%%%%%%%%%%%%
\section{Setting up the black hole bomb boundary  value problem}\label{sec:theory}
%%%%%%%%%%%%%%%%%%%%%%%%%%%%%%%%%%%%%%%%%%%%%%%%%%
%%%%%%%%%%%%%%%%%%%%%%%%%%%%%%%%%%%%%%%%%%%%%%%%%%

The setup of our problem was already discussed in the perturbative analysis of the problem in \cite{Dias:2018yey}. Here, to have a self-contained exposition, we discuss only the key aspects needed to formulate the problem and the strategy to  compute physical quantities without ambiguities. We ask the reader to see \cite{Dias:2018yey} for  details.

%%%%%%%%%%%%%%%%%%%%%
\subsection{Einstein-Maxwell gravity with a confined scalar field}

We consider the action for Einstein$-$Maxwell$-$Klein-Gordon gravity:
\begin{align}\label{action}
S=\frac{1}{16 \pi G_N}\int{\mathrm d^4 x\sqrt{g}\left({\cal R}-\frac 1 2F_{\mu\nu}F^{\mu\nu}-2D_{\mu}\phi(D^{\mu}\phi)^{\dagger}+ m^{2} \phi \phi^{\dagger}\right)},
\end{align}
where ${\cal R}$ is the Ricci scalar,  $A$ is the Maxwell gauge potential, $F=\mathrm d A$,   $D_{\mu}=\nabla_{\mu}-i q A_{\mu}$ is the gauge covariant derivative of the system, and $\phi$ is a complex scalar field with mass $m$ and charge $q$. We consider only massless scalar fields, although it is certainly possible to extend our analysis to $m>0$. We fix Newton's constant $G_N \equiv 1$. 

We want to find the black hole solutions of \eqref{action} that are static, spherically symmetric and asymptotically flat. $U(1)$ gauge transformations  allow us to work with a real scalar field and a gauge potential that vanishes at the horizon. Further choosing the Schwarzschild gauge, an {\it ansatz} with the desired symmetries is then
\begin{align}\label{fieldansatz}
\mathrm d s^2=-f(r)\mathrm d t^2+g(r)\mathrm d r^2+r^2\mathrm d \Omega_2^2, \qquad A_{\mu}\mathrm d x^{\mu}=A(r)\mathrm d t, \qquad \phi=\phi^{\dagger}=\phi(r),
\end{align} 
with $\mathrm d \Omega_2^2$ being the metric for the unit 2-sphere (expressed in terms of the polar and azimuthal angles $x=\cos\theta$ and $\varphi$).  
The scalar field is forced to be confined inside a box of radius $L$. The system then has a scaling symmetry that  allows us to normalize coordinates ($T=t/L, R=r/L$) and thermodynamic quantities in units of $L$, and place the box at radius $R=1$ \cite{Dias:2018yey}. 

In these conditions, the equations of motion that follow from \eqref{action} can be found in~\cite{Dias:2018yey}. These are a set of three ordinary differential equations for the fields $f(R),\ A(R)$ and $\phi(R)$, and an algebraic equation that expresses $g(R)$ as a function of the other 3 fields and their first derivatives. 
Well-posedness of the boundary value problem requires that we give boundary conditions at the horizon and asymptotic boundary of our spacetime. Additionally, we must specify Israel junction conditions at the timelike hypersurface $\Sigma=R-1=0$ where the box is located. Again, our hairy BHs have vanishing scalar field at and outside this box, $\phi(R\geq 1)=0$.

The horizon, with radius $R=R_+=\frac{r_+}{L}$ is the locus $f(R_+)=0$. We have three second order ODEs and thus there are six free parameters when we do a Taylor expansion about the horizon. Regularity demands Dirichlet boundary conditions that set three of these parameters to zero. We are thus left with only three constants $f_0,A_0,\phi_0$ (say) such that the regular fields have the Taylor expansion around the origin:
\begin{align}\label{BChorizon}
\begin{split}
& f(R_+)=f_0(R-R_+)+\mathcal O((R-R_+)^2),\\
& A(R_+)=A_0(R-R_+)+\mathcal O((R-R_+)^2),\qquad \phi(R_+)=\phi_0+\mathcal O((R-R_+)^2).
\end{split}
\end{align}
Consider now the asymptotic boundary of our spacetime, $R\to\infty$. The scalar field vanishes outside the box, $\phi=0$, and the equations of motion have the solution: $f^{out}(R)=c_f-\frac{M_0}{R}+\frac{\rho^2}{2 R^2},$ $A_t^{out}(R)=c_A+\frac{\rho}{R}$ and $g^{out}(R)=c_f/f^{out}(R)$ (onwards, the superscript~$^{out}$  represents fields outside the cavity). Here, $c_f,M_0,c_A$ and $\rho$ are arbitrary parameters, \ie  we have an asymptotically flat solution for any value of these constants. However, the theory has a second scaling symmetry  ($e=qL$)
\begin{align}\label{2scalingsym}
\{T,R,x,\varphi\}\to\{\lambda_2 T, R,x,\varphi\} , \quad \{f,g,A_t,\varphi\}\to \{\lambda_2^{-2} f,g,\lambda_2^{-1} A_t,\varphi\}, \quad \{e, R_+\}\to \{e, R_+\},
\end{align}
that we use to set $c_f=1$ so that $f |_{r\to\infty}=1$ (and $g^{out}=1/f^{out}$)~\cite{Dias:2018yey}.
Outside the box the solution to the equations of motion is then 
\begin{equation}\label{BCinfinity}
f^{out}(R)\big|_{R\geq 1}=1-\frac{M_0}{R}+\frac{\rho^2}{2 R^2}\,,\qquad A_t^{out}(R)\big|_{R\geq 1}=c_A+\frac{\rho}{R}\,,\qquad \phi^{out}(R)\big|_{R\geq1}=0\,.
\end{equation}
As required by Birkhoff's theorem for the Einstein-Maxwell theory \cite{WILTSHIRE198636,inverno:1992}, this is the Reissner-Nordstr\"om (RN) solution. The leftover free constants in \eqref{BCinfinity}, $M_0,c_A,\rho$,  will be determined only after we have the solution inside the cavity and specify the Israel junctions conditions at the latter. 

Some of the parameters in \eqref{BCinfinity} are related to the ADM  conserved charges  \cite{Arnowitt:1962hi}. Indeed, the  dimensionless ADM mass and electric charge of the system are ($G_N\equiv 1$):\footnote{
Note that the Maxwell term in action \eqref{action} is $\frac{1}{2}F^2$, not the perhaps more common $F^2$ term. It follows that the extremal RN BH satisfies the  ADM relation $M=\sqrt{2}|Q |$, instead of $M=|Q|$ that holds when the Maxwell term in the action is $F^2$. Extremal RN BHs have $\mu = \sqrt{2}$ where $\mu$ is the chemical potential of the BH, and RN BHs exist for $0< \mu \leq \sqrt{2}$.}
\begin{equation}\label{eq:ADM}
M/L=\lim_{R\to \infty}\frac{R^2 f'(R)}{2\sqrt{f(R)  g(R)}}=\frac{M_0}{2}, \qquad \qquad
Q/L=\lim_{R\to \infty}\frac{R^2 A_t'(R)}{2 f(R) g(R)}=-\frac{\rho}{2}.
\end{equation}
These ADM conserved charges measured at the asymptotic boundary include the contribution from the energy-momentum content of the cavity that confines the scalar hair. In the next subsection we discuss the properties of this box.

In these conditions, hairy BHs of the theory are a 2-parameter family of solutions that we can take to be the horizon radius $R_+$ and the value of the (interior) derivative of the scalar field at the box,  $\epsilon\equiv \phi^{\prime\:in} \big|_{R=1^-}$.

As mentioned in section \ref{sec:summary}, it follows from Birkhoff's theorem that  in the asymptotic region our solutions are necessarily described by the RN solution. Therefore, the ADM mass $M$ and charge $Q$ cannot be used to distinguish the several solutions of the theory. It is thus natural to instead use the Brown-York quasilocal mass $\mathcal{M}$ and charge $\mathcal{Q}$, measured at the box, to display our solutions in a phase diagram of the theory \cite{Brown:1992br}.
From section II.C of \cite{Dias:2018yey}, the Brown-York quasilocal mass and charge contained inside a 2-sphere with radius $R=1$ are ($G_N\equiv 1$)\footnote{The Brown-York quasilocal quantities reduce to the ADM ones  when we evaluate the former at $R\to\infty$.}
\begin{equation}\label{BYmassCharge}
{\cal M}/L= R\left( 1-\frac{1}{\sqrt{g(R)}}\right)\Big|_{R=1}\,,\qquad \qquad
{\cal Q}/L = \frac{R^2A_t'(R)}{2 \sqrt{g(R)} f(R)}\Big|_{R=1}.
\end{equation}
The thermodynamic description of our solutions is complete after defining the chemical potential, temperature and entropy: 
\begin{align}\label{furtherThermo}
\mu=A(1)-A(R_+)\,, \qquad T_H L=\lim_{R\to R_+}\frac{f'(R)}{4\pi\sqrt{f(R) g(R)}},\qquad S/L^2=\pi R_+^2,
\end{align}
where we work in the gauge $A(R_+)=0$.
These quantities must satisfy the quasilocal form of the first law of thermodynamics: 
\begin{equation}\label{1stlawBH}
\mathrm d {\cal M}=T_H \,\mathrm d S+\mu \, \mathrm d {\cal Q}, 
\end{equation}
which is used to check our solutions.

As explained before, for reference we will often compare the hairy families of solutions against extremal RN BHs. RN BHs confined in a cavity can be parametrized by the horizon radius $R_+$ and the chemical potential $\mu$, and their quasilocal mass and charge are  \cite{Dias:2018yey}
\begin{align}\label{quasilocalRN}
{\cal M}/L \big|_{RN}= 1 - \frac{\sqrt{2}(1-R_+)}{\sqrt{2-(2-\mu^2)R_+}}, \qquad \mathcal{Q} /L \big|_{RN}= \frac{\mu R_+}{\sqrt{2} \sqrt{2-(2-\mu^2)R_+}}.
\end{align}
where $0<R_+\leq 1$ (for the horizon to be confined inside the box) and $0\leq \mu\leq \mu_{\rm ext}$, with extremality reached at $\mu_{\rm ext}=\sqrt{2}$. Note that at extremality one has ${\cal M}/L = R_+$ and ${\cal Q}/L =R_+/\sqrt{2}$. On the other hand, for any $\mu$, when $R_+=1$ one has ${\cal M}/L = 1$ and ${\cal Q}/L =2^{-1/2}$. 

\subsection{Description of the box: Israel junction conditions and stress tensor}

So far, we discussed the boundary conditions at the horizon and asymptotic boundaries.
However, hairy BHs are solutions that join an interior spacetime ($R<1$; with superscript~$^{in}$) with the known RN exterior background  \eqref{BCinfinity} ($R>1$; with superscript~$^{out}$). So, in practice we simply need to find the interior solution. For that, we need to integrate our equations of motion in the domain $R\in [R_+,1]$. Therefore, we must specify appropriate conditions at the outer boundary of our integration domain, \ie at $R=1$. Next, we detail the procedures required to do this.

At and outside the box, \ie for $R\geq 1$, the scalar field must vanish. However, its derivative when approaching the cavity from inside, \ie as $R\to 1^-$, is finite (except for the trivial RN solution) and we will label this quantity $\epsilon$:\footnote{Our theory has the symmetry $\phi\to -\phi$ so we can focus our attention only on the case $\epsilon>0$.}
\begin{equation}\label{def:epsilon} 
\phi^{in} \big|_{R=1}=\phi^{out} \big|_{R=1}=0,\qquad \phi^{out}(R)=0, \qquad \phi^{\prime\:in} \big|_{R=1}\equiv \epsilon.
\end{equation}
That is, inside the box the scalar field is forced to have the Taylor expansion $\phi\big|_{R=1^-}=\epsilon (R-1)+\mathcal{O}\left((R-1)^2\right)$. 
We are forcing a jump in the derivative of the scalar field normal to the cavity  timelike hypersurface $\Sigma$. This can be done only if we further impose junction conditions at $\Sigma$ on the gravitoelectric fields as discussed next.

The box is the timelike hypersurface  $\Sigma=R-1=0$. It has outward unit normal $n_{\mu}=\partial_\mu\Sigma/|\partial \Sigma|$ ($n_\mu n^\mu=1$) and coordinates $\xi^a=(\tau, \theta, \varphi)$. An observer in the interior of $\Sigma$ measures the time $T^{\rm in}(\tau) = \tau$ and the induced line element and gauge 1-form at $\Sigma$ are
\begin{eqnarray}\label{SigmaIn}
{\mathrm d}s^2|_{\Sigma^{in}}&=&h_{ab}^{in} \,{\mathrm d}\xi^a {\mathrm d}\xi^b =- f^{in}|_{R=1}{\mathrm d}\tau^2+{\mathrm d}\Omega^2_2\,,\nonumber \\
 A_t |_{\Sigma^{in}}&=& a_{a}^{in} {\mathrm d}\xi^a = A_t^{in}|_{R=1}{\mathrm d}\tau\,,
\end{eqnarray}
where $h_{ab}^{in}$ is the induced metric in $\Sigma$ and $a_a^{in}$ is the induced gauge potential in $\Sigma$. 
Meanwhile, from the perspective of an observer outside the cavity, $\Sigma$ is parametrically described by $R=1$ and $T^{out}(\tau)=N \tau$ (so,  $N$ is a reparametrization freedom parameter) so that the induced line element and gauge 1-form for this observer are
\begin{eqnarray}\label{SigmaOut}
{\mathrm d}s^2 |_{\Sigma^{out}}&=&h_{ab}^{out} \,{\mathrm d}\xi^a {\mathrm d}\xi^b =- N^2 f^{out}|_{R=1}{\mathrm d}\tau^2+{\mathrm d}\Omega^2_2\,,\nonumber \\
 A_t |_{\Sigma^{out}}&=& a_{a}^{out} {\mathrm d}\xi^a = N A_t^{out}|_{R=1}{\mathrm d}\tau\,,
\end{eqnarray}

 Ideally, we would like to have a smooth crossing of $\Sigma$, whereby the induced gravitational  $h_{ab}$ and gauge $a_a$ fields and their normal derivatives are continuous across $\Sigma$. That is to  say, the Israel junction conditions should be obeyed \cite{Israel:1966rt,Israel404,Kuchar:1968,Barrabes:1991ng}:
\begin{subequations}\label{IsraelJunctionConditions}
\begin{align}
& a_{a}^{in}\big|_{R=1}=a_{a}^{out}\big|_{R=1}\,, \label{eq:Israeljunction1} \\
&h_{ab}^{in}\big|_{R=1}=h_{ab}^{out}\big|_{R=1}\,; \label{eq:Israeljunction2} \\
& f_{aR}^{in}\big|_{R=1}=f_{aR}^{out}\big|_{R=1}\,,\label{eq:Israeljunction3} \\
&K_{ab}^{in}\big|_{R=1}=K_{ab}^{out}\big|_{R=1}; \label{eq:Israeljunction4}
\end{align}
\end{subequations}
where $h_{ab}=g_{ab}-n_a n_b$ is the induced metric at $\Sigma$ and  $K_{ab}=h_a^{\phantom{a}\mu}h_b^{\phantom{a}\nu}\nabla_\mu n_\nu$ is the extrinsic curvature.

In the absence of the scalar condensate, we can set $N=1$ and all the junction conditions \eqref{IsraelJunctionConditions} are satisfied. However, our hairy solutions are continuous but not differentiable at $R=1$: they satisfy the conditions \eqref{eq:Israeljunction1}-\eqref{eq:Israeljunction3} but not \eqref{eq:Israeljunction4}. Since the extrinsic curvature condition is violated, our hairy solutions are singular at $\Sigma$. But this singularity simply signals the presence  of a Lanczos-Darmois-Israel surface stress tensor ${\cal S}_{ab}$ at the hypersurface layer proportional to the difference of the extrinsic curvature across  the hypersurface  \cite{Israel:1966rt,Israel404,Kuchar:1968,Barrabes:1991ng}:
\begin{align}\label{eq:inducedT}
{\cal S}_{ab}=-\frac{1}{8 \pi}\Big([K_{ab}]-[K] h_{ab}\Big),
\end{align}
where $K$ is the trace of the extrinsic curvature and $[K_{ab}]\equiv K_{ab}^{out}\big|_{R=1}-K_{ab}^{in}\big|_{R=1}$. This surface tensor is the pull-back of the energy-momentum tensor integrated over a small region around the hypersurface $\Sigma$, \ie it is obtained integrating the appropriate Gauss-Codazzi equation  \cite{Israel:1966rt,Israel404,Kuchar:1968,Barrabes:1991ng,MTW:1973}. It is also given by the jump across $\Sigma$ of the Brown-York surface tensor \cite{Brown:1992br} (see also discussion in \cite{Dias:2018yey}).
Essentially, \eqref{eq:inducedT} describes the energy-momentum tensor of the cavity (the ``internal structure" of the box) that is needed to confine the scalar field. 
Since the two Maxwell junction conditions \eqref{eq:Israeljunction1}-\eqref{eq:Israeljunction2} are satisfied,  our hairy solutions  will have a surface layer with no electric charge.

The strategy to find the hairy BHs of the theory can now be outlined. The hairy solution inside the box is found integrating numerically the coupled system of three ODEs in the domain $R\in [R_+,1]$. This is done while imposing the boundary conditions \eqref{BChorizon} at the horizon and, at the box, we impose $\phi(1^{-})=0$ and use the scaling symmetry \eqref{2scalingsym} to set $f(1^{-})=1$. After this task, we can compute the quasilocal charges \eqref{BYmassCharge} and the other thermodynamic quantities \eqref{furtherThermo} of the system. To find the solution in the full domain $R\in [R_+,\infty]$ we impose the three junction conditions \eqref{eq:Israeljunction1}-\eqref{eq:Israeljunction3} at the box to match the interior solution with the outer solution  (described by the RN solution \eqref{BCinfinity}). This operation finds the parameters $M_0,C_A,\rho$ in  \eqref{BCinfinity} as a function of the reparametrization freedom parameter  $N$  introduced in \eqref{SigmaOut}. 
The Israel stress tensor ${\cal S}_a^b$ is just a function of $N$ and, if $\phi^{in}\neq 0$, we cannot  choose $N$ to kill all the components of ${\cal S}_a^b$ (there are two non-vanishing components, ${\cal S}_t^t$ and ${\cal S}_\theta^\theta={\cal S}_\varphi^\varphi$). In this process, we have arbitrary freedom to choose $N$. This simply reflects the freedom we have to select the energy-momentum content of the box needed to confine the scalar condensate. We should however, make a selection that respects some or all the energy conditions \cite{Wald:106274}. Once this choice is made, we can finally compute the ADM mass and charge \eqref{eq:ADM} of the hairy solution which, necessarily, includes the contribution from the box.

%%%%%%%%%%%%%%%%%%%%%%%%%%%%%%%%%%%%%%%%%%%%%%%%%
\subsection{Numerical scheme}\label{subsec:method}
%%%%%%%%%%%%%%%%%%%%%%%%%%%%%%%%%%%%%%%%%%%%%%%%%

The hairy BHs we seek are a 2-parameter family of solutions, that we can take to be the horizon radius $R_+$ and the scalar  field amplitude $\epsilon\equiv \phi'(R=1)$ as defined in \eqref{def:epsilon}. In practice, we set up a two dimensional discrete grid  where we march our solutions along these two parameters. In other words, we give $R_+$ and $\epsilon$ as inputs of our numerical code, and in the end of the day we read the horizon parameters $f_0,A_0,\phi_0$ in \eqref{BChorizon}, and the values of the derivative of $f$ and the value of $A$ and its derivative at the box, $R=1$.
Typically, we start near the merger with the RN BH where a good seed (approximation) for the Newton-Raphson method we use is the RN BH itself but with a small perturbation that also excites the scalar field.  

We find it convenient to introduce  a new radial coordinate
\begin{equation}
  y = \frac{R - R_{+}}{1 - R_{+}}
\end{equation}
so that the event horizon is at $y = 0$ and the box at $y = 1$. The equations of motion now depend explicitly on $R_+$. 

Moreover, we also find useful to redefine the fields as
\begin{equation}
  f = y \, q_{1}(y), \qquad A = y \, q_{2}(y), \qquad \phi = -(1-y) \, q_{3}(y)
\end{equation}
which automatically imposes the boundary conditions~\eqref{BChorizon} at the horizon.
We now use the scaling symmetry \eqref{2scalingsym} to set $f(1^{-})=1$ and introduce the scalar amplitude \eqref{def:epsilon} at the box. This can be done through imposing the  boundary conditions 
\begin{equation}
  q_{1}(1) = 1, \qquad\qquad q_{3}(1) = \epsilon.
\end{equation}
The other boundary conditions for $q_{1,2,3}$ are derived boundary conditions in the sense that they follow directly from evaluating the equations of motion at the boundaries $y=0$ and $y=1$ \cite{Dias:2015nua}.
Under these conditions, the hairy BHs are described by smooth functions $q_{1,2,3}$ that we search for numerically.

 To solve numerically our boundary value problem, we use a standard Newton-Raphson algorithm and discretise the coupled system of three ODEs using pseudospectral collocation (with Chebyshev-Gauss-Lobatto nodes). The resulting algebraic linear systems are solved by LU decomposition. These numerical methods are described in detail in the review \cite{Dias:2015nua}. Since we are using pseudospectral collocation, and our functions are analytic, our results must have exponential convergence with the number of grid points. We check this is indeed the case and the thermodynamic quantities that we display have, typically,  8 decimal digit accuracy. We further use the quasilocal first law \eqref{1stlawBH} (typically, obeyed within an error smaller than $10^{-3}\%$) to check our solutions.

%%%%%%%%%%%%%%%%%%%%%%%%%%%%%%%%%%%%%%%%%%%%%%%%%%
%%%%%%%%%%%%%%%%%%%%%%%%%%%%%%%%%%%%%%%%%%%%%%%%%%
\section{Phase diagram of the charged black hole bomb system}\label{sec:phasediag}
%%%%%%%%%%%%%%%%%%%%%%%%%%%%%%%%%%%%%%%%%%%%%%%%%%
%%%%%%%%%%%%%%%%%%%%%%%%%%%%%%%%%%%%%%%%%%%%%%%%%%

The properties of the hairy black holes of the charged black hole bomb system are closely linked to the superradiant/near-horizon instability of RN black holes\footnote{For a general RN black hole the superradiant and near-horizon instabilities are entangled, so we will simply refer to an RN instability, regardless of the origin.}, so we first highlight some features of this instability to provide the context needed to interpret the hairy black hole phase diagram (see~\cite{Dias:2018zjg} for details). 

In the left panel of Fig.~\ref{FIG:Summary_onset} we sketch (from \cite{Dias:2018zjg}) the scalar field instability onset charge $e_{\rm onset} = q_{\rm onset}L$ as a function of $R_+$ for three families of RN black holes with constant chemical potential $\mu$, \ie the minimum scalar charge needed for a black hole with $(R_{+}, \mu)$ to be unstable. We can see that the near-horizon charge $\enh$ is a lower bound for an RN instability, \ie caged RN BHs are always stable when $e < \enh$. Correspondingly, we also find no hairy black holes when $e < \enh$. At the other end, all extremal RN black holes, no matter their $R_+$, are unstable at or above the superradiant charge $\es$. In between these two critical charges $\enh < e < \es$ we have a window of horizon radii $R_{+} \in [R_{+}|_{P}, 1]$ within which sufficiently near-extremal RN black holes are unstable. In equivalent words, for $\enh < e < \es$,  extremal RN BHs are unstable for quasilocal charges in the range    $\mathcal{Q}/L \in [(\mathcal{Q}/L)|_{P}, 2^{-\frac{1}{2}}]$.  In the upcoming phase diagrams we will indicate the instability onset curve  as a yellow curve $P\alpha$ and, when applicable, we will also use a gold diamond point $P$ to identify the minimum charge for instability. The onset curve starts at point $P$ where it intersects the extremal RN curve and terminates at  point $\alpha$ with $\mathcal{Q}/L =2^{-\frac{1}{2}}$ (\ie $R_+=1$) where it intersects again the  extremal RN curve.

In all our plots  $\cal{M}$ and $ \mathcal{Q}$ are the Brown-York quasilocal mass and charge \eqref{BYmassCharge} of the system, measured at the location of the box. Different solutions tend to pile-up in certain regions of the  $\mathcal{Q}$-$\mathcal{M}$ diagram (as illustrated in Fig.~\ref{FIGe2.3:MassCharge}). Thus, the distinction between different solutions becomes clearer if we use instead $\Delta \mathcal{M}=\mathcal{M}-\mathcal{M}\big|_{\rm ext\, RN}$ which is the quasilocal mass difference of a hairy solution with an extremal RN that has the \textit{same} quasilocal charge $ \mathcal{Q}$. Thus, in our $\mathcal{Q}$-$\Delta\mathcal{M}$ plots, the horizontal orange line $O\alpha$ with $\Delta \mathcal{M}=0$ describes the extremal RN BH. It is constrained to have $\mathcal{Q}/L\leq 2^{-\frac{1}{2}}$ (point $\alpha$) in  order to fit inside the box (this extremal line will be represented by a  dark red line in the 3-dimensional plots $\mathcal{Q}$-$\Delta\mathcal{M}$-$S$).  

From the RN quasilocal charges \eqref{quasilocalRN}, in the quasilocal $\mathcal{Q}-\mathcal{M}$ plot, the region that represents RN BHs with horizon radius inside the box is the triangular surface bounded by the  lines $\mathcal{Q}=0$,  $\mathcal{M}=\sqrt{2}\mathcal{Q}$ and 
$\mathcal{M}/L=1$. 
Therefore, in the $\mathcal{Q}-\Delta\mathcal{M}$ plane,  non-extremal RN BHs with $R_+\leq 1$ are those inside the triangular region bounded by $\mathcal{Q}=0$, $\Delta\mathcal{M}=0$ and $\Delta\mathcal{M}/L=1-\sqrt{2}\mathcal{Q}/L$. The boundary $\mathcal{Q}=0$ describes the Schwarzschild limit, $\Delta\mathcal{M}=0$ is the extremal RN boundary and the   latter curve is
\begin{equation}\label{redDashed}
\left(\mathcal{Q}/L,\Delta\mathcal{M}/L\right)=\left(L^{-1} \mathcal{Q}\big|_{\rm ext\,RN},1-L^{-1}\mathcal{M}\big|_{\rm ext\,RN} \right)=\left( \frac{R_+}{\sqrt{2}}, 1-R_+\right)
\end{equation}
where $\mathcal{M}\big|_{\rm ext\,RN}$ and $\mathcal{Q}\big|_{\rm ext\,RN}$ are given by 
 \eqref{quasilocalRN} with $\mu=\mu_{\rm ext}=\sqrt{2}$.
 The red dashed line in the forthcoming $\mathcal{Q}-\Delta\mathcal{M}$ plots is this parametric curve \eqref{redDashed} with $R_+$ {\it allowed to take also values above 1}.   
Indeed, it turns out that the most charged solutions we  find approach this dashed red line  \eqref{redDashed} (in the limit where scalar condensate amplitude approaches infinity).  In this sense, for a given quasilocal mass (smaller than 1), this red dashed line  \eqref{redDashed} represents the maximal quasilocal charge that confined solutions can have, with or without scalar hair.

As discussed in our summary of results (section~\ref{sec:summary}), the charged black hole bomb system has a total of four critical scalar field charges. Besides $e_{\hbox{\tiny NH}}=\frac{1}{2 \sqrt{2}}\sim 0.354$ and  $e_{\hbox{\tiny S}}=\frac{\pi }{\sqrt{2}}\sim 2.221$ discussed above, the two others are $e_{\gamma}\sim 1.13$ and $ \ec \sim 1.8545\pm 0.0005$.
Accordingly, the phase diagram of hairy boson stars and hairy black holes depends on the value of $e$ compared to these four fundamental critical charges of the system. 
Thus, in the next subsections, we describe the properties of hairy solutions in the following four windows of scalar field charge: 1) $e_{\hbox{\tiny NH}} \leq e < e_{\gamma}$, 2) $e_{\gamma} \leq e\leq  \ec$, 3)  $ \ec \leq  e<  e_{\hbox{\tiny S}}$, and 4) $e\geq e_{\hbox{\tiny S}}$. For concreteness, we will display results for a representative value of $e$ for each one of these windows,  namely: 1) $e=1$ (section~\ref{sec:phasediag1}), 2) $e=1.85$ (section~\ref{sec:phasediag2}), 3) $e=2$  (section~\ref{sec:phasediag3}), and 4) $e=2.3$ (section~\ref{sec:phasediag4}).  Altogether, these results (and others not presented) will  allow us to extract the conclusions summarized in section~\ref{sec:summary}.

\begin{figure}[!htb]
  \centerline{
    \includegraphics[width=.48\textwidth]{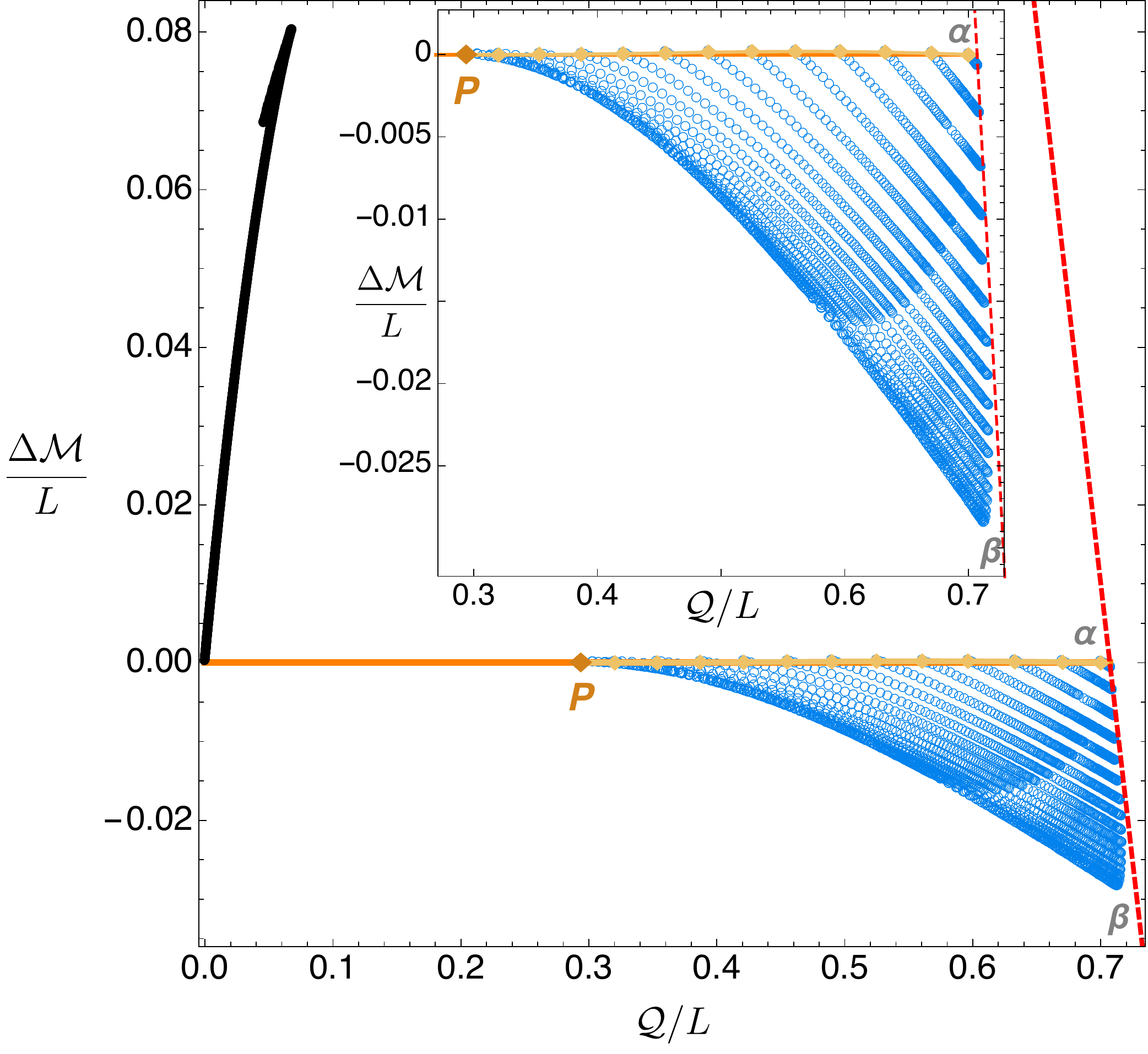}
    \hspace{0.3cm}
    \includegraphics[width=.48\textwidth]{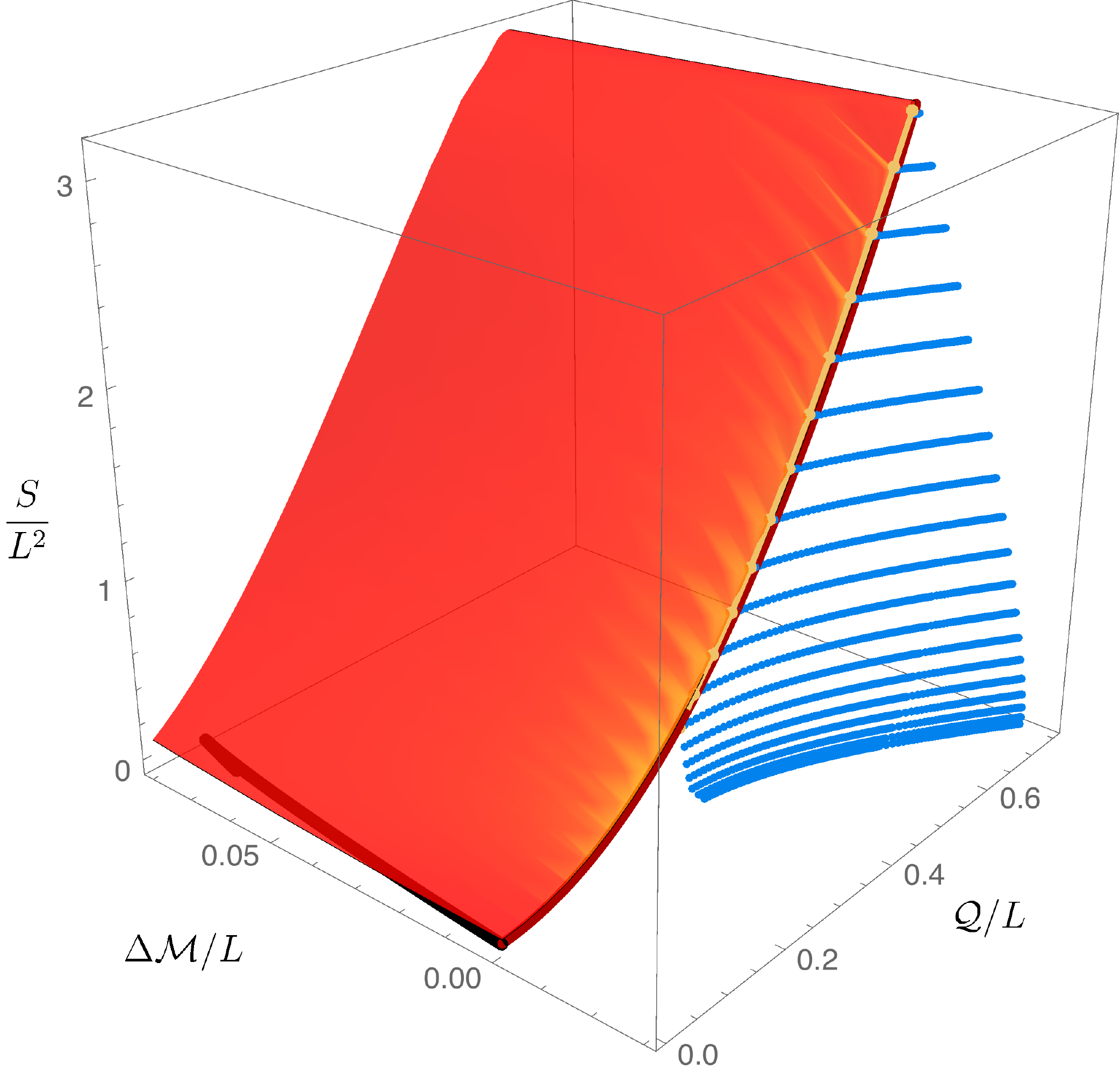}
  }
  \caption{Phase diagrams for Einstein-Maxwell theory with a scalar field charge $e=1$ ($\enh\leq e<\egam$) in a Minkowski box.   {\bf Left panel:} Quasilocal mass difference $\Delta \mathcal{M}/L$ as a  function of the quasilocal charge $\mathcal{Q}/L$.
  The black disk curve is the main/perturbative soliton  family, the orange line is the extremal RN BH (RN black holes exist above it), and  the blue circles describe hairy black holes. The yellow curve is the superradiant onset curve of RN (just above but very close to the extremal RN curve with the two merging at $P$ and $\alpha$). It agrees with the hairy solutions in the limit where these have $\epsilon=0$ (no scalar condensate) and thus merge with RN family.   The red dashed line  with negative slope   signals solutions with $\Delta\mathcal{M}/L=1-\sqrt{2}\mathcal{Q}/L$ {\it i.e.} black holes with horizon radius $R_+=1$ (above this value they do not fit inside the cavity).
  {\bf Right panel:} Dimensionless entropy $S/L^2$ as a function of the quasilocal charge and mass difference. RN BHs are the two parameter red surface with extremality described by  the 1-parameter curve $\Delta\mathcal{M}=0$ (dark red). The instability onset is described  by the yellow curve (very close to extremality) and RN between these two curves are unstable. 
  When they coexist with RN BHs, for a given $(\cQ,\cM)/L$, hairy BHs (blue dots) always have more entropy than RN, \ie they dominate the microcanonical ensemble. For $\enh\leq e<\egam$,  hairy BHs terminate at an extremal BH (\ie with zero temperature) and finite entropy (and divergent horizon curvature). The soliton (black dots) with zero entropy is also shown.}
  \label{FIGe1:MassCharge}
\end{figure}

%%%%%%%%%%%%%%%%%%%%%%%%%%%%%%%%%%%%%%%%%%%%%%%%%%
\subsection{Phase diagram for $\enh\leq e<\egam$}\label{sec:phasediag1}
%%%%%%%%%%%%%%%%%%%%%%%%%%%%%%%%%%%%%%%%%%%%%%%%%%

The left panel of Fig.~\ref{FIGe1:MassCharge} is the phase diagram $\cQ$-$\Delta\cM$ for $e = 1$, representative of the range $\enh\leq e<\egam$. The black disk curve describes the only family of boson stars of the theory for this (range of) $e$ which is the main/perturbative family. This corresponds to the black curve $OABC\cdots$ (already present for $e<e_{\hbox{\tiny NH}}$) with its Chandrasekhar limit $A$ and a series of cusps $A,B,C,\cdots$ and associated zig-zagged branches sketched in the top-left panel of Fig.~\ref{FIG:Summary_sketch}. The properties of this boson star were already studied in much detail in \cite{Dias:2021acy} so we do not expand  further. Our interest here are the hairy BHs. 

The horizontal orange curve $OP\alpha$ with  $\Delta\cM=0$ is the extremal RN BH family with $\cQ/L\leq 2^{-\frac{1}{2}}$ and boxed non-extremal BHs exist above this line and to the left of the red dashed line  \eqref{redDashed} to fit inside the cavity, as detailed above.
The yellow curve $P\alpha$, that intersects and terminates at the extremal RN curve precisely at $P$ and $\alpha$, describes the instability onset curve of RN BHs as computed using linear analysis in  \cite{Dias:2018zjg}. It coincides with the merger line of hairy BHs with RN BHs, as it had to. Indeed, recall that hairy BHs can be parametrized by their horizon radius $R_+$ and the scalar field amplitude $\epsilon$. When $\epsilon =0 $ we recover the 1-parameter family $P\alpha$ of RN BHs at the onset of the instability.
RN BHs are unstable  below this onset curve $P\alpha$ all the way down to the horizontal extremal line also labelled $P\alpha$. This region is extremely small for this value of $e$ but it will be  wider as $e$ increases.
 
Hairy BHs (blue circles) exist inside the closed line $P\alpha\beta$. That is, they exist below the onset curve $P\alpha$ and to the left of the red $\alpha\beta$ dashed line  \eqref{redDashed}, all the way down till they reach a line $P\beta$ where the  Kretschmann curvature scalar evaluated at the horizon $K|_{\mathcal{H}} = R_{abcd} R^{abcd}\big|_{R_+}$ grows large without bound. This occurs at finite $R_+$ and thus at {\it finite entropy}  $S/L^{2} = \pi R_{+}^{2}$, and the {\it temperature  vanishes} along this boundary curve $P\beta$.
The entropy is however not constant along this singular extremal boundary curve (in practice, the last curve we plot has $R_+=0.1$ but it should extend a bit further down in the region close to $\alpha\beta$). 
We typically find that lines of constant $R_+$ extend all the way to the red $\alpha\beta$ dashed line  \eqref{redDashed}, but the latter is only reached in the limit $\epsilon\to \infty$. This makes it harder to extend our solution to regions even closer to $\alpha\beta$ (a fixed step in $\epsilon$ corresponds to an increasingly smaller progression in $\cQ$ as $\alpha\beta$ is approached).  
Hairy BHs do not exist for $\cQ< \cQ|_{P}$, in agreement with the linear analysis of the left panel of Fig.~\ref{FIG:Summary_onset}, and there is clearly no relation between the hairy BHs and the boson star of the theory when $e = 1$ and, more generically, for $\enh\leq e<\egam$.

Because point $P$ does not coincide with the origin $O$, hairy BHs with $\enh\leq e<\egam$ were not found in the perturbative analysis of \cite{Dias:2018yey}. Indeed, this perturbative analysis only captures hairy BHs that have small mass and charge. 

The right panel of Fig.~\ref{FIGe1:MassCharge} plots the same phase diagram as the left panel but this time with the entropy $S/L^{2}$ on the extra vertical axis. The latter is the appropriate  thermodynamic potential to discuss the preferred thermal phases  of the microcanonical ensemble: for a given quasilocal mass $\cM/L$ and charge $\cQ/L$ fixed,  the dominant phase is the one with the largest entropy.
The red surface represents (a subset\footnote{We just plot the portion of the RN surface with $\Delta \cM<0.085$ that covers the region where the boson star also exists.}) of RN BHs, both stable and unstable with the boundary line of stability being again the yellow dotted curve, here very close to the extremal RN BH (dark red with $\Delta\cM=0$). In the $S=0$ plane we find the perturbative boson star (black curve).
 The blue dots fill the 2-dimensional surface that describe hairy BHs (which merge with RN along the yellow line). Again we see (not very clearly but it will be more clear for higher $e$) that hairy BHs coexist with RN black holes in the region between the onset and extremal RN curves. In this case, we find that hairy black holes always have a larger entropy than the corresponding RN BHs with same $\cM/L$ and $\cQ/L$. 
So they are the thermodynamically preferred phase in the microcanonical ensemble. 
 
Hence,  it follows from the second law of thermodynamics that hairy BHs with $(\cQ,\cM)$ between the RN onset and extremality curves are natural candidates for the endpoint of the RN superradiant/near-horizon instability when we do a time evolution  of the instability where we preserve the mass and charge of the system.

%%%%%%%%%%%%%%%%%%%%%%%%%%%%%%%%%%%%%%%%%%%%%%%%%%
\subsection{Phase diagram for $\egam \leq e < \ec$}\label{sec:phasediag2}  
%%%%%%%%%%%%%%%%%%%%%%%%%%%%%%%%%%%%%%%%%%%%%%%%%%

\begin{figure}[!htb]
  \centerline{
    \includegraphics[width=.48\textwidth]{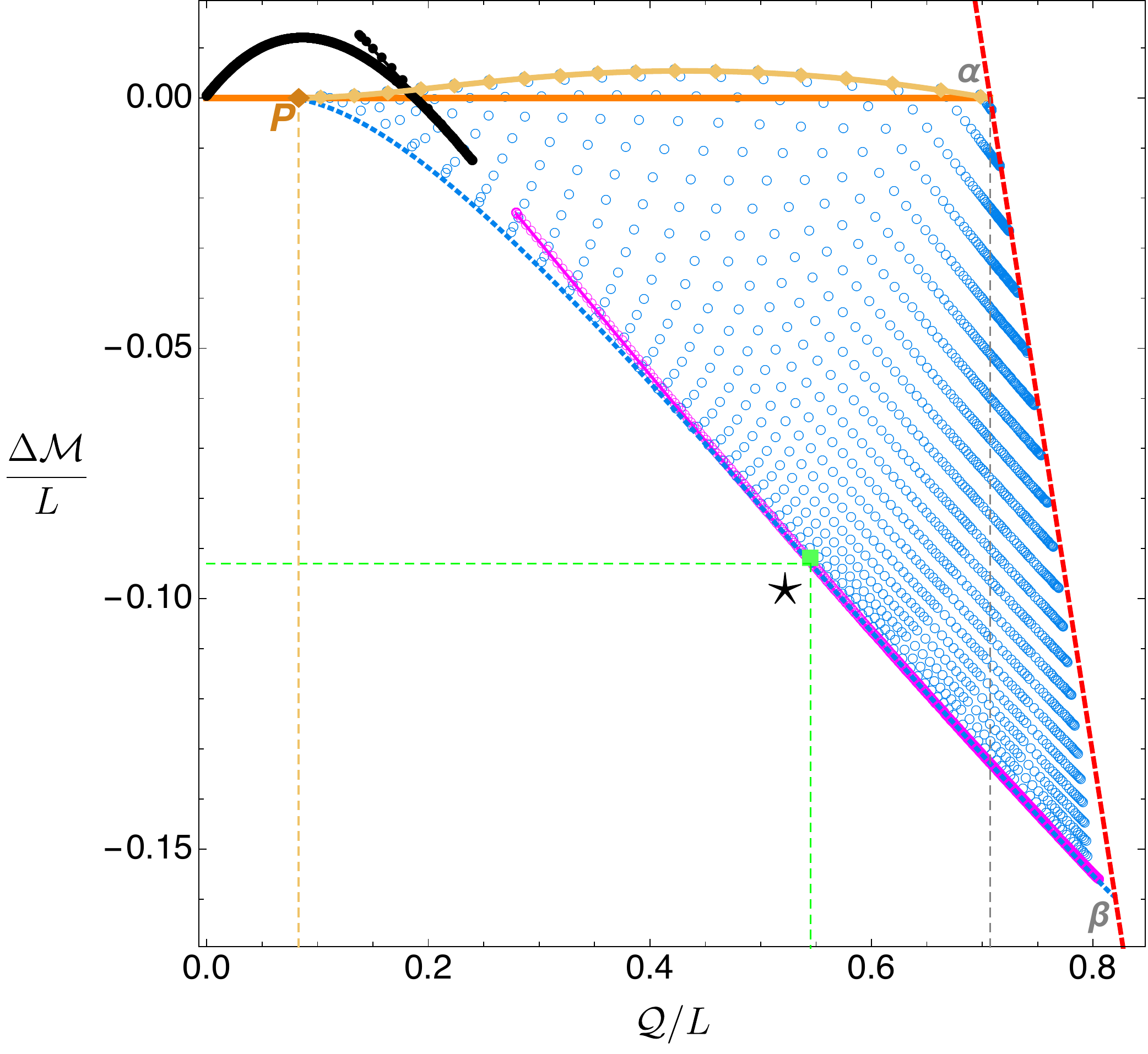}
    \hspace{0.3cm}
    \includegraphics[width=.48\textwidth]{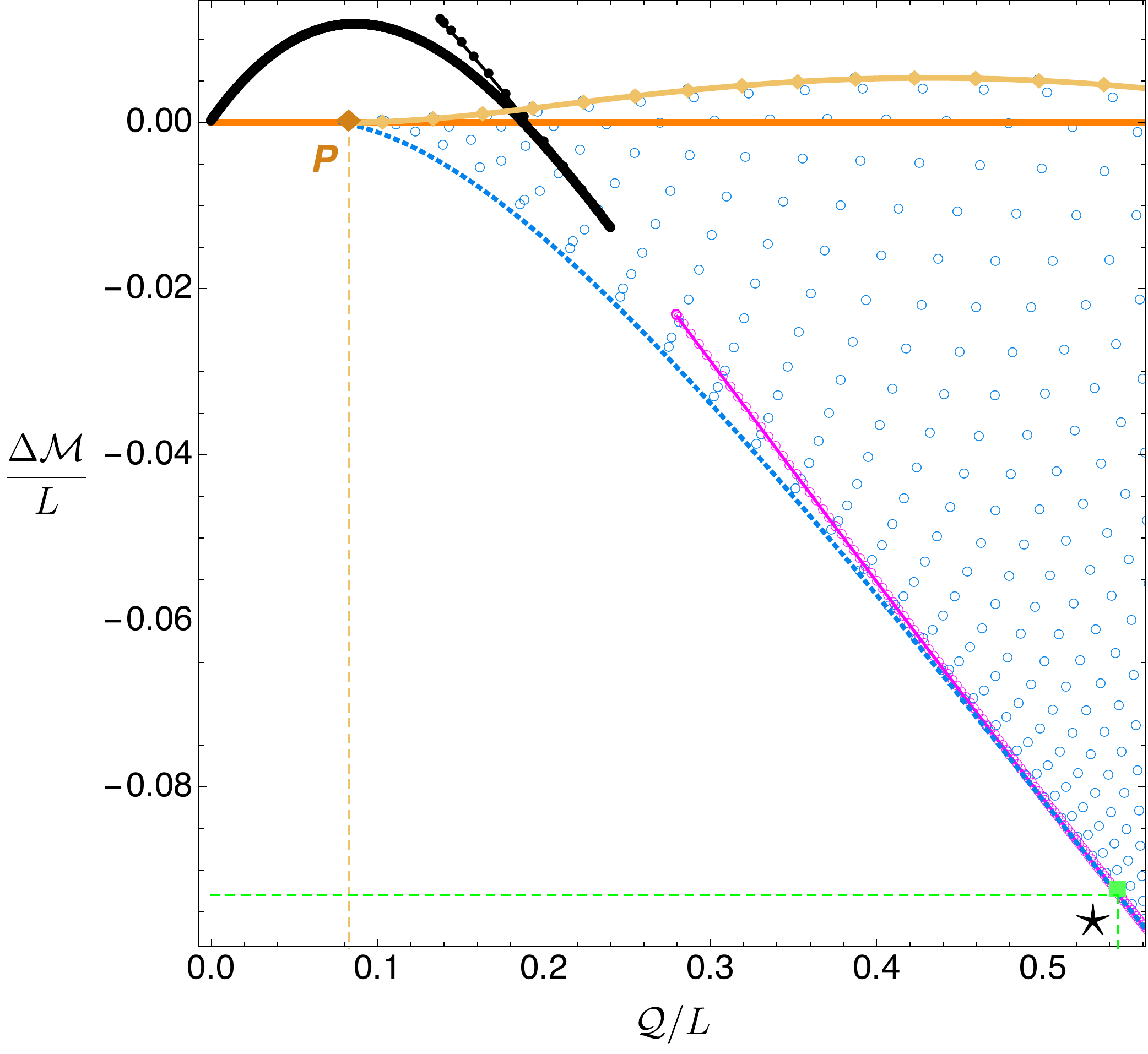}
  }
  \caption{
    Phase diagram for Einstein theory with a scalar field charge $e=1.85$ ($\egam \leq e < \ec$) in a Minkowski box. The blue circles describe hairy black holes, the black disk (magenta circle) curve is the soliton main (secondary) family, and the orange line is the extremal RN BH (non-extremal RN BHs exist above it). The yellow curve is the superradiant onset curve of RN. As it could not be otherwise, it agrees with the hairy solutions in the limit where these have $\epsilon=0$ and thus merge with RN family.  The dashed vertical line is at $\cQ=2^{-1/2}$ which is the maximum local charge that an extremal RN BH can have while fitting inside a box with radius $R=1$.  The red dashed line \eqref{redDashed} describes the boundary for black holes that can fit inside the cavity with radius $R=1$. The green solid square labelled with a star ($\star$) has $(\cQ_\star,\cM_\star, \Delta\cM_\star)\sim(0.545, 0.678, -0.093)$.
    The auxiliary blue dotted curve $P\star\beta$ in the bottom describes the line where hairy BHs terminate with unbounded horizon curvature. Hairy BHs that terminate in the trench $P\star$ of this auxiliary curve do so at finite entropy and vanishing temperature. On the other hand, hairy BHs that terminate in the trench segment $\star\beta$ (that coincides with magenta soliton line) do so at zero entropy and large (possibly infinite) temperature.
  }
  \label{FIGe1.85:MassCharge}

\end{figure}

\begin{figure}[!htb]
  \centerline{
 \includegraphics[width=.50\textwidth]{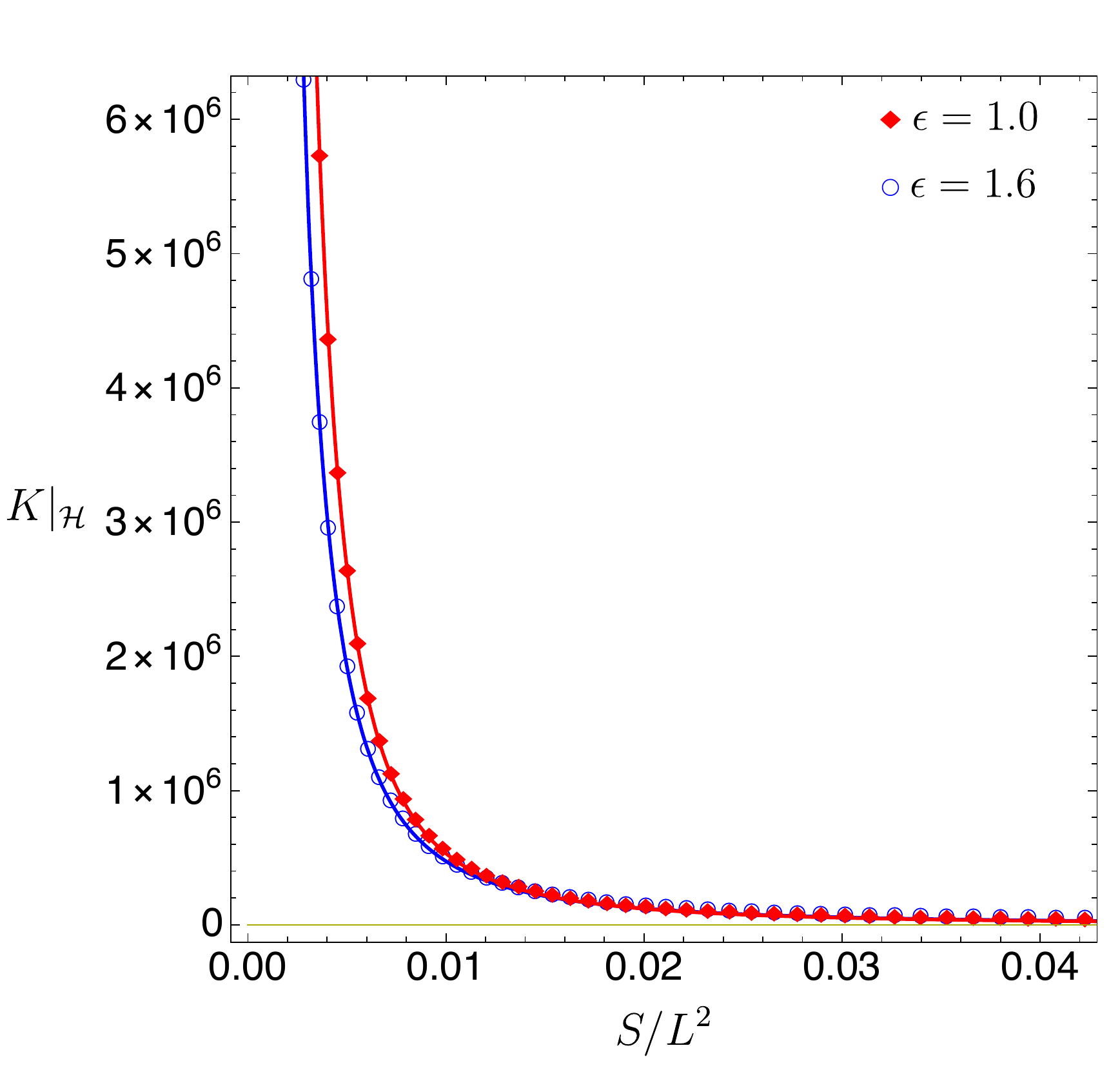}
    \hspace{0.3cm}
        \includegraphics[width=.48\textwidth]{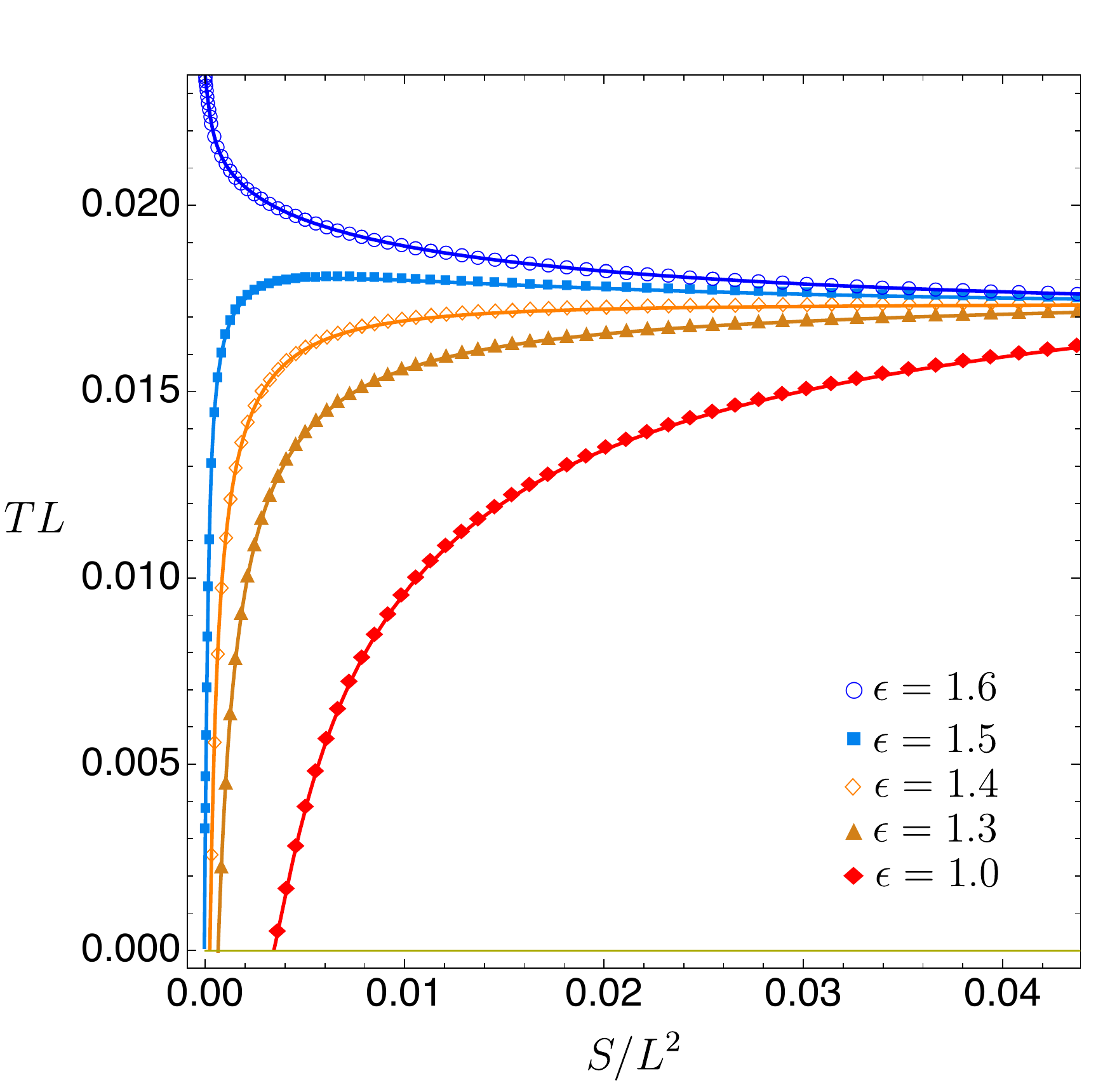}
  }
  \caption{Kretschmann curvature at the horizon (left panel) and temperature  (right panel) as a function of the entropy ($S/L^{2}=\pi R_+^2$) for several hairy BH families with constant scalar amplitude $\epsilon$ and scalar field charge $e=1.85$ ($\egam \leq e < \ec$).}
  \label{FIGe1.85:tempCurv}
\end{figure}
\begin{figure}[!htb]
  \centerline{
    \includegraphics[width=.7\textwidth]{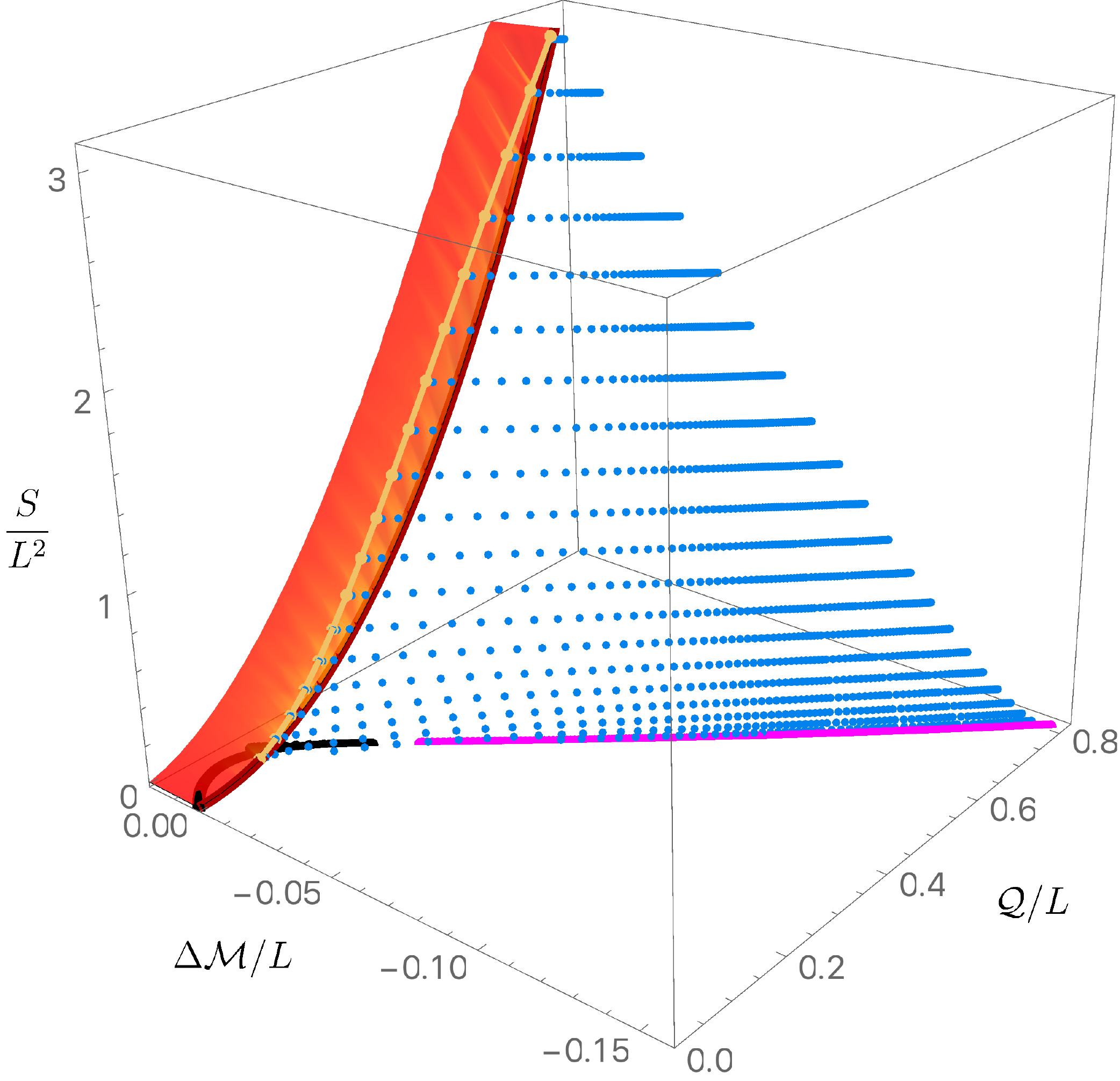}
  }
  \caption{Entropy as a function of the quasilocal charge and mass difference for Einstein theory with a scalar field charge  $e=1.85$ ($\egam \leq e < \ec$) in a Minkowski box. The red surface  represents RN BHs in the range $0\leq \Delta\cM<0.02$ (they extend for higher $\Delta\cM$) with the dark red line with $\Delta\cM=0$ being the extremal RN BH family. The yellow line  describes the merger line between RN BHs and hairy BHs, and RN BHs between this line and the dark red extremal line are unstable. The blue disks describe hairy BHs and the black (magenta) lines with $S=0$ describe the main (secondary) soliton family.
  When they coexist with RN BHs, for a given $(\cQ,\cM)/L$, hairy BHs always have more entropy than RN, \ie they dominate the microcanonical ensemble.
  }
  \label{FIGe1.85:entropy}
\end{figure}

In Fig.~\ref{FIGe1.85:MassCharge} we display the phase diagram when $e = 1.85$, which is representative of the range $\egam \leq e < \ec$ that we sketched in the top-right panel of Fig.~\ref{FIG:Summary_sketch}. As a first observation we note that, besides the main or perturbative boson star family (black curve) already present for $e<\egam$, the diagram now  also has the magenta line that starts at finite $\cQ$, passes though point  $\star$, and terminates at point  $\beta$ on the red dashed line. This is the secondary or non-perturbative family of boson stars. On its left side,  this family has itself a series of cusps and zig-zagged secondary branches denoted as $B,C,\cdots$ in the sketch of the top-right panel of Fig.~\ref{FIG:Summary_sketch} (not displayed in  Fig.~\ref{FIGe1.85:MassCharge}). These details are not relevant here, and we ask the reader to see \cite{Dias:2021acy} for an exhaustive study of boson stars' properties. It is however important to emphasize that this secondary/non-perturbative family exists (as a ground state family) only for  $\egam \leq e < \ec$, thus explaining the origin of the critical charges $\egam$ and  $\ec$. At $e=\ec$ the magenta line of Fig.~\ref{FIGe1.85:MassCharge} merges with the black line (see section~\ref{sec:phasediag3}). On the other hand, as we decrease  $e$ below $\ec$ one finds that the gap $\Delta\cQ$ between the black and magenta families increases, and the ``length'' of the magenta line decreases because the left endpoint of this curve approaches $\beta$. It keeps doing so till it only exists on a very small neighbourhood of the red dashed line and, at $e=\egam$, this line shrinks to the single point $\beta$. Below $\egam$, the non-perturbative family ceases to exist (as seen in section ~\ref{sec:phasediag1}). Essentially because it no longer fits inside the cavity. This discussion  is better illustrated in  the right panel of Fig.~\ref{FIG:Summary_onset}:  1) if  we collect all non-perturbative solitons in a single plot, we find that they exist  only in the window $\egam \leq e < \ec$ and they fill the area bounded by the auxiliary dashed lines $ a_c\beta_c\gamma$; 2) very close to $\ec$ the non-perturbative soliton is almost on top of the auxiliary curve $ a_c\beta_c$; and 3) on the opposite end, as $e\to\egam$, the perturbative soliton line shrinks to the point $\gamma$ on the red dashed line. 

What are the consequences of these boson star discussions for the hairy BHs? 
Hairy BHs with  $e = 1.85$ are the blue circles in Fig.~\ref{FIGe1.85:MassCharge}. 
As before, they exist in the area bounded by $P\alpha\beta$, where $P\alpha$ is the merger yellow line with RN BHs and coincides with the instability onset curve of \cite{Dias:2018zjg}, and $\alpha\beta$  is a segment of the red dashed line \eqref{redDashed}. Starting  at the onset curve $P\alpha$ and moving down, \eg along constant $\cQ$ lines, we find that that hairy BHs terminate at the  line $P\beta$ (or $P\star\beta$). This is the blue dashed line in Fig.~\ref{FIGe1.85:MassCharge} which describes hairy BHs with minimum entropy/horizon radius for a given charge.
Along this line, the Kretschmann curvature scalar evaluated at the horizon $K|_{\mathcal{H}}$ grows very large (most probably, $K|_{\mathcal{H}}\to \infty$). 
To illustrate this, in the left panel of Fig.~\ref{FIGe1.85:tempCurv} we plot  $K|_{\mathcal{H}}$ as a function of the entropy $S/L^{2} = \pi R_{+}^{2}$ as we approach the line $P\beta$ (at small $S$) along curves of constant scalar amplitude $\epsilon$ (shown in the legends). Indeed, for small $S/L^2$ the curvature is growing very large. 

So far the phase diagram of hairy BHs looks similar to the one for $\enh\leq e<\egam$ (section~\ref{sec:phasediag1}). However, for $\egam \leq e < \ec$ we now find  that the way  hairy BHs terminate along the singular curve differs substantially depending on whether it ends to the left or  to the right  of the green square point $\star$ in Fig.~\ref{FIGe1.85:MassCharge} (with $\cQ_{\star}/L \simeq 0.545$ for $e=1.85$). When the hairy BHs terminate along $P\star$, they do so at {\it finite entropy} and {\it vanishing temperature}. On the other hand, hairy BHs that terminate along $\star\beta$ do so at {\it vanishing entropy} and {\it  large (possibly infinite)  temperature}.
To illustrate this, in the right panel of Fig.~\ref{FIGe1.85:tempCurv} we plot the temperature $T L$ as a function of the entropy $S/L^2= \pi R_{+}^{2}$ as we follow hairy BH families that approach the singular line $P\beta$ at (different; see legends) constant scalar amplitude $\epsilon$. Point $\star$ has  $(\cQ_\star,\Delta\cM_\star)\sim(0.545, -0.093)$ which corresponds to $(R_+,\epsilon)\big|_\star= (0,1.55 \pm 0.05)$. Hairy BHs with $\epsilon< \epsilon_\star$ terminate at $P\star$, while hairy BHs with  $\epsilon> \epsilon_\star$ end at $\star\beta$. The right panel of Fig.~\ref{FIGe1.85:tempCurv} indeed shows that hairy BHs with $\epsilon< \epsilon_\star$ approach $P\star$ at finite $S/L^2$ and with $TL\to 0$ (like all hairy BHs of section~\ref{sec:phasediag1}), while those with $\epsilon> \epsilon_\star$  approach $\star\beta$  with $S\to 0$ and  $TL\to \infty$. 

Another important conclusion that emerges from Fig.~\ref{FIGe1.85:MassCharge}, is that hairy BHs which have a zero horizon radius limit terminate precisely along the segment $\star \beta$ of the secondary/non-perturbative soliton family. This means that hairy BHs terminate with the {\it same} $\cQ$ and $\cM$ as the non-perturbative soliton (but the gravitoelectric and scalar fields of the two solutions are different).
On the other hand, those that end at $P\star$ do so in a manner that is very similar to the way the hairy BHs with $\enh\leq e<\egam$ terminate (section~\ref{sec:phasediag1}). 

We find that the critical charge  $\cQ_{\star}(e)$ decreases as $e$ grows from $\egam$ till $\ec$.
As explained when discussing the right plot of Fig.~\ref{FIG:Summary_onset},  the non-perturbative soliton line shrinks to the point $\beta$ when $e \to \egam$. Thus, our expectation is that the critical charge $\cQ_{\star}$ also reaches $\cQ\big|_\beta$ when $e \to \egam^+$. That is to say, we expect that hairy black holes are connected to the non-perturbative soliton as soon as it exists. However, determining  numerically $\cQ_{\star}$ in this limit is very difficult, since hairy BHs near $\beta$ have very large values of $\epsilon$. 

The hairy BHs with $\egam\leq e<\ec$ we find were not captured by  the perturbative analysis of \cite{Dias:2018yey} because they do not extend to arbitrarily small mass and charge. 

In Fig.~\ref{FIGe1.85:entropy}, we plot the thermodynamic potential of the microcanonical ensemble $-$ the entropy $S/L^2$ $-$ as a function of $\cQ$ and $\Delta\cM$. In the $S=0$ plane we  find the perturbative boson star (black curve) and, for larger $\cQ$ and after a gap, the non-perturbative boson star (magenta curve). As before, the red surface describes the RN BH family parametrized  by $R_+$ and $\mu$ as in \eqref{quasilocalRN} and with $S/L^2= \pi R_{+}^{2}$. It terminates at the dark red extremal curve with $\Delta\cM=0$. We only plot the portion of the RN surface with $\Delta \cM<0.02$ that covers the region where the perturbative boson star also exists. 
Unstable RN BHs are those  between the instability onset (yellow dotted curve) and the extremal RN dark red curve.  The blue dots fill the 2-dimensional surface that describes hairy BHs. It merges with RN BHs along the yellow dotted curve and then extends to lower $\Delta\cM$ with an entropy that is always larger the the RN BH with the same $\cQ$ and $\cM$ (when they coexist).
Therefore, hairy BHs are the thermodynamically dominant phase in the microcanonical ensemble. 
Consequently, from  the second law of thermodynamics, hairy BHs with $(\cQ,\cM)$ between the RN onset and extremality curves are  candidates for the endpoint of the RN superradiant/near-horizon instability when in  a time evolution of the RN instability at constant mass and charge.

%%%%%%%%%%%%%%%%%%%%%%%%%%%%%%%%%%%%%%%%%%%%%%%%%%
\subsection{Phase diagram for $\ec\leq e<\es$}\label{sec:phasediag3}
%%%%%%%%%%%%%%%%%%%%%%%%%%%%%%%%%%%%%%%%%%%%%%%%%%

%\clearpage
\begin{figure}[!b]
\centerline{
  \includegraphics[width=.48\textwidth]{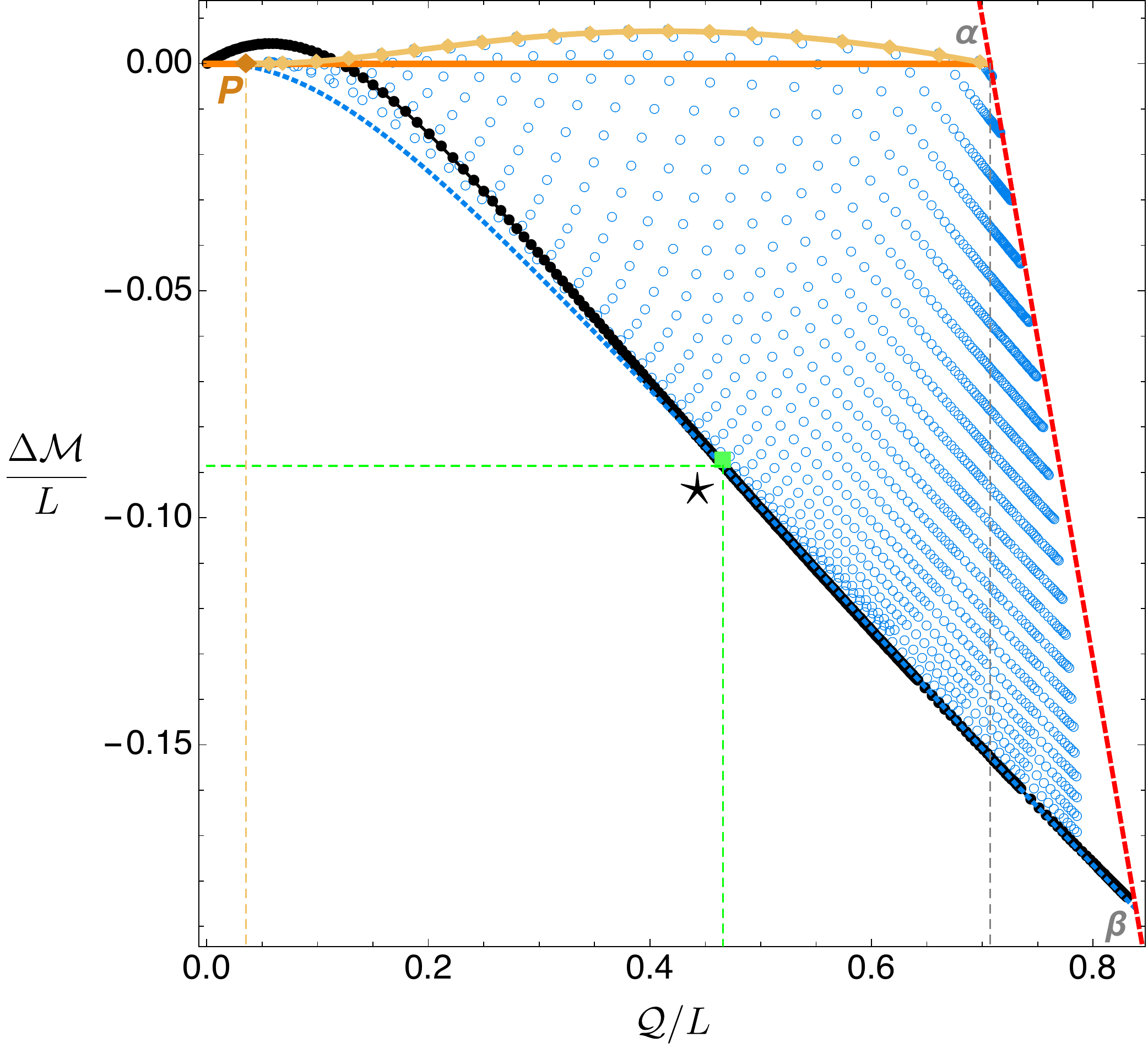}
  \hspace{0.3cm}
  \includegraphics[width=.48\textwidth]{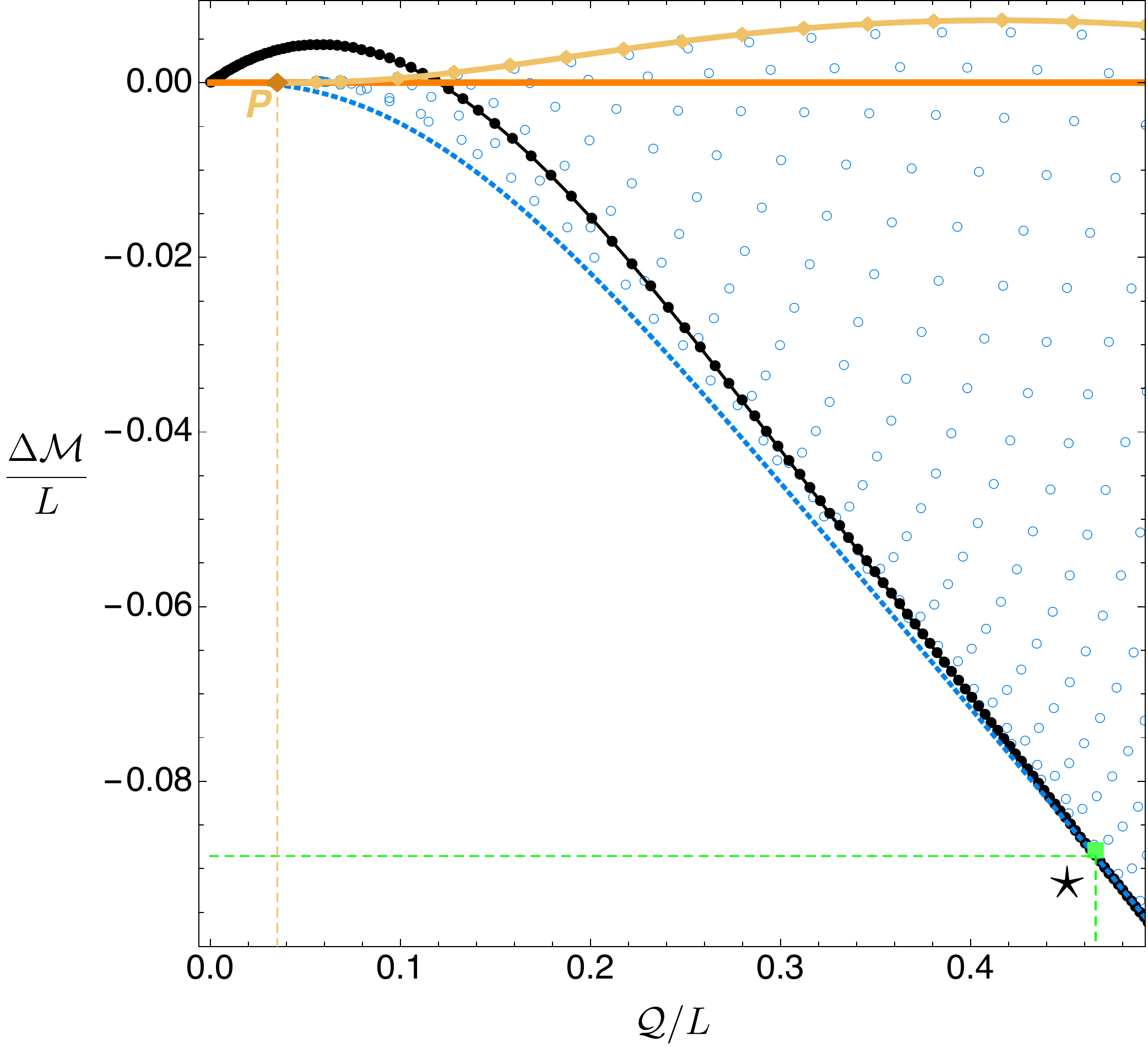}
}
\caption{Phase diagram for Einstein theory with a scalar field charge $e=2$ ($\ec\leq e<\es$) in a Minkowski box. As before, the blue circles describe hairy black holes, the black disk curve is the soliton main family, and the orange line is the extremal RN BH (RN black holes exist above it). The gray and red dashed curves have the same interpretation as in Fig.~\ref{FIGe1.85:MassCharge}.
The green solid square labelled with a star ($\star$) has $(\cQ_\star,\cM_\star, \Delta\cM_\star)\sim(0.466, 0.659, -0.0886)$.
The auxiliary blue dotted curve $P\star\beta$ in the bottom describes the line where hairy BHs terminate with unbounded horizon curvature. Hairy BHs that terminate in the trench $P\star$ of this auxiliary curve have zero temperature ($T= 0$)  and finite entropy $S/L=\pi R_+^2$. On the other hand, hairy BHs that terminate in the trench segment $\star\beta$ (that coincides with the black soliton line) have zero entropy and large (possibly infinite) temperature.
Note that these $\star\beta$ terminal hairy BHs have the same $(\cQ,\Delta\cM)$ as the main soliton family with $\cQ> \cQ_\star$.
}
\label{FIGe2:MassCharge}
\end{figure}

In Fig.~\ref{FIGe2:MassCharge} we give the phase diagram for $e = 2$. This is the case we choose to illustrate the solution spectra in the range $\egam \leq e < \ec$ that we sketched in the bottom-left panel of Fig.~\ref{FIG:Summary_sketch}. 
Comparing with the diagram of Fig.~\ref{FIGe1.85:MassCharge} we immediately notice that the magenta line representing the non-perturbative soliton family is no longer present in Fig.~\ref{FIGe2:MassCharge}. This is because at $e=\ec$, the perturbative and non-perturbative boson star families (\ie the black and magenta lines of Fig.~\ref{FIGe1.85:MassCharge}) merge and for $e\geq \ec$ the main or perturbative boson  star family no longer has a Chandrasekhar mass limit and now extends from the origin $O$ all  the way to $\beta$ in the red dashed line. This merger at $\ec$ occurs in an interesting elaborated manner. In particular, going back to top-right sketch of Fig.~\ref{FIG:Summary_sketch}, at $e=\ec$ the secondary zig-zagged branches $\cdots CBA$ of the perturbative (black) soliton also merge with  the  secondary zig-zagged branches $abc\cdots$ of the non-perturbative (magenta) soliton. As a consequence, for $e\geq \ec$ there is also a secondary soliton $\cdots CBbc\cdots$ (purple line in bottom-left of Fig.~\ref{FIG:Summary_sketch}) that has higher energy than the perturbative (black) soliton $P\beta$. This secondary family is not displayed in Fig.~\ref{FIGe2:MassCharge} because it plays no role on the discussion of hairy BHs of the theory. The reader can find a detailed discussion of soliton's properties across the transition  at $e=\ec$ in \cite{Dias:2021acy}.

\begin{figure}[!b]
  \centerline{
      \includegraphics[width=.53\textwidth]{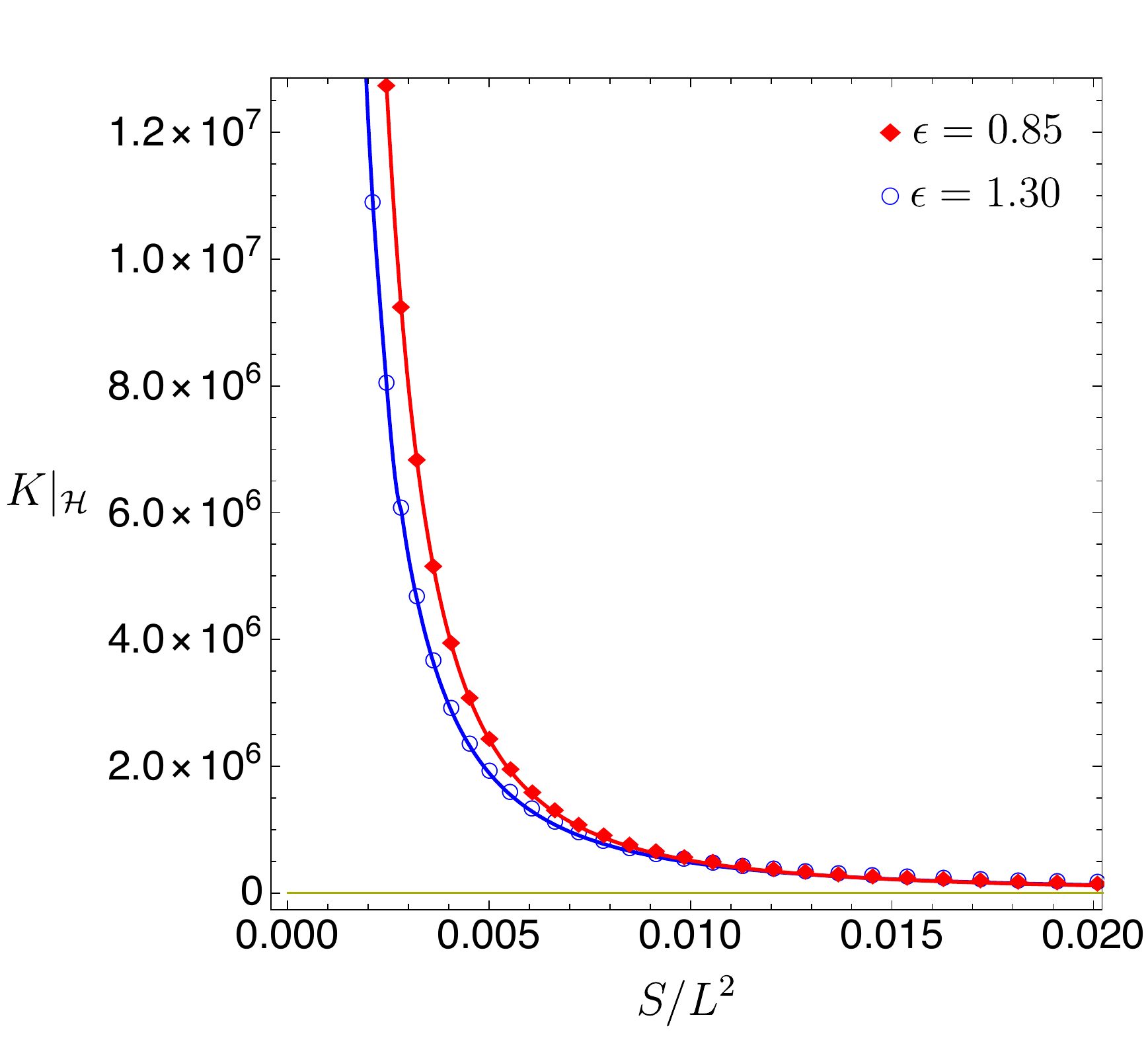}
          \hspace{0.3cm}
    \includegraphics[width=.47\textwidth]{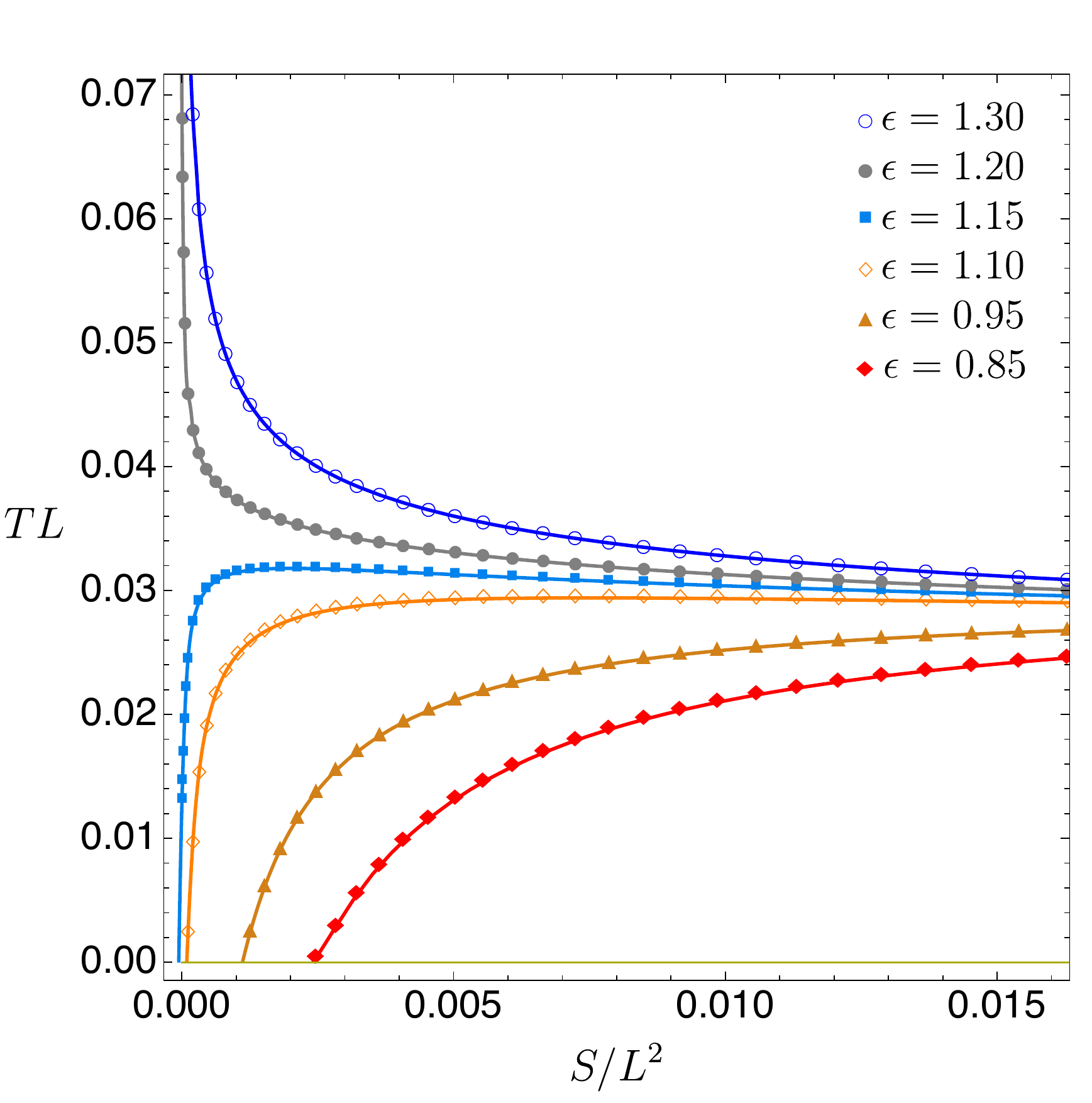}
  }
  \caption{ Kretschmann curvature at the horizon (left panel) and temperature (right panel) as a function of the entropy ($S/L^{2}=\pi R_+^2$)  for hairy BH families with fixed $\epsilon$ and scalar field charge  $e=2$ ($\ec\leq e<\es$).}
  \label{FIGe2:tempCurv}
\end{figure}

Since the colour code and associated labelling in Fig.~\ref{FIGe2:MassCharge} is the same as in  Figs.~\ref{FIGe1:MassCharge} and \ref{FIGe1.85:MassCharge} we can now immediately discuss  the hairy BHs. Again they exist in the area enclosed by $P\alpha\beta$ filled with the blue circles. They merge with the RN family  along the yellow dotted line $P\alpha$ when the scalar condensate vanishes, which agrees with  the RN instability curve found in  \cite{Dias:2018zjg}. The hairy BHs then exist  all the way down to the blue dashed line $P\beta$  (or $P\star\beta$) which, for a given charge, identifies the hairy BH that has minimum entropy/horizon radius.
The Kretschmann curvature evaluated at the horizon $K|_{\mathcal{H}}$ diverges. For a given charge, $P\beta$ identifies the hairy BHs with minimum entropy/horizon radius and $K|_{\mathcal{H}}$  grows very large  along it. This is confirmed in the left panel of Fig.~\ref{FIGe2:tempCurv}: as we approach $P\beta$ (at small $S$)  along lines of of constant scalar amplitude $\epsilon$ (identified in the legends),  $K|_{\mathcal{H}}$ is growing very large. 

Point $\star$ with charge  $\cQ_{\star} \simeq 0.466$ describes a transition point. Hairy BHs that end to the left of this point along $P\star$ do so at finite $S$ with $T\to  0$. However, one has $S\to 0$ and $T\to \infty$ when the hairy BHs  terminate along $\star\beta$ with $\cQ>\cQ_{\star}$.
This is confirmed in the right panel of Fig.~\ref{FIGe2:tempCurv} where we plot the temperature $T L$ as a function of the entropy $S/L^2= \pi R_{+}^{2}$ as we follow different families of constant scalar amplitude hairy BHs that approach the singular line $P\beta$. Point~$\star$ has  $(\cQ_\star,\Delta\cM_\star)\sim(0.466, -0.0886)$ which corresponds to $(R_+,\epsilon)\big|_\star= (0,1.175 \pm 0.005)$. Hairy BHs with $\epsilon< \epsilon_\star$ have  $\cQ<\cQ_{\star}$ and terminate at $P\star$, while hairy BHs with  $\epsilon> \epsilon_\star$ have $\cQ>\cQ_{\star}$ and end at $\star\beta$.

From Fig.~\ref{FIGe2:MassCharge} and the right panel of Fig.~\ref{FIGe2:tempCurv}, it should not go without notice that the hairy BHs that have a zero horizon radius limit terminate along the trench $\star\beta$ of the perturbative soliton family. That is, {\it when} the hairy BHs  have zero entropy, they have the same charge $\cQ$ and mass $\cM$ as the perturbative soliton.
In a nutshell, hairy BHs with $\ec\leq e<\es$ have a behaviour that is qualitatively similar to those of $\egam \leq e < \ec$ (section~\ref{sec:phasediag2}). However, the  zero entropy BHs now terminate on top of the perturbative soliton in the $\cQ$-$\cM$ phase diagram instead of ending on the non-perturbative soliton (which is now an excited solution $\cdots CBbc\cdots$ in the bottom-left panel of Fig.~\ref{FIG:Summary_sketch}). We also find that the critical charge  $\cQ_{\star}(e)$ decreases and  approaches $\cQ_P$ as $e$ grows from $\ec$ to $ \es$. Moreover, we find that $\cQ_{\star}\to \cQ_P \to 0$  as $e\to \es$.

  \begin{figure}[!hb]
  \centerline{
    \includegraphics[width=.7\textwidth]{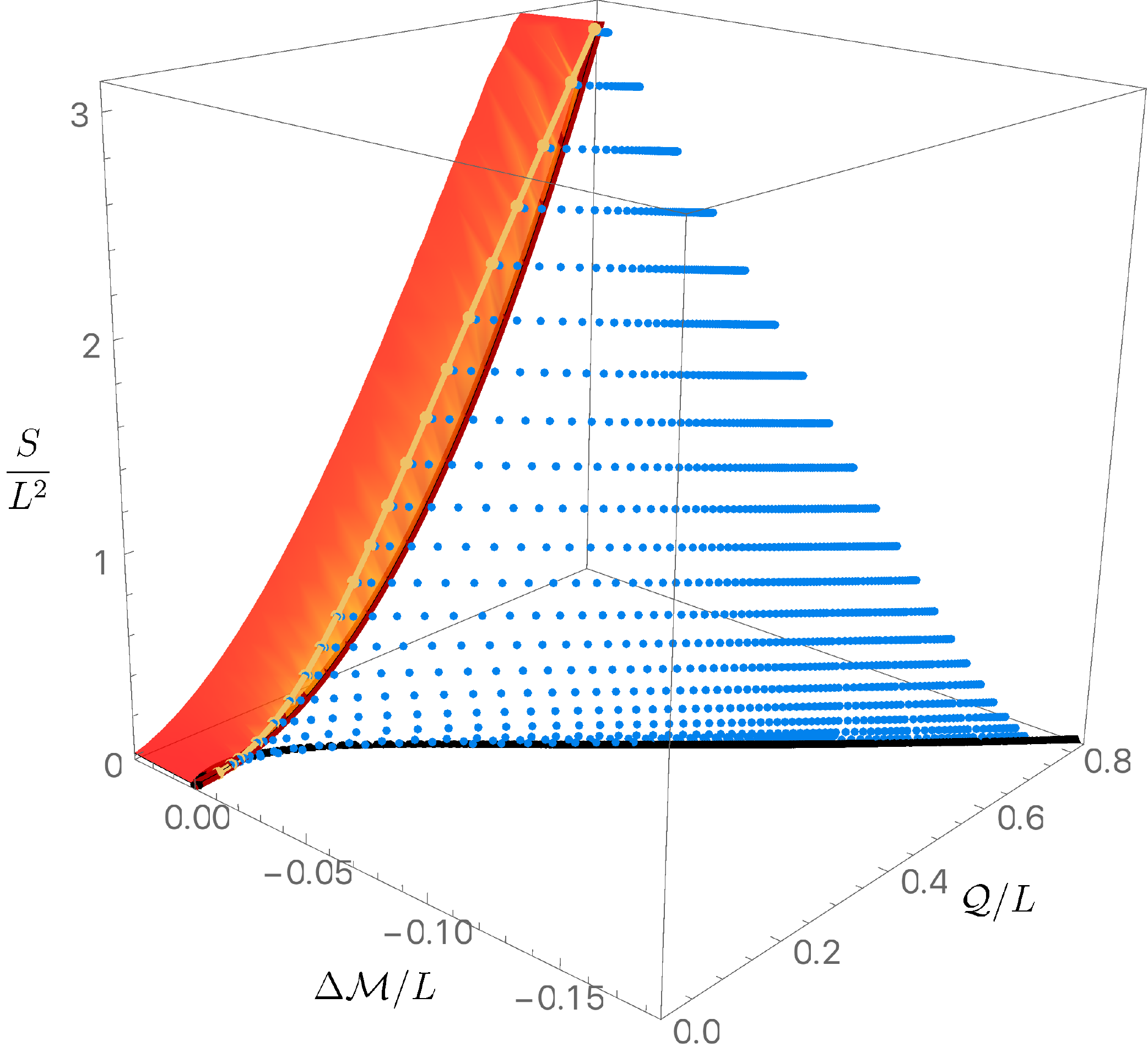}
  }
  \caption{Entropy as a function of the quasilocal charge and mass difference for Einstein theory with a scalar field charge  $e=2$ ($\ec\leq e<\es$) in a Minkowski box. When they coexist with RN BHs, for a given $(\cQ,\cM)/L$, hairy BHs always have more entropy than RN, \ie they dominate the microcanonical ensemble. For $\ec\leq e<\es$, when $\cQ>\cQ_\star(e)$, hairy BHs have a  zero entropy limit where they coincide with the soliton (black disk) curve in the sense that they have the same  $(\cQ,\cM)/L$ as the soliton (the temperature and horizon curvature diverges). However, when $0<\cQ<\cQ_\star(e)$, hairy BHs terminate at an extremal BH (\ie with zero temperature) and finite entropy (and divergent horizon curvature) along a line that does not coincide with the black disk one for the soliton.
  }
  \label{FIGe2:entropy}
\end{figure}

Fig.~\ref{FIGe2:entropy}, displays the phase diagram of the microcanonical ensemble for $e=2$: the entropy $S/L^2$ as a function of $\cQ$ and $\Delta\cM$. The colour code of this diagram is the same as Fig.~\ref{FIGe1.85:entropy}. Because $e$ is bigger than the cases considered before, we see that the region between the onset yellow curve and the extremal RN dark red curve  where RN BHs are unstable is now quite visible. Again  we find that the hairy BHs (blue circles) that bifurcate from the yellow onset curve always have higher entropy that the RN BHs with the same $(\cQ/L,\cM/L)$ when they coexist. 
It follows that also for $\ec\leq e< \es $, hairy BHs are the preferred thermodynamic phase in the microcanonical ensemble. As expected from Fig.~\ref{FIGe2:MassCharge},  for $\cQ\geq \cQ_{\star} \simeq 0.466$, the hairy  BHs terminate with zero entropy  on top of the perturbative boson star (black curve).

It is natural to expect that the hairy BHs we find should be the endpoint of the RN instability if we perturb an RN BH  in the unstable region (where they always coexist with hairy BHs) and do a time evolution at constant charge and mass. The system would evolve to a final configuration that is stable against the original perturbation while respecting the second law of thermodynamics.
Finally note that the hairy BHs with $\ec\leq e<\es$ described in this section were not studied in  the perturbative analysis of \cite{Dias:2018yey} because the  latter can only capture solutions that have  a zero mass and charge limit.

%%%%%%%%%%%%%%%%%%%%%%%%%%%%%%%%%%%%%%%%%%%%%%%%%%
\subsection{Phase diagram for $e \geq \es$}\label{sec:phasediag4}
%%%%%%%%%%%%%%%%%%%%%%%%%%%%%%%%%%%%%%%%%%%%%%%%%%

The critical charge  $e=e_{\hbox{\tiny S}}=\frac{\pi }{\sqrt{2}}\sim 2.221$ is special for two main (related) reasons. First, it is the minimal charge above which scalar fields can drive {\it arbitrarily small} extremal RN BHs unstable via superradiance, as observed in the instability onset charge plot of  the left panel of Fig.~\ref{FIG:Summary_onset}. Indeed, the extremal onset curve $e_{\hbox{\tiny onset}}(R_+)$ reaches $e=\es$ as $R_+\to 0$. The value of $\es$ can be predicted analytically as done in Section III.A of \cite{Dias:2018yey}. For $e>\es$, we can also have near-extremal BHs unstable for arbitrarily small $R_+$ or, equivalently, for arbitrarily small mass and charge.

\begin{figure}[!htb]
  \centerline{
  \includegraphics[width=.485\textwidth]{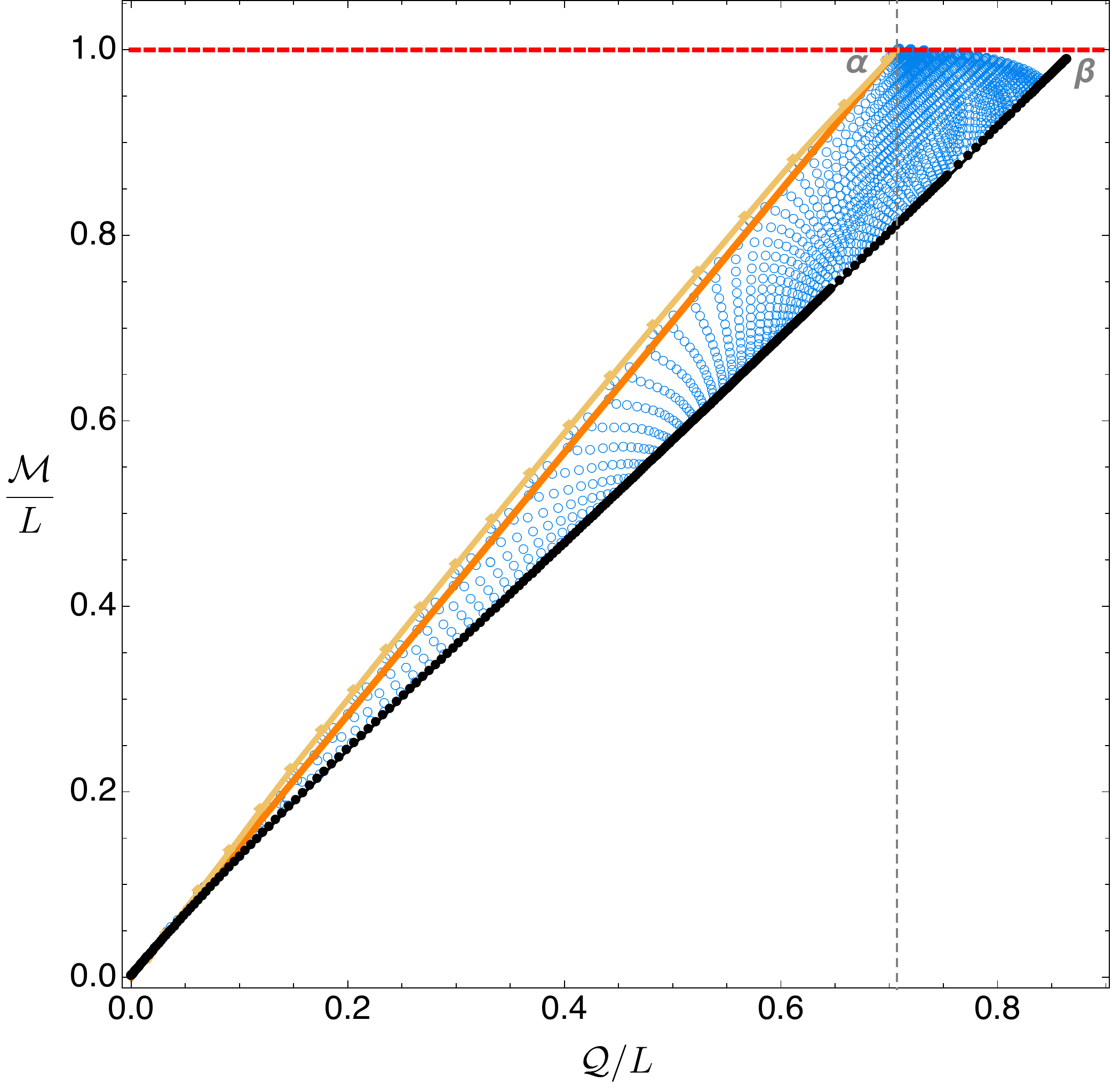}
  \hspace{0.3cm}
  \includegraphics[width=.52\textwidth]{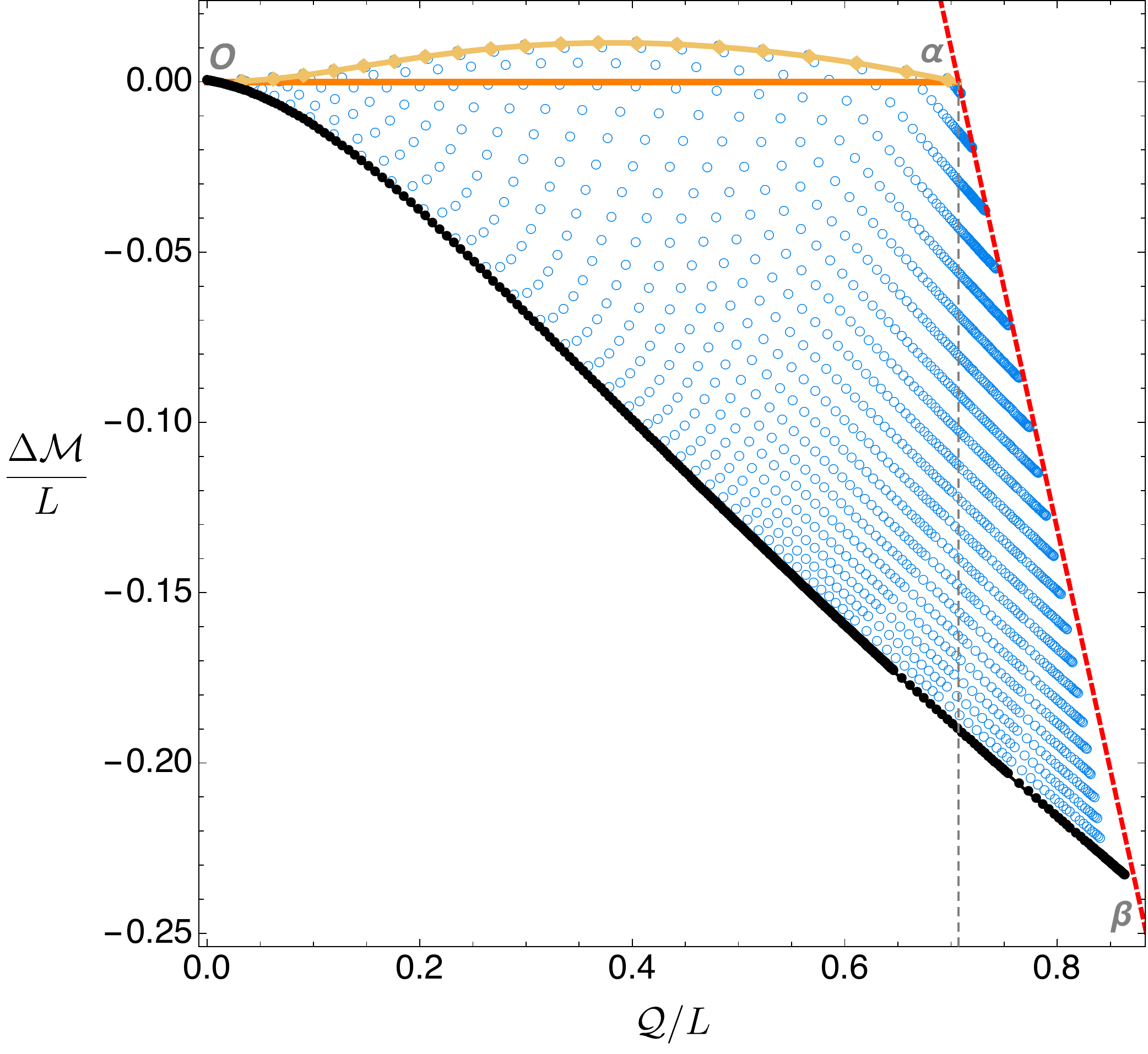}
  }
  \caption{
    Phase diagram for Einstein theory with a scalar field charge $e=2.3$ ($e>\es$) in a Minkowski box. In the left panel we have the $\cQ$-$\cM$ phase diagram: we see that the solutions pile up and this is why we have instead been plotting the phase diagram $\cQ$-$\Delta\cM$ (right panel). 
    The blue circles describe hairy black holes, the black disk curve is the perturbative soliton family and the orange line is the extremal RN BH (RN black holes exist above it). The yellow curve is the superradiant onset curve of RN. As it could not be otherwise, it agrees with the hairy solutions in the limit where these have $\epsilon=0$ and thus merge with RN family.  The gray and red dashed curves have the same interpretation as in Fig.~\ref{FIGe1.85:MassCharge}.
    For $e>\es$, the zero entropy limit of the hairy BH is the soliton (black disk curve) in the sense that they have the same  $(\cQ,\cM)/L$ as the soliton.}
  \label{FIGe2.3:MassCharge}
\end{figure}
\begin{figure}[!b]
  \centerline{
    \includegraphics[width=.48\textwidth]{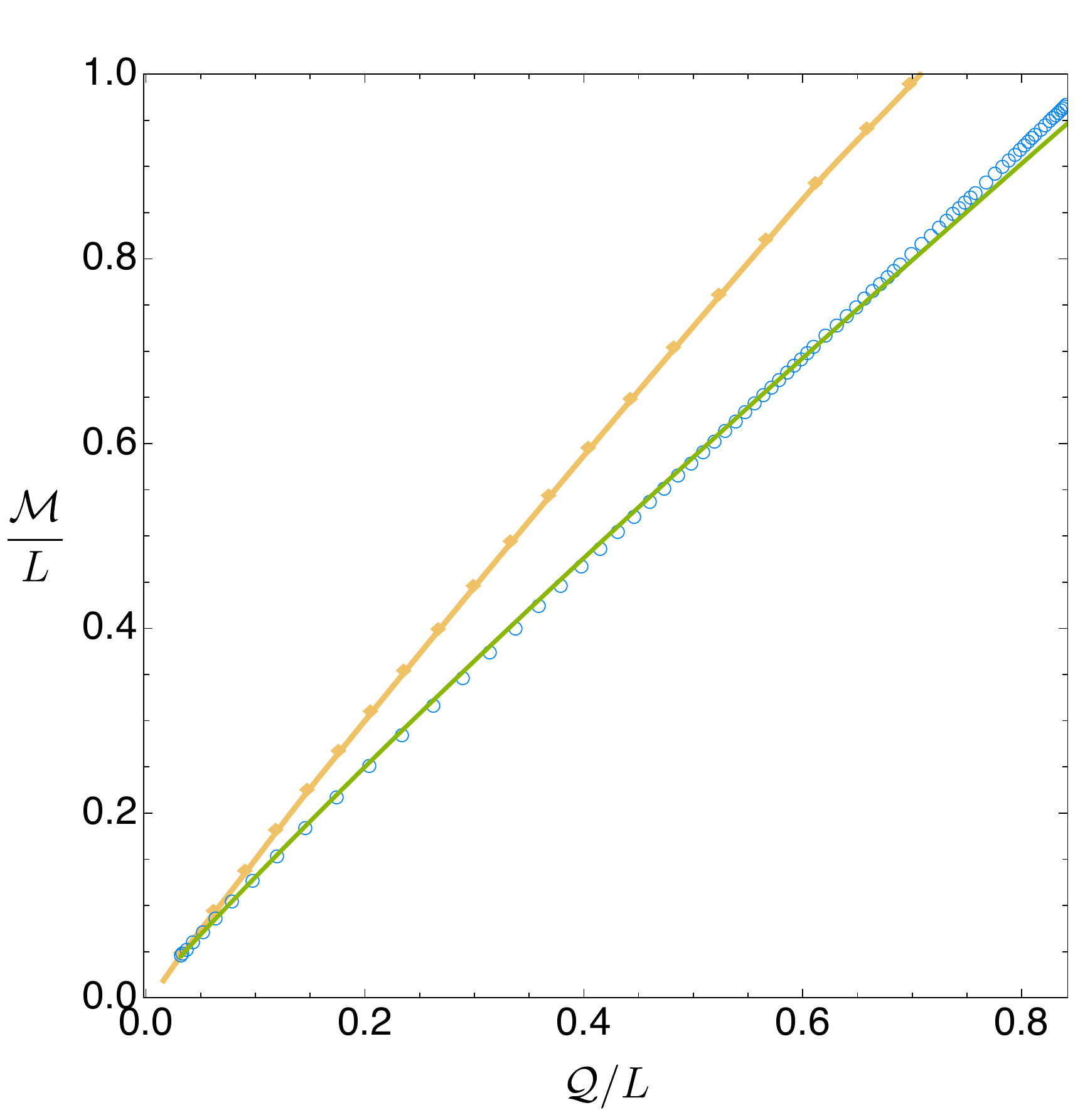}
  }
  \caption{Comparing the exact numerical results (blue circles) with the perturbative analytical predictions \eqref{ThermoHairyBH1}-\eqref{ThermoHairyBH2} (green curve) for a family of black holes with constant $R_+=0.05$ and $e=2.3$. As expected, the perturbative analysis matches the exact results only for small $R_+$ and small $\epsilon$ (\ie close to the origin and in the neighbourhood of the merger, yellow diamond, line which has $\epsilon=0$). That is to say, for the $R_+=0.05$ family shown, good agreement occurs for small $\cQ$, say $\cQ\lesssim 0.2$.
  }
  \label{FIGe2.3:perturbative}
\end{figure}

This scalar charge $\es$ is also special because at $e=\es$ the slope of the perturbative soliton at the origin vanishes, \ie $\frac{\d \Delta\cM}{\d \cQ}\big|_{\cQ=0}=0$. For $e<\es$, this slope is positive and we always have (some) perturbative solitons with higher quasilocal mass than the extremal RN (for sufficiently small $\cQ$). On the other hand, for  $e>\es$ the slope is always negative, $\frac{\d \Delta\cM}{\d \cQ}\big|_{\cQ=0}<0$, and thus perturbative solitons never coexist with RN BHs.

Ultimately as a consequence of these two properties, two important changes occur in the phase diagram of Fig.~\ref{FIGe2:MassCharge} as we follow its evolution across $\es$ and land on  Fig.~\ref{FIGe2.3:MassCharge}. First, the minimal charge for  instability $-$ that we have been denoting as $\cQ_P$ $-$ approaches zero as $e\to \es^-$ and  $\cQ_P=0$ for $e\geq \es$. This is illustrated in Fig.~\ref{FIGe2.3:MassCharge} for the case $e=2.3$. Second, we find that the hairy BHs (blue circles inside $O\alpha\beta$) now {\it always} terminate on top of the perturbative boson star (black line $O\beta$) as we move down, \eg at constant $\cQ$, from the onset curve $O\alpha$. That is to say, one also has $\cQ_\star = 0$ for $e\geq \es$.  As hairy BHs  approach this perturbative soliton curve, the   Kretschmann curvature at the horizon, the entropy and temperature have the same behaviour as the one observed in Fig.~\ref{FIGe1.85:MassCharge} for BHs terminating along $\star\beta$: $K|_{\mathcal{H}}\to \infty$, $S\to 0$ and $T\to \infty$.

Since for $e\geq \es$ the hairy BHs exist all the way down to $(\cQ,\cM)\to(0,0)$ one might expect that their properties can be captured by a perturbative analysis (to higher orders) around Minkowski space with gauge field in a box. This is indeed the case and such analysis was performed in  \cite{Dias:2018yey}. This is a double expansion perturbation theory with the expansion parameters being the horizon radius $R_+$ and the scalar amplitude $\epsilon$. Of course, here one assumes that $R_+\ll 1$ and $\epsilon \ll 1$ which translates into $\cQ\ll 1$ and $\cM\ll 1$. The analysis of \cite{Dias:2018yey} culminates with explicit expansions for the thermodynamic quantities of the hairy BHs, which are listed in (5.27) of \cite{Dias:2018yey}. In particular, the expansion for the quasilocal mass and charge are:
{\small
\begin{align}\label{ThermoHairyBH1}
{\cal M}/L=&\Bigg[\frac{R_+}{4} \left(\frac{\pi ^2}{e^2}+2\right) +\frac{R_+^2}{32 e^4} \Bigg(\pi ^4 \Big(8 [\text{Ci}(2 \pi )- \gamma-\ln (2 \pi )] +5\Big)+4 \left(e^2+\pi ^2\right) e^2\Bigg)+\mathcal{O}(R_+^3)\Bigg]\nonumber\\
&+\epsilon ^2 \Bigg[\frac{1}{2}+\frac{R_+}{12 \pi  e^2} \Bigg( 9 \pi ^3 \Big[\gamma-\text{Ci}(2 \pi ) -2+\ln (2 \pi )\Big]+\left(8 \pi ^2-3 e^2\right) \Big[ 2 \text{Si}(2 \pi )-\text{Si}(4 \pi )\Big] \Bigg)\nonumber\\
&+\mathcal{O}(R_+^2)\Bigg]+\epsilon ^4 \Bigg[\frac{15 \pi ^2-6 e^2+16 \pi  \big[\text{Si}(4 \pi )-2 \text{Si}(2 \pi )\big]}{24 \pi ^2}+\mathcal{O}(R_+)\Bigg]+\mathcal{O}(\epsilon^6),
\end{align}
\begin{align}\label{ThermoHairyBH2}
{\cal Q}/L=&\Bigg[\frac{\pi  R_+}{2 e}+\frac{R_+^2}{8 e^3} \Bigg(\pi ^3 \Big(2 [\text{Ci}(2 \pi )-\gamma - \ln (2 \pi )]+1\Big)+2 \pi  e^2\Bigg)+\mathcal{O}(R_+^3)\Bigg]+\epsilon ^2 \Bigg[\frac{e}{2 \pi }\nonumber\\
&\hspace{-0.4cm}+\frac{R_+}{12 \pi ^2 e} \Bigg(12 \pi ^3 \Big(\gamma-\text{Ci}(2 \pi ) +\ln (2 \pi )-\frac{7}{4}\Big)+\left(8 \pi ^2-3 e^2\right) \big[2 \text{Si}(2 \pi )-\text{Si}(4 \pi )\big]\Bigg)+\mathcal{O}(R_+^2)\Bigg]\nonumber\\
&\hspace{-0.4cm}-\Bigg[\epsilon ^4\frac{e \left(\left(8 \pi ^2-e^2\right) (2 \text{Si}(2 \pi )-\text{Si}(4 \pi ))+4 \pi  e^2-8 \pi ^3\right)}{8 \pi ^4}+\mathcal{O}(R_+)\Bigg]+\mathcal{O}(\epsilon^6),
\end{align}
}
where $\text{Ci}(x)=-\int_x^{\infty}\frac{\cos z}{z}\mathrm d z$ and $\text{Si}(x)=\int_0^x \frac{\sin z}{z}\mathrm d z$  are the cosine and sine integral functions, respectively, and $\gamma\sim 0.577216$ is Euler's constant. This perturbation scheme assumes that $R_+$ and $\epsilon$ do not have a hierarchy of scales. When $R_+=0$, \eqref{ThermoHairyBH1}-\eqref{ThermoHairyBH2}  reduces to the soliton thermodynamics and, when $\epsilon=0$,  \eqref{ThermoHairyBH1}-\eqref{ThermoHairyBH2}  yields the expansion of the caged RN BH thermodynamics. In \cite{Dias:2018yey} it was argued that  \eqref{ThermoHairyBH1}-\eqref{ThermoHairyBH2} should provide a good approximation (as monitored by the first law) for $\epsilon \lesssim 0.1, R_{+} \lesssim 0.1$. Now that we have the exact (numerical) results for the hairy BHs in all their domain of existence we can use  \eqref{ThermoHairyBH1}-\eqref{ThermoHairyBH2} to further check our numerics while, simultaneously, testing the regime of validity of  \eqref{ThermoHairyBH1}-\eqref{ThermoHairyBH2} .  As an example of this exercise, in Fig.~\ref{FIGe2.3:perturbative} we compare the perturbative prediction  \eqref{ThermoHairyBH1}-\eqref{ThermoHairyBH2} $-$ the green curve$-$ to our exact numerical results (blue circles) for the 1-parameter family of hairy BHs with $R_{+} = 0.05$ parametrized by $\epsilon$ (with $\epsilon =0$ at the merger with RN; the yellow curve). As expected, we observe good agreement for $\cQ \lesssim 0.2$, say. Of course, the fact that the perturbative analysis does not differ much from the exact results for higher values of $\cQ$ is to be seen as accidental; the perturbative is certainly not valid for such high charges.

\begin{figure}[!htb]
  \centerline{
    \includegraphics[width=.6\textwidth]{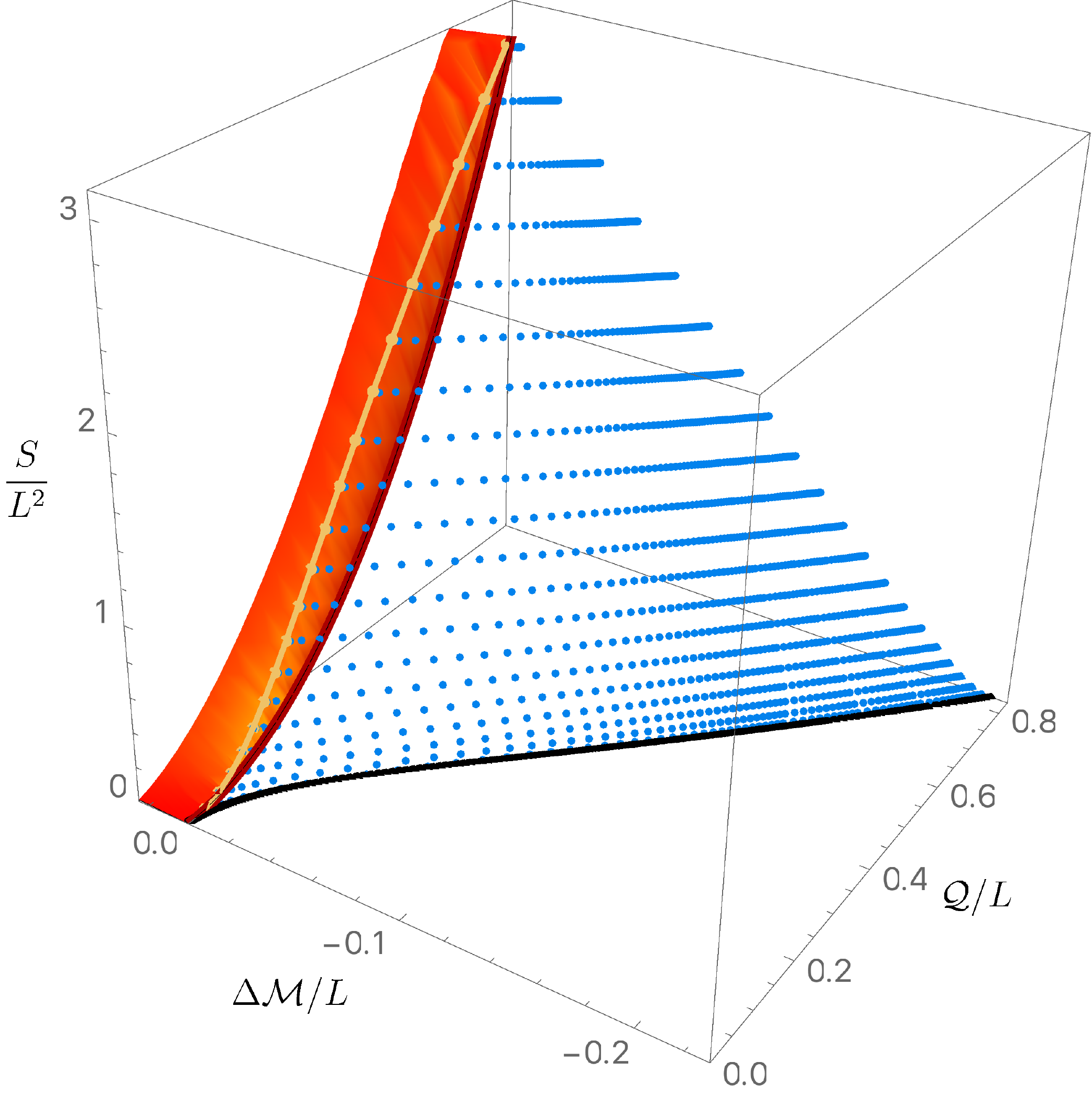}
  }
  \caption{Entropy as a function of the quasilocal charge and mass difference for Einstein theory with a scalar field charge $e=2.3$ ($e>\es$)  in a Minkowski box. When they coexist with RN BHs, for a given $(\cQ,\cM)/L$, hairy BHs always have more entropy than RN, \ie they dominate the microcanonical ensemble. For $e>\es$, the zero entropy limit of the hairy BH is the soliton (black disk curve) in the sense that they have the same  $(\cQ,\cM)/L$ as the soliton (the temperature and horizon curvature diverges).
  }
  \label{FIGe2.3:entropy}
\end{figure}

As in the previous cases, we end our discussion of the $e\geq \es$ case with the plot of Fig. \ref{FIGe2.3:entropy} of the entropy as a function of the charge and mass. The colour coding is the same as in previous cases so it suffices to emphasize that again the hairy BHs (blue circles) are the preferred phase in the microcanonical ensemble. Indeed, in the region between the onset yellow curve and the extremal RN dark red curve with $\Delta\cM=0$ where they coexist with (unstable) RN BHs, hairy BHs always have higher dimensionless entropy for a given charge $\cQ/L$ and mass $\cM/L$. It further follows from the second law, that the unstable RN BHs should evolve in time towards the hairy BH we find with the same $\cQ/L$ and $\cM/L$.

%%%%%%%%%%%%%%%%%%%%%%%%%%%%%%%%%
\section{Conclusions and discussion}\label{sec:Boxstructure}
 
Recapping what we did so far, we integrated  the equations of motion in the domain $R\in [R_+,1]$ subject to regular boundary conditions at horizon and vanishing scalar field at the box.
This is all we need to get the quasilocal phase diagrams of the previous section. But the description of the solution is only complete once we give the full solution all the way up to the asymptotically flat boundary.
 
Studies of scalar fields confined in a Minkowski cavity are already available in the literature: 1) at the linear level \cite{Herdeiro:2013pia,Hod:2013fvl,Degollado:2013bha,Hod:2014tqa,Li:2014gfg,Hod:2016kpm,Fierro:2017fky,Li:2014xxa,Li:2014fna,Li:2015mqa,Li:2015bfa}, 2) within a higher order perturbative analysis of the elliptic problem \cite{Dias:2018zjg,Dias:2018yey}, 3) as a nonlinear elliptic problem (without having asymptotically flat boundary conditions \cite{Dolan:2015dha,Ponglertsakul:2016wae,Ponglertsakul:2016anb} or without matching with the exterior solution \cite{Basu:2016srp}), and 4) as an initial-value problem \cite{Sanchis-Gual:2015lje,Sanchis-Gual:2016tcm,Sanchis-Gual:2016ros}. 
However, with the exception of the perturbative analysis of \cite{Dias:2018yey}, the properties of the ``internal structure" of the cavity required to confine the scalar field and its contribution to the ADM mass and charge that ultimately describe, by Birkhoff's theorem, the exterior RN solution are not discussed. 

However, having the interior solution, we can compute the Lanczos-Darmois-Israel surface stress tensor  \eqref{eq:inducedT} that describes the energy-momentum of the box $\Sigma$. 
We  impose the three Israel junction conditions  \eqref{eq:Israeljunction1}-\eqref{eq:Israeljunction3}  on the gravitoelectric fields on the surface layer $\Sigma$. The fields $f,A_t,\phi$ are then continuous across $\Sigma$ and the component of the electric field orthogonal to $\Sigma$ is also $C^0$ across $\Sigma$. The latter  means that we can confine the charged scalar condensate without needing to have a surface electric charge density on $\Sigma$. The three conditions  \eqref{eq:Israeljunction1}-\eqref{eq:Israeljunction3} permit us to match the interior and exterior solutions, \ie they fix the parameters $M_0,c_A,\rho$ in  \eqref{BCinfinity} as a function of the reparametrization freedom parameter  $N$  introduced in \eqref{SigmaOut}:
\begin{equation}\label{matching}
M_0=\frac{1}{N^2}\left(1-\frac{A_t'(1)^2}{2}\right)-1,\qquad c_A=\frac{A_t'(1)+A_1(1)}{N}\,, \qquad \rho =-\frac{A_t'(1)}{N}\,.
\end{equation} 
Effectively, these conditions fix the exterior RN solution as a function of the interior solution and of the box's energy-momentum. Not surprisingly, we have a 1-parameter freedom ($N$) to choose the box content that is able to confine the scalar condensate. Several cavities can do the job.

Ideally, we would fix  $N$ requiring that the gravitational field is not only $C^0$ but also  differentiable across the box. That is, the fourth junction condition \eqref{eq:Israeljunction4} would also be obeyed and thus the extrinsic curvature
\begin{equation}\label{Kab}
K^{t}_{\phantom{t}t}=-\frac{f'(R)}{2f(R)\sqrt{g(R)}}\,,
 \qquad K^{i}_{\phantom{i}j}=\frac{1}{R\sqrt{g}}\,\delta^{i}_{\phantom{i}j} \,,\quad (i,j)=(\theta,\varphi)\,,\\
\end{equation} 
 would also be continuous across the box. But, except when $\phi(R)=0$, no choice of $N$  allows us to simultaneously set $[K_t^t]=0$ and $[K_i^i]=0$.
 All we can do is to fix $N$ requiring that $[K_t^t]=0$ (at the expense of having $[K_i^i]\neq 0$) or vice-versa, or any other combination.

A choice of $N$ fixes the energy density and pressure of the box since its stress tensor can be written in the perfect fluid form, ${\cal S}_{(a)(b)}= \mathcal{E} u_{(a)}u_{(b)}+ \mathcal{P}(h_{(a)(b)}+u_{(a)}u_{(b)})$, with $u=f^{-1/2}\partial_t$ and local energy density $ \mathcal{E}$ and pressure $ \mathcal{P}$ given by
\begin{equation}\label{densitiesbox}
\mathcal{E}=-S^t_{\ t}\,,\qquad \mathcal{P}=S^x_{\ x}=S^{\phi}_{\ \phi}\,.
 \end{equation}
We are further constrained to make a choice such that  relevant energy conditions are obeyed. Ultimately, failing these would mean that we cannot build the necessary box with the available materials. Different versions of these energy conditions read $(i=\theta,\varphi)$ \cite{Wald:106274}:
\begin{eqnarray}\label{energyconditions}
\hbox{Weak energy condition:} && {\cal E}\geq 0 \quad \land \quad {\cal E}+{\cal P}_{i}\geq 0  \,;\\
\hbox{Strong energy condition:} & &{\cal E}+{\cal P}_{i}\geq 0 \quad \land \quad {\cal E}+\sum_{i=1}^2 {\cal P}_{i}\geq 0\,;\\
\hbox{Null energy condition:} & &{\cal E}+{\cal P}_{i}\geq 0\,;\\
\hbox{Dominant energy condition:}& & {\cal E}+|{\cal P}_{i}|\geq 0\,.
\end{eqnarray}

We have experimented with different choices of $N$ and found that  are many selections that indeed satisfy \eqref{energyconditions} (and equally many others that don't). An example of this exercise is given in \cite{Dias:2021acy} for the boson star case. Given that there seems to be no preferred choice, we do not do a further aleatory  illustration here. Instead, we approach the problem from an experimental perspective. That is to say, in practice, we are given a cavity (that obeys the energy conditions or else it could not have been built with available materials). In principle, we can identify its stress tensor and hence compute $N$. We then insert this  into the Israel matching conditions \eqref{matching} to find the exact RN exterior solution and, in particular, the asymptotic ADM charges.
We end up with an asymptotically flat static black hole solution (or boson star \cite{Dias:2021acy}) that is regular everywhere except across the box (where the extrinsic curvature  has a discontinuity) and that describes confined scalar radiation floating above the horizon and in thermodynamic equilibrium with it. That is to say, we have established that the configuration originally envisioned (in the rotating case) by Zel'dovich \cite{Zeldovich:1971},  Press-Teukolsky  \cite{Press:1972zz} and  \cite{Hawking:1976de,Gibbons:1976pt,Hawking:1979ig,Page:1981,Hawking:1982dh,Braden:1990hw} using linear considerations indeed exists as a non-linear equilibrium solution of the Einstein-Maxwell-scalar equations. And we further established that this is the thermal phase that dominates the microcanonical ensemble. In an ongoing programme, we are extending the current analysis to the rotating BH bomb system.

The hairy BHs we find are stable to the RN instabilities (since they merge with RN precisely at the onset of the original instability; see also \cite{Dolan:2015dha,Ponglertsakul:2016wae}) and have higher entropy than the RN BHs.  It follows from this and the second law of thermodynamics that the charged black hole bomb does not need to break apart: in a time evolution at fixed energy and charge, the unstable RN BH should simply evolve towards the hairy BH we find.  It would be interesting to confirm this doing time evolutions along the lines of those performed in \cite{Sanchis-Gual:2015lje,Sanchis-Gual:2016tcm,Sanchis-Gual:2016ros} in the precise setup we described.

Not less interestingly, Minkowski space in a box (no horizon) with a scalar perturbation is itself non-linearly unstable to the formation of a BH for arbitrarily small amplitude \cite{Maliborski:2013jca}, very much alike the pure global AdS spacetime  \cite{Dafermos2006,DafermosHolzegel2006,Bizon:2011gg,Dias:2011ss,Dias:2012tq,Buchel:2012uh,Buchel:2013uba,Choptuik:2017cyd}. The weakly turbulent phenomenon is responsible for this instability \cite{Dafermos2006,DafermosHolzegel2006,Bizon:2011gg,Dias:2011ss,Dias:2012tq,Dias:2016ewl,Rostworowski:2016isb,Dias:2017tjg}. It would be interesting to study this non-linear  instability when the scalar field is charged. Unlike the neutral case, for certain windows of charge and energy,  there are now two possible families of BHs and not just the RN one. Therefore a time evolution of the non-linear instability along the lines of \cite{Bizon:2011gg,Buchel:2012uh,Buchel:2013uba,Maliborski:2013jca,Balasubramanian:2014cja,Bizon:2014bya,daSilva:2014zva,Balasubramanian:2015uua,Choptuik:2017cyd} should lead in some cases to gravitational collapse into an RN BH and in others into a hairy BH (there should also be a wide class  of initial data for which no BH should form at all). Accordingly, the evolution details should differ in these different cases.

 %%%%%%%%%%%%%%%%%%%%%%%%%%%%%%%%%%%%%
%%%%%%%%%%%%%%%%%%%%%%%%%%%%%%%%%%%%%
\vskip .5cm
\centerline{\bf Acknowledgements}
\vskip .2cm
We acknowledge Ramon Masachs for contributions on the earlier stages of this project. OD acknowledges financial support from the STFC ``Particle Physics Grants Panel (PPGP) 2016" Grant No.~ST/P000711/1 and the STFC ``Particle Physics Grants Panel (PPGP) 2018" Grant No.~ST/T000775/1. The authors further acknowledge the use of the IRIDIS High Performance Computing Facility, and associated support services at the University of Southampton, in the completion of this work.

%%%%%%%%%%%%%%%%%%%%%%%%%%%%%%%%%%%%%
%%%%%%%%%%%%%%%%%%%%%%%%%%%%%%%%%%%%%

%\begin{appendix}
%%%%%%%%%%%%%%%%%%%%%%%%%%%%%%%%%%%%%%%%%%%%%%%%%%
%%%%%%%%%%%%%%%%%%%%%%%%%%%%%%%%%%%%%%%%%%%%%%%%%%
%\section{xxxxxxx}
%%%%%%%%%%%%%%%%%%%%%%%%%%%%%%%%%%%%%%%%%%%%%%%%%%
%%%%%%%%%%%%%%%%%%%%%%%%%%%%%%%%%%%%%%%%%%%%%%%%%%
%%%%%%%%%%%%%%%%%%%%%%%%%%%%%%%%%%%%%%%%%%%%%%%%%%
%%%%%%%%%%%%%%%%%%%%%%%%%%%%%%%%%%%%%%%%%%%%%%%%%%
%\end{appendix}

%%%%%%%%%%%%%%%%%%%%%%%%%%%%%%%%%%%%%%%%%%%%%%%%%%
\bibliography{refs_Box}{}

\providecommand{\href}[2]{#2}\begingroup\raggedright\begin{thebibliography}{10}

\bibitem{Zeldovich:1971}
Y.~B. Zel'dovich, \emph{{Generation of Waves by a Rotating Body}}, {\emph{JETP
  Lett.} {\bf 14} (1971) 180}.

\bibitem{Press:1972zz}
W.~H. Press and S.~A. Teukolsky, \emph{{Floating Orbits, Superradiant
  Scattering and the Black-hole Bomb}},
  \href{http://dx.doi.org/10.1038/238211a0}{\emph{Nature} {\bf 238} (1972)
  211--212}.

\bibitem{Cardoso:2004nk}
V.~Cardoso, O.~J.~C. Dias, J.~P.~S. Lemos and S.~Yoshida, \emph{{The Black hole
  bomb and superradiant instabilities}},
  \href{http://dx.doi.org/10.1103/PhysRevD.70.049903,
  10.1103/PhysRevD.70.044039}{\emph{Phys. Rev.} {\bf D70} (2004) 044039},
  [\href{https://arxiv.org/abs/hep-th/0404096}{{\tt hep-th/0404096}}].

\bibitem{Penrose:1969pc}
R.~Penrose, \emph{{Gravitational collapse: The role of general relativity}},
  \href{http://dx.doi.org/10.1023/A:1016578408204}{\emph{Riv. Nuovo Cim.} {\bf
  1} (1969) 252--276}.

\bibitem{PhysRevLett.25.1596}
D.~Christodoulou, \emph{Reversible and irreversible transformations in
  black-hole physics},
  \href{http://dx.doi.org/10.1103/PhysRevLett.25.1596}{\emph{Phys. Rev. Lett.}
  {\bf 25} (Nov, 1970) 1596--1597}.

\bibitem{Dias:2011at}
O.~J.~C. Dias, G.~T. Horowitz and J.~E. Santos, \emph{{Black holes with only
  one Killing field}},
  \href{http://dx.doi.org/10.1007/JHEP07(2011)115}{\emph{JHEP} {\bf 07} (2011)
  115}, [\href{https://arxiv.org/abs/1105.4167}{{\tt 1105.4167}}].

\bibitem{Dias:2015rxy}
O.~J.~C. Dias, J.~E. Santos and B.~Way, \emph{{Black holes with a single
  Killing vector field: black resonators}},
  \href{http://dx.doi.org/10.1007/JHEP12(2015)171}{\emph{JHEP} {\bf 12} (2015)
  171}, [\href{https://arxiv.org/abs/1505.04793}{{\tt 1505.04793}}].

\bibitem{Choptuik:2017cyd}
M.~W. Choptuik, O.~J.~C. Dias, J.~E. Santos and B.~Way, \emph{{Collapse and
  Nonlinear Instability of AdS Space with Angular Momentum}},
  \href{http://dx.doi.org/10.1103/PhysRevLett.119.191104}{\emph{Phys. Rev.
  Lett.} {\bf 119} (2017) 191104},
  [\href{https://arxiv.org/abs/1706.06101}{{\tt 1706.06101}}].

\bibitem{Ishii:2018oms}
T.~Ishii and K.~Murata, \emph{{Black resonators and geons in AdS5}},
  \href{http://dx.doi.org/10.1088/1361-6382/ab1d76}{\emph{Class. Quant. Grav.}
  {\bf 36} (2019) 125011}, [\href{https://arxiv.org/abs/1810.11089}{{\tt
  1810.11089}}].

\bibitem{Ishii:2020muv}
T.~Ishii, K.~Murata, J.~E. Santos and B.~Way, \emph{{Superradiant instability
  of black resonators and geons}},
  \href{http://dx.doi.org/10.1007/JHEP07(2020)206}{\emph{JHEP} {\bf 07} (2020)
  206}, [\href{https://arxiv.org/abs/2005.01201}{{\tt 2005.01201}}].

\bibitem{Ishii:2021xmn}
T.~Ishii, K.~Murata, J.~E. Santos and B.~Way, \emph{{Multioscillating black
  holes}},  \href{https://arxiv.org/abs/2101.06325}{{\tt 2101.06325}}.

\bibitem{Herdeiro:2014goa}
C.~A.~R. Herdeiro and E.~Radu, \emph{{Kerr black holes with scalar hair}},
  \href{http://dx.doi.org/10.1103/PhysRevLett.112.221101}{\emph{Phys. Rev.
  Lett.} {\bf 112} (2014) 221101}, [\href{https://arxiv.org/abs/1403.2757}{{\tt
  1403.2757}}].

\bibitem{Denardo:1973pyo}
G.~Denardo and R.~Ruffini, \emph{{On the energetics of Reissner Nordstr\"om
  geometries}},
  \href{http://dx.doi.org/10.1016/0370-2693(73)90198-6}{\emph{Phys. Lett.} {\bf
  B45} (1973) 259--262}.

\bibitem{Israel:1966rt}
W.~Israel, \emph{{Singular hypersurfaces and thin shells in general
  relativity}}, \href{http://dx.doi.org/10.1007/BF02710419,
  10.1007/BF02712210}{\emph{Nuovo Cim.} {\bf B44S10} (1966) 1}.

\bibitem{Israel404}
W.~Israel, \emph{Discontinuities in spherically symmetric gravitational fields
  and shells of radiation},
  \href{http://dx.doi.org/10.1098/rspa.1958.0252}{\emph{Proceedings of the
  Royal Society of London A: Mathematical, Physical and Engineering Sciences}
  {\bf 248} (1958) 404--414},
  [\href{https://arxiv.org/abs/http://rspa.royalsocietypublishing.org/content/royprsa/248/1254/404.full.pdf}{{\tt
  http://rspa.royalsocietypublishing.org/content/royprsa/248/1254/404.full.pdf}}].

\bibitem{Kuchar:1968}
K.~Kuchar, \emph{{Charged shells in general relativity and their gravitational
  collapse}}, {\emph{Czech. J. Phys. B} {\bf B18} (1968) 435}.

\bibitem{Barrabes:1991ng}
C.~Barrabes and W.~Israel, \emph{{Thin shells in general relativity and
  cosmology: The Lightlike limit}},
  \href{http://dx.doi.org/10.1103/PhysRevD.43.1129}{\emph{Phys. Rev.} {\bf D43}
  (1991) 1129--1142}.

\bibitem{Wald:106274}
R.~M. Wald, \emph{{General relativity}}.
\newblock Chicago Univ. Press, Chicago, IL, 1984.

\bibitem{WILTSHIRE198636}
D.~Wiltshire, \emph{Spherically symmetric solutions of einstein-maxwell theory
  with a gauss-bonnet term},
  \href{http://dx.doi.org/https://doi.org/10.1016/0370-2693(86)90681-7}{\emph{Physics
  Letters B} {\bf 169} (1986) 36 -- 40}.

\bibitem{inverno:1992}
R.~D'Inverno, \emph{{Introducing Einstein's Relativity}}.
\newblock Clarendon Press, 1992.

\bibitem{Dias:2018zjg}
O.~J. Dias and R.~Masachs, \emph{{Charged black hole bombs in a Minkowski
  cavity}}, \href{http://dx.doi.org/10.1088/1361-6382/aad70b}{\emph{Class.
  Quant. Grav.} {\bf 35} (2018) 184001},
  [\href{https://arxiv.org/abs/1801.10176}{{\tt 1801.10176}}].

\bibitem{Breitenlohner:1982jf}
P.~Breitenlohner and D.~Z. Freedman, \emph{{Stability in Gauged Extended
  Supergravity}},
  \href{http://dx.doi.org/10.1016/0003-4916(82)90116-6}{\emph{Annals Phys.}
  {\bf 144} (1982) 249}.

\bibitem{Gubser:2008px}
S.~S. Gubser, \emph{{Breaking an Abelian gauge symmetry near a black hole
  horizon}}, \href{http://dx.doi.org/10.1103/PhysRevD.78.065034}{\emph{Phys.
  Rev.} {\bf D78} (2008) 065034}, [\href{https://arxiv.org/abs/0801.2977}{{\tt
  0801.2977}}].

\bibitem{Dias:2010ma}
O.~J.~C. Dias, R.~Monteiro, H.~S. Reall and J.~E. Santos, \emph{{A Scalar field
  condensation instability of rotating anti-de Sitter black holes}},
  \href{http://dx.doi.org/10.1007/JHEP11(2010)036}{\emph{JHEP} {\bf 11} (2010)
  036}, [\href{https://arxiv.org/abs/1007.3745}{{\tt 1007.3745}}].

\bibitem{Dias:2011tj}
O.~J.~C. Dias, P.~Figueras, S.~Minwalla, P.~Mitra, R.~Monteiro and J.~E.
  Santos, \emph{{Hairy black holes and solitons in global $AdS_5$}},
  \href{http://dx.doi.org/10.1007/JHEP08(2012)117}{\emph{JHEP} {\bf 08} (2012)
  117}, [\href{https://arxiv.org/abs/1112.4447}{{\tt 1112.4447}}].

\bibitem{Dias:2018yey}
O.~J. Dias and R.~Masachs, \emph{{Evading no-hair theorems: hairy black holes
  in a Minkowski box}},
  \href{http://dx.doi.org/10.1103/PhysRevD.97.124030}{\emph{Phys. Rev. D} {\bf
  97} (2018) 124030}, [\href{https://arxiv.org/abs/1802.01603}{{\tt
  1802.01603}}].

\bibitem{Liebling:2012fv}
S.~L. Liebling and C.~Palenzuela, \emph{{Dynamical Boson Stars}},
  \href{http://dx.doi.org/10.12942/lrr-2012-6}{\emph{Living Rev. Rel.} {\bf 20}
  (2017) 5}, [\href{https://arxiv.org/abs/1202.5809}{{\tt 1202.5809}}].

\bibitem{Dias:2021acy}
O.~J.~C. Dias, R.~Masachs and P.~Rodgers, \emph{{Boson stars and solitons
  confined in a Minkowski box}},  \href{https://arxiv.org/abs/2101.01203}{{\tt
  2101.01203}}.

\bibitem{Brown:1992br}
J.~D. Brown and J.~W. York, Jr., \emph{{Quasilocal energy and conserved charges
  derived from the gravitational action}},
  \href{http://dx.doi.org/10.1103/PhysRevD.47.1407}{\emph{Phys. Rev.} {\bf D47}
  (1993) 1407--1419}, [\href{https://arxiv.org/abs/gr-qc/9209012}{{\tt
  gr-qc/9209012}}].

\bibitem{Hawking:1976de}
S.~W. Hawking, \emph{{Black Holes and Thermodynamics}},
  \href{http://dx.doi.org/10.1103/PhysRevD.13.191}{\emph{Phys. Rev. D} {\bf 13}
  (1976) 191--197}.

\bibitem{Gibbons:1976pt}
G.~W. Gibbons and M.~J. Perry, \emph{{Black Holes and Thermal Green's
  Functions}}, \href{http://dx.doi.org/10.1098/rspa.1978.0022}{\emph{Proc. Roy.
  Soc. Lond.} {\bf A358} (1978) 467--494}.

\bibitem{Hawking:1979ig}
S.~W. Hawking and W.~Israel, \emph{{Penrose, R. (Chapter 12) at} {General
  Relativity}: {An Einstein Centenary Survey}}.
\newblock Univ. Pr., Cambridge, UK, 1979.

\bibitem{Page:1981}
D.~Page, \emph{{Black hole formation in a box}},
  \href{http://dx.doi.org/10.1007/BF00759861}{\emph{Gen Relat Gravit.} {\bf 13}
  (1981) 1117--1126}.

\bibitem{Hawking:1982dh}
S.~W. Hawking and D.~N. Page, \emph{{Thermodynamics of Black Holes in anti-De
  Sitter Space}}, \href{http://dx.doi.org/10.1007/BF01208266}{\emph{Commun.
  Math. Phys.} {\bf 87} (1983) 577}.

\bibitem{Braden:1990hw}
H.~W. Braden, J.~D. Brown, B.~F. Whiting and J.~W. York, Jr., \emph{{Charged
  black hole in a grand canonical ensemble}},
  \href{http://dx.doi.org/10.1103/PhysRevD.42.3376}{\emph{Phys. Rev.} {\bf D42}
  (1990) 3376--3385}.

\bibitem{Andrade:2015gja}
T.~Andrade, W.~R. Kelly, D.~Marolf and J.~E. Santos, \emph{{On the stability of
  gravity with Dirichlet walls}},
  \href{http://dx.doi.org/10.1088/0264-9381/32/23/235006}{\emph{Class. Quant.
  Grav.} {\bf 32} (2015) 235006}, [\href{https://arxiv.org/abs/1504.07580}{{\tt
  1504.07580}}].

\bibitem{Basu:2010uz}
P.~Basu, J.~Bhattacharya, S.~Bhattacharyya, R.~Loganayagam, S.~Minwalla and
  V.~Umesh, \emph{{Small Hairy Black Holes in Global AdS Spacetime}},
  \href{http://dx.doi.org/10.1007/JHEP10(2010)045}{\emph{JHEP} {\bf 10} (2010)
  045}, [\href{https://arxiv.org/abs/1003.3232}{{\tt 1003.3232}}].

\bibitem{Bhattacharyya:2010yg}
S.~Bhattacharyya, S.~Minwalla and K.~Papadodimas, \emph{{Small Hairy Black
  Holes in $AdS_5 x S^5$}},
  \href{http://dx.doi.org/10.1007/JHEP11(2011)035}{\emph{JHEP} {\bf 11} (2011)
  035}, [\href{https://arxiv.org/abs/1005.1287}{{\tt 1005.1287}}].

\bibitem{Gentle:2011kv}
S.~A. Gentle, M.~Rangamani and B.~Withers, \emph{{A Soliton Menagerie in AdS}},
  \href{http://dx.doi.org/10.1007/JHEP05(2012)106}{\emph{JHEP} {\bf 05} (2012)
  106}, [\href{https://arxiv.org/abs/1112.3979}{{\tt 1112.3979}}].

\bibitem{Arias:2016aig}
R.~Arias, J.~Mas and A.~Serantes, \emph{{Stability of charged global AdS$_{4}$
  spacetimes}}, \href{http://dx.doi.org/10.1007/JHEP09(2016)024}{\emph{JHEP}
  {\bf 09} (2016) 024}, [\href{https://arxiv.org/abs/1606.00830}{{\tt
  1606.00830}}].

\bibitem{Markeviciute:2016ivy}
J.~Markeviciute and J.~E. Santos, \emph{{Hairy black holes in AdS$_{5}$
  $\times$ S$^{5}$}},
  \href{http://dx.doi.org/10.1007/JHEP06(2016)096}{\emph{JHEP} {\bf 06} (2016)
  096}, [\href{https://arxiv.org/abs/1602.03893}{{\tt 1602.03893}}].

\bibitem{Markeviciute:2018cqs}
J.~Markeviciute, \emph{{Rotating Hairy Black Holes in AdS$_5\times$S$^5$}},
  \href{http://dx.doi.org/10.1007/JHEP03(2019)110}{\emph{JHEP} {\bf 03} (2019)
  110}, [\href{https://arxiv.org/abs/1809.04084}{{\tt 1809.04084}}].

\bibitem{Dias:2016pma}
O.~J. Dias and R.~Masachs, \emph{{Hairy black holes and the endpoint of AdS$_4$
  charged superradiance}},
  \href{http://dx.doi.org/10.1007/JHEP02(2017)128}{\emph{JHEP} {\bf 02} (2017)
  128}, [\href{https://arxiv.org/abs/1610.03496}{{\tt 1610.03496}}].

\bibitem{Arnowitt:1962hi}
R.~L. Arnowitt, S.~Deser and C.~W. Misner, \emph{{The Dynamics of general
  relativity}}, \href{http://dx.doi.org/10.1007/s10714-008-0661-1}{\emph{Gen.
  Rel. Grav.} {\bf 40} (2008) 1997--2027},
  [\href{https://arxiv.org/abs/gr-qc/0405109}{{\tt gr-qc/0405109}}].

\bibitem{Sanchis-Gual:2015lje}
N.~Sanchis-Gual, J.~C. Degollado, P.~J. Montero, J.~A. Font and C.~Herdeiro,
  \emph{{Explosion and Final State of an Unstable Reissner-Nordstr\"om Black
  Hole}}, \href{http://dx.doi.org/10.1103/PhysRevLett.116.141101}{\emph{Phys.
  Rev. Lett.} {\bf 116} (2016) 141101},
  [\href{https://arxiv.org/abs/1512.05358}{{\tt 1512.05358}}].

\bibitem{Sanchis-Gual:2016tcm}
N.~Sanchis-Gual, J.~C. Degollado, C.~Herdeiro, J.~A. Font and P.~J. Montero,
  \emph{{Dynamical formation of a Reissner-Nordstr\"om black hole with scalar
  hair in a cavity}},
  \href{http://dx.doi.org/10.1103/PhysRevD.94.044061}{\emph{Phys. Rev.} {\bf
  D94} (2016) 044061}, [\href{https://arxiv.org/abs/1607.06304}{{\tt
  1607.06304}}].

\bibitem{Sanchis-Gual:2016ros}
N.~Sanchis-Gual, J.~C. Degollado, J.~A. Font, C.~Herdeiro and E.~Radu,
  \emph{{Dynamical formation of a hairy black hole in a cavity from the decay
  of unstable solitons}},
  \href{http://dx.doi.org/10.1088/1361-6382/aa7d1f}{\emph{Class. Quant. Grav.}
  {\bf 34} (2017) 165001}, [\href{https://arxiv.org/abs/1611.02441}{{\tt
  1611.02441}}].

\bibitem{MTW:1973}
C.~W. Misner, K.~S. Thorne and J.~A. Wheeler, \emph{{Gravitation}}.
\newblock W.H.~Freeman and Co., San Francisco, 1973.

\bibitem{Dias:2015nua}
O.~J.~C. Dias, J.~E. Santos and B.~Way, \emph{{Numerical Methods for Finding
  Stationary Gravitational Solutions}},
  \href{http://dx.doi.org/10.1088/0264-9381/33/13/133001}{\emph{Class. Quant.
  Grav.} {\bf 33} (2016) 133001}, [\href{https://arxiv.org/abs/1510.02804}{{\tt
  1510.02804}}].

\bibitem{Herdeiro:2013pia}
C.~A.~R. Herdeiro, J.~C. Degollado and H.~F. R\'unarsson, \emph{{Rapid growth
  of superradiant instabilities for charged black holes in a cavity}},
  \href{http://dx.doi.org/10.1103/PhysRevD.88.063003}{\emph{Phys. Rev.} {\bf
  D88} (2013) 063003}, [\href{https://arxiv.org/abs/1305.5513}{{\tt
  1305.5513}}].

\bibitem{Hod:2013fvl}
S.~Hod, \emph{{Analytic treatment of the charged black-hole-mirror bomb in the
  highly explosive regime}},
  \href{http://dx.doi.org/10.1103/PhysRevD.88.064055}{\emph{Phys. Rev.} {\bf
  D88} (2013) 064055}, [\href{https://arxiv.org/abs/1310.6101}{{\tt
  1310.6101}}].

\bibitem{Degollado:2013bha}
J.~C. Degollado and C.~A.~R. Herdeiro, \emph{{Time evolution of superradiant
  instabilities for charged black holes in a cavity}},
  \href{http://dx.doi.org/10.1103/PhysRevD.89.063005}{\emph{Phys. Rev.} {\bf
  D89} (2014) 063005}, [\href{https://arxiv.org/abs/1312.4579}{{\tt
  1312.4579}}].

\bibitem{Hod:2014tqa}
S.~Hod, \emph{{Resonance spectra of caged black holes}},
  \href{http://dx.doi.org/10.1140/epjc/s10052-014-3137-3}{\emph{Eur. Phys. J.}
  {\bf C74} (2014) 3137}, [\href{https://arxiv.org/abs/1410.4567}{{\tt
  1410.4567}}].

\bibitem{Li:2014gfg}
R.~Li, J.-K. Zhao and Y.-M. Zhang, \emph{{Superradiant Instability of
  D-Dimensional Reissner-Nordstr\"m Black Hole Mirror System}},
  \href{http://dx.doi.org/10.1088/0253-6102/63/5/569}{\emph{Commun. Theor.
  Phys.} {\bf 63} (2015) 569--574},
  [\href{https://arxiv.org/abs/1404.6309}{{\tt 1404.6309}}].

\bibitem{Hod:2016kpm}
S.~Hod, \emph{{The charged black-hole bomb: A lower bound on the charge-to-mass
  ratio of the explosive scalar field}},
  \href{http://dx.doi.org/10.1016/j.physletb.2016.02.009}{\emph{Phys. Lett.}
  {\bf B755} (2016) 177--182}, [\href{https://arxiv.org/abs/1606.00444}{{\tt
  1606.00444}}].

\bibitem{Fierro:2017fky}
O.~Fierro, N.~Grandi and J.~Oliva, \emph{{Superradiance of charged black holes
  in Einstein-Gauss-Bonnet Gravity}},
  \href{https://arxiv.org/abs/1708.06037}{{\tt 1708.06037}}.

\bibitem{Li:2014xxa}
R.~Li and J.~Zhao, \emph{{Superradiant instability of charged scalar field in
  stringy black hole mirror system}},
  \href{http://dx.doi.org/10.1140/epjc/s10052-014-3051-8}{\emph{Eur. Phys. J.}
  {\bf C74} (2014) 3051}, [\href{https://arxiv.org/abs/1403.7279}{{\tt
  1403.7279}}].

\bibitem{Li:2014fna}
R.~Li and J.~Zhao, \emph{{Numerical study of superradiant instability for
  charged stringy black hole mirror system}},
  \href{http://dx.doi.org/10.1016/j.physletb.2014.12.007}{\emph{Phys. Lett.}
  {\bf B740} (2015) 317--321}, [\href{https://arxiv.org/abs/1412.1527}{{\tt
  1412.1527}}].

\bibitem{Li:2015mqa}
R.~Li, Y.~Tian, H.-b. Zhang and J.~Zhao, \emph{{Time domain analysis of
  superradiant instability for the charged stringy black hole mirror system}},
  \href{http://dx.doi.org/10.1016/j.physletb.2015.09.073}{\emph{Phys. Lett.}
  {\bf B750} (2015) 520--527}, [\href{https://arxiv.org/abs/1506.04267}{{\tt
  1506.04267}}].

\bibitem{Li:2015bfa}
R.~Li, J.~Zhao, X.~Wu and Y.~Zhang, \emph{{Scalar clouds in charged stringy
  black hole-mirror system}},
  \href{http://dx.doi.org/10.1140/epjc/s10052-015-3370-4}{\emph{Eur. Phys. J.}
  {\bf C75} (2015) 142}, [\href{https://arxiv.org/abs/1501.07358}{{\tt
  1501.07358}}].

\bibitem{Dolan:2015dha}
S.~R. Dolan, S.~Ponglertsakul and E.~Winstanley, \emph{{Stability of black
  holes in Einstein-charged scalar field theory in a cavity}},
  \href{http://dx.doi.org/10.1103/PhysRevD.92.124047}{\emph{Phys. Rev.} {\bf
  D92} (2015) 124047}, [\href{https://arxiv.org/abs/1507.02156}{{\tt
  1507.02156}}].

\bibitem{Ponglertsakul:2016wae}
S.~Ponglertsakul, E.~Winstanley and S.~R. Dolan, \emph{{Stability of
  gravitating charged-scalar solitons in a cavity}},
  \href{http://dx.doi.org/10.1103/PhysRevD.94.024031}{\emph{Phys. Rev.} {\bf
  D94} (2016) 024031}, [\href{https://arxiv.org/abs/1604.01132}{{\tt
  1604.01132}}].

\bibitem{Ponglertsakul:2016anb}
S.~Ponglertsakul and E.~Winstanley, \emph{{Effect of scalar field mass on
  gravitating charged scalar solitons and black holes in a cavity}},
  \href{http://dx.doi.org/10.1016/j.physletb.2016.10.073}{\emph{Phys. Lett.}
  {\bf B764} (2017) 87--93}, [\href{https://arxiv.org/abs/1610.00135}{{\tt
  1610.00135}}].

\bibitem{Basu:2016srp}
P.~Basu, C.~Krishnan and P.~N.~B. Subramanian, \emph{{Hairy Black Holes in a
  Box}}, \href{http://dx.doi.org/10.1007/JHEP11(2016)041}{\emph{JHEP} {\bf 11}
  (2016) 041}, [\href{https://arxiv.org/abs/1609.01208}{{\tt 1609.01208}}].

\bibitem{Maliborski:2013jca}
M.~Maliborski and A.~Rostworowski, \emph{{Time-Periodic Solutions in an
  Einstein AdS-Massless-Scalar-Field System}},
  \href{http://dx.doi.org/10.1103/PhysRevLett.111.051102}{\emph{Phys. Rev.
  Lett.} {\bf 111} (2013) 051102}, [\href{https://arxiv.org/abs/1303.3186}{{\tt
  1303.3186}}].

\bibitem{Dafermos2006}
M.~Dafermos, \emph{The black hole stability problem},  in \emph{Talk at the
  Newton Institute}, Available at: \href{http://www-
  old.newton.ac.uk/webseminars/pg+ws/2006/gmx/1010/dafermos/}{http://www-
  old.newton.ac.uk/webseminars/pg+ws/2006/gmx/1010/dafermos/}, University of
  Cambridge, 2006.

\bibitem{DafermosHolzegel2006}
M.~Dafermos and G.~Holzegel, \emph{Dynamic instability of solitons in 4+1
  dimensional gravity with negative cosmological constant},  in \emph{Seminar
  at DAMTP}, Available at:
  \href{https://www.dpmms.cam.ac.uk/~md384/ADSinstability.pdf}{https://www.dpmms.cam.ac.uk/$\sim$md384/ADSinstability.pdf},
  University of Cambridge, 2006.

\bibitem{Bizon:2011gg}
P.~Bizon and A.~Rostworowski, \emph{{On weakly turbulent instability of anti-de
  Sitter space}},
  \href{http://dx.doi.org/10.1103/PhysRevLett.107.031102}{\emph{Phys. Rev.
  Lett.} {\bf 107} (2011) 031102}, [\href{https://arxiv.org/abs/1104.3702}{{\tt
  1104.3702}}].

\bibitem{Dias:2011ss}
O.~J.~C. Dias, G.~T. Horowitz and J.~E. Santos, \emph{{Gravitational Turbulent
  Instability of Anti-de Sitter Space}},
  \href{http://dx.doi.org/10.1088/0264-9381/29/19/194002}{\emph{Class. Quant.
  Grav.} {\bf 29} (2012) 194002}, [\href{https://arxiv.org/abs/1109.1825}{{\tt
  1109.1825}}].

\bibitem{Dias:2012tq}
O.~J.~C. Dias, G.~T. Horowitz, D.~Marolf and J.~E. Santos, \emph{{On the
  Nonlinear Stability of Asymptotically Anti-de Sitter Solutions}},
  \href{http://dx.doi.org/10.1088/0264-9381/29/23/235019}{\emph{Class. Quant.
  Grav.} {\bf 29} (2012) 235019}, [\href{https://arxiv.org/abs/1208.5772}{{\tt
  1208.5772}}].

\bibitem{Buchel:2012uh}
A.~Buchel, L.~Lehner and S.~L. Liebling, \emph{{Scalar Collapse in AdS}},
  \href{http://dx.doi.org/10.1103/PhysRevD.86.123011}{\emph{Phys. Rev.} {\bf
  D86} (2012) 123011}, [\href{https://arxiv.org/abs/1210.0890}{{\tt
  1210.0890}}].

\bibitem{Buchel:2013uba}
A.~Buchel, S.~L. Liebling and L.~Lehner, \emph{{Boson stars in AdS spacetime}},
  \href{http://dx.doi.org/10.1103/PhysRevD.87.123006}{\emph{Phys. Rev.} {\bf
  D87} (2013) 123006}, [\href{https://arxiv.org/abs/1304.4166}{{\tt
  1304.4166}}].

\bibitem{Dias:2016ewl}
O.~J.~C. Dias and J.~E. Santos, \emph{{AdS nonlinear instability: moving beyond
  spherical symmetry}},  \href{https://arxiv.org/abs/1602.03890}{{\tt
  1602.03890}}.

\bibitem{Rostworowski:2016isb}
A.~Rostworowski, \emph{{Comment on "AdS nonlinear instability: moving beyond
  spherical symmetry" [Class. Quantum Grav. 33 23LT01 (2016)]}},
  \href{https://arxiv.org/abs/1612.00042}{{\tt 1612.00042}}.

\bibitem{Dias:2017tjg}
O.~J.~C. Dias and J.~E. Santos, \emph{{AdS nonlinear instability: breaking
  spherical and axial symmetries}},
  \href{http://dx.doi.org/10.1088/1361-6382/aad514}{\emph{Class. Quant. Grav.}
  {\bf 35} (2018) 185006}, [\href{https://arxiv.org/abs/1705.03065}{{\tt
  1705.03065}}].

\bibitem{Balasubramanian:2014cja}
V.~Balasubramanian, A.~Buchel, S.~R. Green, L.~Lehner and S.~L. Liebling,
  \emph{{Holographic Thermalization, Stability of Anti-de Sitter Space, and the
  Fermi-Pasta-Ulam Paradox}},
  \href{http://dx.doi.org/10.1103/PhysRevLett.113.071601}{\emph{Phys. Rev.
  Lett.} {\bf 113} (2014) 071601}, [\href{https://arxiv.org/abs/1403.6471}{{\tt
  1403.6471}}].

\bibitem{Bizon:2014bya}
P.~Bizon and A.~Rostworowski, \emph{{Comment on Holographic Thermalization,
  Stability of Anti-de Sitter Space, and the Fermi-Pasta-Ulam Paradox?}},
  \href{http://dx.doi.org/10.1103/PhysRevLett.115.049101}{\emph{Phys. Rev.
  Lett.} {\bf 115} (2015) 049101}, [\href{https://arxiv.org/abs/1410.2631}{{\tt
  1410.2631}}].

\bibitem{daSilva:2014zva}
E.~da~Silva, E.~Lopez, J.~Mas and A.~Serantes, \emph{{Collapse and Revival in
  Holographic Quenches}},
  \href{http://dx.doi.org/10.1007/JHEP04(2015)038}{\emph{JHEP} {\bf 04} (2015)
  038}, [\href{https://arxiv.org/abs/1412.6002}{{\tt 1412.6002}}].

\bibitem{Balasubramanian:2015uua}
V.~Balasubramanian, A.~Buchel, S.~R. Green, L.~Lehner and S.~L. Liebling,
  \emph{{Reply to Comment on Holographic Thermalization, Stability of Anti-de
  Sitter Space, and the Fermi-Pasta-Ulam Paradox?}},
  \href{http://dx.doi.org/10.1103/PhysRevLett.115.049102}{\emph{Phys. Rev.
  Lett.} {\bf 115} (2015) 049102},
  [\href{https://arxiv.org/abs/1506.07907}{{\tt 1506.07907}}].

\end{thebibliography}\endgroup
\bibliographystyle{JHEP}

\end{document}